\documentclass[useAMS,usenatbib]{mn2e}
\input{epsf.tex}
\newcommand{\simgt}{\lower.5ex\hbox{$\; \buildrel > \over \sim \;$}}
\newcommand{\simlt}{\lower.5ex\hbox{$\; \buildrel < \over \sim \;$}}
\newcommand{\xx}{\mbox{\boldmath$x$}}
\newcommand{\vv}{\mbox{\boldmath$v$}}
\newcommand{\re}{r_e}

\newcommand{\ue}{u_e}
\newcommand{\ve}{v_e}

\begin{document}
%
\title{Antonov Problem and Quasi-Equilibrium States in N-body System}
%
%
%
%
%
%
%
%
\date{\today}
\pubyear{2004} \volume{000} \pagerange{1} \onecolumn
%
%
%
\author[A. Taruya  and  M. Sakagami]
{Atsushi Taruya$^1$, Masa-aki Sakagami$^2$\\
$^1$ Research Center for the Early Universe(RESCEU), School of Science, 
University of Tokyo, Tokyo 113-0033, Japan\\
$^2$ Graduate School of Human and Environmental Studies, Kyoto University, 
Kyoto 606-8501, Japan}
\maketitle \label{firstpage}
%
%
%
%
%
%
%
%
%
%
\begin{abstract} 
In this paper, a quantitative characterization for the evolutionary 
sequence of stellar self-gravitating system is investigated, 
focusing on the pre-collapse stage of the long-term 
dynamical evolution. In particular, we consider the quasi-equilibrium 
behaviors of the $N$-body systems in the setup of the so-called 
Antonov problem, i.e., self-gravitating $N$-body system confined in an 
adiabatic wall and try to seek a possible connection with 
thermostatistics of self-gravitating systems. For this purpose, 
a series of long-term $N$-body simulations with various initial conditions 
are performed. We found that a quasi-equilibrium sequence away from the 
thermal equilibrium can be characterized by the 
one-parameter family of the stellar models. Especially, 
the stellar polytropic distribution satisfying the effective 
equation of state $P\propto\rho^{1+1/n}$ provides an excellent approximation 
to the evolutionary sequence of the $N$-body system. 
Based on the numerical results, we discuss 
a link between the quasi-equilibrium state and the 
generalized thermostatistics by means of the non-extensive entropy.  
\end{abstract}
\begin{keywords}
stellar dynamics -- methods: numerical -- celestial mechanics -- 
globular clusters: general
\end{keywords}
%
%
%
%
%
%
%
%
%
%
%
%
%
%
\section{Introduction}
\label{sec: intdo}
%
%
%
%
%
 \subsection{Self-gravitating $N$-body systems and thermostatistics}
\label{subsec:history}

The long-term dynamical evolution of stellar self-gravitating system 
driven by the two-body relaxation is an old problem with rich history 
in astronomy and astrophysics and even in statistical physics. The 
problem, in nature, involves the long-range attractive nature of 
gravity and because of its complexity and peculiarity as well as 
the physical reality, astronomers and statistical physicists have 
attracted much attention on this subject.

Historically, an important consequence from the thermodynamical 
arguments had arisen in the 1960s. \citet{Antonov1962} 
first pointed out that no stable equilibrium state exists 
for a high-dense clusters. 
To prove this, he considered a very idealized situation called 
{\it Antonov problem}, i.e., a 
stellar self-gravitating system confined in a spherical cavity with 
radius $r_e$ (see Fig.\ref{fig: antonov}). 
Then, under keeping the energy $E$ and the mass $M$ fixed, 
the standard statistical mechanical approach based on the 
maximum entropy principle leads to the conclusion that no stable 
equilibrium state exists for the larger radius 
$r_e > \lambda_{\rm crit}\, (-E/GM^2)$, where $\lambda_{\rm crit}$ is 
the critical value 
\citep[for pedagogical reviews]{Padmanabhan1989,Padmanabhan1990}. 
Note that the numerical value of 
$\lambda_{\rm crit}$ depends on the choice of the entropy and 
Antonov obtained $\lambda_{\rm crit}=0.335$ in the case adopting the 
Boltzmann-Gibbs entropy: 
\begin{equation}
  \label{eq: BG_entropy}
  S_{\rm BG}=-\int d^3\xx d^3\vv\,\,f(\xx,\vv)\,\ln\,f(\xx,\vv),
\end{equation}
where $f(\xx,\vv)$ is the one-particle distribution 
function in phase-space. Later, \citet{LW1968} 
re-examined this issue and showed that the unstable 
thermal state found by Antonov 
can be explained by the thermodynamic instability arising from the 
negative specific heat. They especially called the 
instability {\it gravothermal catastrophe}.

Since the 1960s, the gravothermal instability discovered by Antonov 
has become a standard notion of the stellar dynamics and  
the role of the instability has been extensively discussed. 
Thanks to the unprecedented development of the computer facility 
as well as the sophisticated numerical techniques based on the $N$-body 
simulation and the Fokker-Planck calculation,  
our view of the late-time phase of the stellar gravitating system 
has dramatically improved \citep[e.g.,][]{MH1997,HH2003}. 
Among various theoretical developments, one 
important landmark would be a discovery of the 
{\it gravothermal oscillation}, 
which was originally suggested by \citet{SB1983} \citep[see also][]{BS1984} 
and was later confirmed by numerical simulation \citep[]{M1996}. With a 
great advantage of a special purpose hardware, GRAPE 
\citep[e.g.,][]{SCMIEU1990,HM1999},  
the core-collapse triggered by gravothermal instability was 
shown to be terminated by the formation of binary 
as a result of three-body interaction  and the 
oscillatory behaviors of the core motion has been clearly revealed.   
The gravothermal oscillation is thus thought to be one of the  
most fundamental stellar dynamical processes and provides 
a basis to understand the dynamical history of globular clusters 
as real astronomical system.

So far the theory of long-term evolution in stellar self-gravitating 
systems has been developed without any recourse to the 
statistical mechanics or thermodynamics except for 
the seminal works by \citet{Antonov1962} and \citet{LW1968}. 
This is essentially because the issues under consideration 
are, in nature, non-equilibrium problem 
with long-range interaction and usual sense of the thermal equilibrium 
becomes inadequate. Indeed, as \citet{Antonov1962} 
emphasized, no strict meaning of the thermal equilibrium exists 
in the case of an isolated stellar system without boundary. 
Nevertheless, thermostatistical point-of-view is helpful 
in predicting the fate of the unstable system. This would be even 
true in the non-equilibrium situation as long as the collisional 
evolution during the relaxation 
timescales is concerned. In this sense, there might be some 
possibilities to recast the non-equilibrium self-gravitating systems 
in terms of an extended view of the thermostatistics. 
The present paper attempts to address these issues by considering 
an old problem of stellar dynamics from a somewhat new point-of-view. 
Especially, we wish to focus on the thermostatistical characterization 
of non-equilibrium self-gravitating system by means of the 
new thermostatistical formalism and discuss the validity of it.

Indeed, the self-gravitating system may be the best testing ground 
for the recent postulated introduction of a non-extensive generalization 
of Boltzmann-Gibbs statistics, originally proposed by \citet{Tsallis1988}. 
In contrast to the Boltzmann-Gibbs entropy (\ref{eq: BG_entropy}) 
which is properly an extensive variable, the entropy used in 
the 'Tsallis' formalism is pseudo-additive and/or non-extensive one
\footnote{Strictly speaking, equation (\ref{eq: Tsallis entropy}) 
is not exactly the same 
entropy as originally introduced by Tsallis in a sense that the entropy 
(\ref{eq: Tsallis entropy}) is defined in $\mu$-space, not in  
$\Gamma$-space. Nevertheless, we keep to call it Tsallis entropy 
because the thermostatistical framework with (\ref{eq: Tsallis entropy}) 
that is constructed 
self-consistently is quite similar to the 
standard Tsallis formalism \citep[see][]{TS2003b}. } : 
\begin{equation}
         S_q=-\frac{1}{q-1}\,\int d^3\xx d^3\vv \,
 \left[\left\{p(\xx,\vv)\right\}^q-p(\xx,\vv)\right],
 \label{eq: Tsallis entropy}
\end{equation}
where the probability $p(\xx,\vv)$ is not one-particle 
distribution function but a sort of primitive phase-space 
distribution that satisfies the normalization condition:  
 \begin{equation}
   \label{eq: normalization_p(x,v)}
   \int\,\,d^3\xx d^3\vv \,\,p(\xx,\vv) = 1.
 \end{equation}
Note that the Boltzmann-Gibbs entropy 
(\ref{eq: BG_entropy}) is recovered in the limit $q \rightarrow 1$  
and in this limit, the primitive distribution is identified with the 
one-particle distribution.

Because of its non-extensive nature, the non-extensive 'Tsallis' 
formalism might have a potential to deal with 
the long-range systems including the self-gravitating systems. 
Further, it is expected to deal with a variety of interesting 
non-equilibrium problems such as quasi-steady state or quasi-equilibrium 
state far from the thermal equilibrium, to which the 
standard Boltzmann-Gibbs statistics cannot be applied 
\citep[][for comprehensive reviews]{Tsallis1999,AO2001}. 
Nevertheless, most of the works on this subject are concerned with 
construction of a consistent formal framework and little works  
have been known concerning the physical realization of the 
non-extensive statistics. Hence, it seems interesting to address 
the issues on the reality of non-extensive statistics.

 \subsection{Non-extensive generalization of the 
thermostatistical description of self-gravitating system} 
\label{subsec:non-extensive}

Among various issues on the physical reality of the non-extensive 
thermostatistics,  we have recently 
re-examined the classic problem considered by \citet{Antonov1962}
by means of the non-extensive thermostatistics with Tsallis' 
generalized entropy 
\citep[][for a review]{TS2002,TS2003a,TS2003b,TS2003c,ST2004}.  
According to the framework using normalized $q$-expectation values 
by \citet{TMP1998}, 
the one-particle distribution $f(\xx,\vv)$ 
is expressed in terms of the primitive distribution $p(\xx,\vv)$ by 
\citep[For the cases using standard linear-mean values, see][]{PP1993,TS2002}: 
 \begin{equation}
 f(\xx,\vv)= M \frac{\displaystyle \left\{p(\xx,\vv)\right\}^q}
   {\displaystyle \int d^3\xx d^3\vv \,\left\{p(\xx,\vv)\right\}^q}.   
   \label{eq: escort_dist}
 \end{equation}
Then, the standard analysis of statistical mechanics 
based on the Tsallis entropy (\ref{eq: Tsallis entropy}) 
reveals that the extremum state of the entropy 
is reduced to the power-law distribution \citep[][]{TS2003b}: 
\begin{equation}
 f(\xx,\vv)=A
\left[ \Phi_0-\,\,\frac{1}{2} v^2 -\Phi(x)\,\,\right]^{q/(1-q)},
\label{eq:q-exp_func}
\end{equation}
where $A$ and $\Phi_0$ are the numerical constants and $\Phi(x)$ is the 
gravitational potential. In terms of  the specific energy 
of a particle defined by $\epsilon=v^2/2+\Phi(\xx)$, 
this gives $f\propto (\Phi_0-\epsilon)^{q/(1-q)}$.   
Equation (\ref{eq:q-exp_func}) is 
the famous one-parameter family of the stellar 
models referred to as the {\it stellar polytropic distribution} 
\citep[e.g.,][]{BT1987}
and the $q$-parameter is related to the 
polytrope index given by \citep[][]{TS2003b}:
\begin{equation}
n=\frac{1}{2}+\frac{1}{1-q}. 
\label{eq:n-q_relation}
\end{equation}
Similar to the case of the 
Boltzmann-Gibbs entropy (\ref{eq: BG_entropy}),  
the statistical mechanical analysis based on the 
non-extensive entropy (\ref{eq: Tsallis entropy}) also showed that 
there exists the thermodynamic instability, which can be 
consistently explained from the presence of the negative specific 
heat in terms of the non-extensive thermodynamics. Thus, the stellar 
polytropes as extremum states of Tsallis entropy 
are consistently characterized by the non-extensive thermostatistics 
and may play a role of the thermal equilibrium and/or 
quasi-equilibrium like the extremum state 
of the Boltzmann-Gibbs entropy. Nevertheless, concerning their reality, 
it is still unclear 
whether the consistent thermostatistical results really 
imply the existence of quasi-equilibrium state or not. 
In order to get a further insight into this issue, apart from the 
thermostatistical analysis, dynamical and/or kinematical aspects 
of the self-gravitating system should be investigated in details 
using the $N$-body simulation, which we will discuss here. 
Some of the work we report in this paper has already 
appeared in Letter form \citep[]{TS2003c}. This paper 
further addresses new additional $N$-body experiments and an improved 
analysis, together with some careful discussions.

 \subsection{Outline of this paper}
\label{subsec:outline}

In section \ref{sec:Antonov}, 
within a setup of the classic issue considered by 
\citet{Antonov1962} and \citet{LW1968},  
we first address some important properties of the 
quasi-equilibrium distribution characterized by non-extensive entropy 
by comparing with those of the standard Boltzmann-Gibbs statistics. 
We then move to discuss 
the $N$-body simulations. In section \ref{sec: simulation}, 
the $N$-body treatment of the Antonov problem is considered and the 
initial conditions are summarized. Section \ref{sec: results} 
is a main part of this paper, which describes the results of $N$-body 
experiments in details. After checking the $N$-body code 
in section \ref{subsec:isothermal},  we discuss the quasi-equilibrium 
behaviors of the long-term evolution starting with the stellar polytropic  
distribution in section \ref{subsec:polytrope} and try to fit 
their evolutionary sequences by the stellar polytropes, which are shown 
to be a remarkably good fit.  The results are also compared with an 
alternative one-parameter family of 
stellar model, i.e.,  the King model. 
In section \ref{subsec:Tremaine}, 
we attempt to clarify the condition for quasi-equilibrium 
state characterized by the stellar polytropes in other class of initial 
conditions, i.e., the stellar models with cusped density 
profile. Finally, section \ref{sec: conclusion} is devoted to 
the discussion and the conclusion. 

 \begin{figure}
   \begin{center}
     \epsfxsize=10cm
     \epsfbox{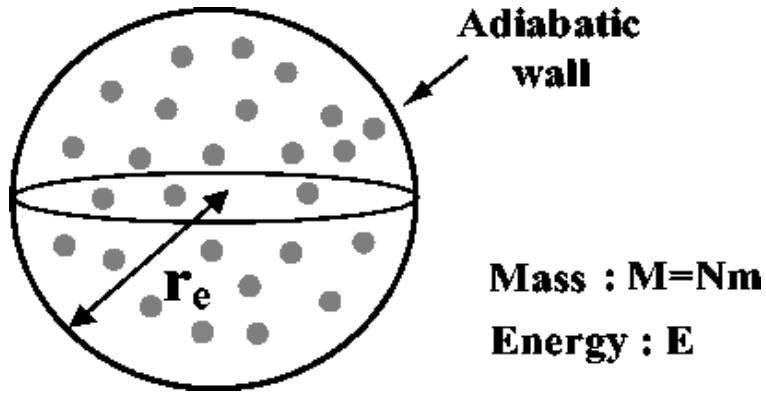}
   \end{center}
     \caption{Setup of the Antonov problem. 
     \label{fig: antonov} }
 \end{figure}
%
%
%
%
%
%
%
%
\section{Antonov problem and 
its non-extensive generalization} 
\label{sec:Antonov}
%
%
%
%
%
In this section, we briefly mention the Antonov problem and the basic 
properties of the equilibrium states characterized by the extensive and 
the non-extensive entropies.

First recall the setup of the so-called Antonov problem. In this problem, 
we consider the many-body particle system confined in an adiabatic wall 
(see Fig.\ref{fig: antonov}). The radius of the wall is given by $r_e$ and 
we assume that the total mass 
$M$ and  the total energy $E$ are kept fixed. All the particles in this 
system interact via Newton gravity and bounce elastically from the wall. 
For further simplification, each particle is assumed to have the same mass 
$m$. Under these conditions, we consider a fully relaxed system and clarify 
the nature of stability or instability for such a system. Note that there are 
two characteristic time-scales in self-gravitating system. One is the 
dynamical time $t_{\rm dyn}$ and the other is the relaxation time 
$t_{\rm relax}$. Usually, for a system containing a large number of 
particles $N\gg1$, these time-scales scale as 
$t_{\rm relax}\simeq (0.1 N/\ln{N})\,t_{\rm dyn}$ \citep[e.g.,][]{BT1987}. 
Accordingly, the equilibrium 
states considered here should be a long-lived state much longer than 
$t_{\rm relax}$ and thus the thermostatistical consideration becomes helpful.

In a language of thermostatistics, a fully relaxed equilibrium state 
corresponds to the state that maximizes the entropy. 
If we adopt the standard Boltzmann-Gibbs entropy (\ref{eq: Tsallis entropy}), 
the equilibrium distribution is obtained by extremizing the 
entropy $S_{\rm BG}$ under keeping the mass $M$ and the energy $E$ fixed:  
 \begin{eqnarray}
 M&=& \int d^3\xx d^3\vv\,\, f(\xx,\vv),
 \label{eq: def_M}
 \\
 E&=& \int d^3\xx d^3\vv\,\,\left[\frac{1}{2}\,v^2+\frac{1}{2}\,
     \Phi(\xx)\right]f(\xx,\vv)~~; 
~~~\Phi(\xx)=-G\int\,d^3\xx' d^3\vv'\,\,\frac{f(\xx',\vv')}{|\xx-\xx'|}.
 \label{eq: def_E}
 \end{eqnarray}
Then the first variation of entropy with respect to $f$,  i.e., 
$\delta[ S_{\rm BG}-\alpha\,\, M-\beta\,E] =0$ with $\alpha$ and $\beta$ being 
Lagrange multipliers yields \citep[][]{Antonov1962,LW1968,Padmanabhan1989}:
 \begin{equation}
 f(\xx,\vv)=\left(\frac{\beta}{2\pi}\right)^{3/2}\,\rho(\xx)\,\,
 e^{-\beta\,v^2/2} 
 \label{eq:isothermal_dist}
 \end{equation}
 with $\rho$ being the mass density of the system defined by 
 \begin{equation}
 \rho(\xx) \equiv  \int d^3\vv\,\,f(\xx,\vv). 
\label{eq: def_rho}
 \end{equation}
Computing the pressure, one can deduce that equation of state 
is indeed isothermal in the case of the isotropic 
velocity distribution:  
\begin{equation}
P(\xx) \equiv  \int d^3\vv\,\,\frac{1}{3} v^2 f(\xx,\vv) ,  
\label{eq: def_P}
\end{equation}
which leads to $P(\xx)=\beta^{-1}\rho(\xx)$.

On the other hand, if one uses the non-extensive entropy 
(\ref{eq: Tsallis entropy}), the variational problem 
with respect to the primitive distribution $p(\xx,\vv)$ not the 
one-particle distribution 
$f(\xx,\vv)$ finally leads to the 
stellar polytropic distribution (\ref{eq:q-exp_func}).  
With a help of the relation (\ref{eq:n-q_relation}),  
this distribution can be rewritten with 
 \begin{eqnarray}
 f(\xx,\vv)&=& \frac{1}{4\sqrt{2}\pi\,\,B(3/2,n-1/2)}\,\,\,
 \frac{\rho(\xx)}{ \{ (n+1)\,P(\xx)/\rho(\xx)\}^{3/2}}  
 \,\,\,\left\{1-\frac{v^2/2}{(n+1)\,P(\xx)/\rho(\xx)}\right\}^{n-3/2}, 
   \label{eq:fedist_poly}
 \end{eqnarray}
where the function $B(a,b)$ represents the beta function. 
Note that we used the following relations to derive the above expression:  
\begin{eqnarray}
\rho(r) &=& 4\sqrt{2}\pi \,B\left(\frac{3}{2},\frac{1}{1-q}\right)
   \,A\,\,[\Phi_0-\Phi(r)]^{1/(1-q)+1/2},
 \label{eq: density}
\\
 P(r) &=& \frac{8\sqrt{2}\pi }{3}\,B\left(\frac{5}{2},\frac{1}{1-q}\right)
   \,A\,\,[\Phi_0-\Phi(r)]^{1/(1-q)+3/2}  
 \label{eq: pressure}
 \end{eqnarray}
with $r$ being the radius $r=|\xx|$. 
These two equations lead to the polytropic equation of state:  
\begin{equation}
   \label{eq: polytrope}
   P(r) = K_n\rho ^{1+1/n}(r). 
\end{equation}
The explicit form of the dimensional constant $K_n$ is 
given in \citet{TS2003b} (Eq.(21) of their paper).

Comparing the distribution function (\ref{eq:fedist_poly}) 
with (\ref{eq:isothermal_dist}), the stellar polytropic distribution can be 
reduced to the isothermal distribution when taking the limit $n\to+\infty$, 
and only in this limit, the isothermal equation of state is recovered. 
Basically, the equilibrium properties of the stellar polytropes are similar 
to those of the isothermal distribution. But, focusing on their thermodynamical 
stabilities, there exists some distinctive features. Before showing this, 
we first note that the one-particle distribution function 
(\ref{eq:fedist_poly}) does not yet completely specify 
the equilibrium configuration, because functional forms of the density and the 
pressure are not determined.   
To determine the equilibrium configuration, we need to 
specify both $\rho(r)$ and $P(r)$, or equivalently the gravitational 
potential $\Phi(r)$. 
This can be achieved by solving the Poisson equation explicitly. Under 
the spherically symmetric configuration, the Poisson equation becomes
\begin{equation}
\label{eq: poisson_eq}  
 \frac{1}{r^2}\frac{d}{dr}\left\{r^2\frac{d\Phi(r)}{dr}\right\}=
  4\pi G \rho(r).
\end{equation}
or the condition of hydrostatic equilibrium: 
\begin{eqnarray}
\frac{dP(r)}{dr}\,=\,\,-\frac{Gm(r)}{r^2}\,\rho(r), \quad
\frac{dm(r)}{dr}\,=\,\,4\pi\rho(r)\,r^2,   
\label{eq: hydro}
\end{eqnarray}
where the quantity $m(r)$ denotes the mass inside the radius $r$. 
Denoting the central density and pressure by 
$\rho_c$ and $P_c$, we then introduce the dimensionless quantities: 
\begin{equation}
\label{eq: dimensionless}
\rho=\rho_c\,\left[\theta(\xi)\right]^n,\,\,\,\,\,\,
r=\left\{\frac{(n+1)P_c}{4\pi G\rho_c^2}\right\}^{1/2}\,\xi, 
\end{equation}
which yields the following ordinary differential equation: 
\begin{equation}
  \theta''+\frac{2}{\xi}\theta'+\theta^n=0,
 \label{eq: Lane-emden_eq}
 \end{equation}
where prime denotes the derivative with respect to $\xi$. 
To obtain the physically relevant solution of (\ref{eq: Lane-emden_eq}), 
we put the following boundary condition:
\begin{equation}
  \theta(0)=1, \,\,\,\,\,\,\,\theta'(0)=0.    
 \label{eq: boundary}
\end{equation}
A family of solutions satisfying (\ref{eq: boundary}) is 
the {\it Emden solution}, which is well-known in the subject of 
stellar structure \citep[e.g., see chapter IV of][]{Chandrasekhar1939}.  
To characterize the equilibrium properties of Emden solutions,  
it is convenient to introduce the following 
set of variables, referred to as homology invariant variables  
\citep[e.g.,][]{Chandrasekhar1939,KW1990}: 
 \begin{eqnarray}
  u &\equiv& \frac{d\ln m(r)}{d\ln r}=
 \frac{4\pi r^3\rho(r)}{m(r)}=-\frac{\xi\theta^n}{\theta'},
 \label{eq: def_u}
 \\
 \nonumber\\
  v &\equiv&  - \frac{d\ln P(r)}{d\ln r}=
 \frac{\rho(r)}{P(r)}\,\,\frac{Gm(r)}{r}
 =-(n+1)\frac{\xi\theta'}{\theta},  
 \label{eq: def_v}
 \end{eqnarray}
which reduce the degree of equation (\ref{eq: Lane-emden_eq}) 
from two to one. One can evaluate the total energy of the confined  
stellar system in terms of the pressure $P_e$ ,the density 
$\rho_e$ at the boundary $\re$ and the total mass $M$:  
 \begin{eqnarray}
   E &=& \frac{3}{2}\,\int_{0}^{\re}dr\,4\pi r^2\,P(r) 
   -\int_0^{\re}\,dt\,\frac{Gm(r)}{r}\,\frac{dm}{dr}
 \nonumber \\
 &=& -\frac{1}{n-5}\,
 \left[\,\frac{3}{2}\left\{ \frac{GM^2}{\re}-(n+1)\frac{MP_e}{\rho_e}\right\}
 +(n-2)\,4\pi\,\re^3\,P_e\,\right], 
 \nonumber 
\end{eqnarray}
by which the dimensionless quantity $\lambda$ can be 
expressed as a function of the homology invariant variables 
at the wall \citep{TS2002,TS2003a,TS2003b}: 
\begin{equation}
\lambda \equiv - \frac{E r_e}{GM^2} = - \frac{1}{n-5}\,
\left[\,\frac{3}{2}\left\{1-\frac{n+1}{\ve}\right\}+(n-2)\frac{\ue}{\ve}
\,\right].
  \label{eq: energy_uv}
\end{equation}

\begin{figure}
  \begin{center}
    \epsfxsize=8.5cm
    \epsfbox{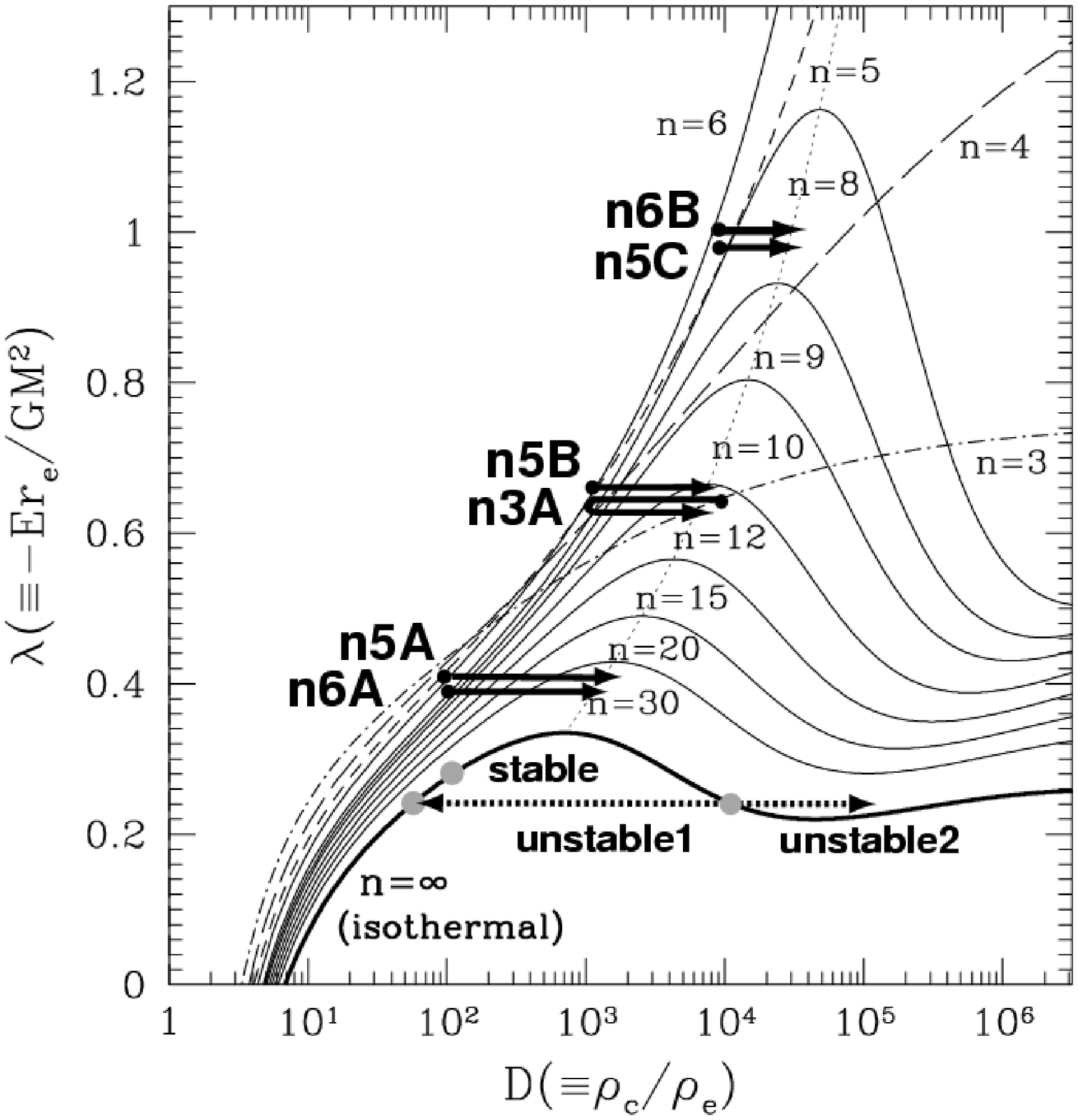}
  \end{center}
    \caption{Energy-density contrast relationship of stellar polytropes
 and evolutionary tracks of the quasi-equilibrium states observed in 
the $N$-body experiments. Each curve represents equilibrium sequence of
 stellar polytrope with a different polytrope index $n$. The thin dotted
 line which traces the turning points for each equilibrium sequence 
 is the marginal stability boundary inferred from the Tsallis entropy 
 $S_q$. The thick arrows show the evolutionary tracks of $N$-body 
 simulation starting with 
 stellar polytropic and/or isothermal distributions  
 (see table \ref{tab:model_iso} and \ref{tab:model_poly}). 
    \label{fig: lambda_poly} }
\end{figure}

In figure \ref{fig: lambda_poly} the dimensionless quantity $\lambda$ 
is plotted as a function of the density contrast $D=\rho_c/\rho_e$, i.e., 
the ratio of the central density to that at the boundary. 
Each curves in $(\lambda,\,D)$-plane corresponds to the equilibrium sequence 
of stellar polytrope with a different value of polytrope index. 
For comparison, we also plot the equilibrium sequence of isothermal distribution 
in thick solid line. Figure \ref{fig: lambda_poly} shows that 
the equilibrium sequences can be classified as two types. One is the 
the equilibrium sequences with indices $n\leq5$. In this case, 
the $\lambda$-curves monotonically increase with density contrast. 
On the other hand, the other type of equilibrium sequences with $n>5$ 
have peaks in each trajectory. At a certain value of the density contrast 
$D_{\rm crit}$, 
the trajectory reaches a maximum $\lambda_{\rm crit}$ 
and beyond this, the monotonicity 
is lost. This is even true in the limit $n\to+\infty$ and the maximum 
value of $\lambda$ and the critical value of $D$ can be read off 
\citep[][]{Antonov1962,LW1968,Padmanabhan1989}: 
\begin{equation}
    \lambda_{\rm crit}=0.335,\quad
    D_{\rm crit}=709.
\end{equation}
In the case of Boltzmann-Gibbs entropy, it had been already known that, 
along the curve $\lambda(D)$ derived from the condition 
$\delta S_{\rm BG} = 0$, all states with $D >  D_{\rm crit}$ are 
unstable. This can be proved by the turning-point 
analysis for a linear series of equilibria 
\citep[][]{LW1968,Padmanabhan1989,Katz1978,Katz1979}. 
Also, the same results can be obtained by explicitly evaluating 
the eigenmodes for the second variation 
of the entropy $\delta^2 S_{\rm BG}$
\citep[][]{Padmanabhan1990,Padmanabhan1989}. The absence of stable thermal 
equilibria in this regime clearly indicates the instability, referred to as the 
{\it the gravothermal catastrophe}.

In similar manner to the isothermal distribution, 
it has been recently shown 
that the stellar polytrope confined in an adiabatic wall 
exhibits the gravothermal instability in the case of the polytropic index 
$n>5$. The evaluation of eigenvalues for 
the second variation of the Tsallis entropy $\delta^2 S_q $ also derives   
the same results as the above turning-point analysis in terms of 
the $\lambda$-curve \citep{TS2002,TS2003a,TS2003b}. 
Note that the gravothermal instability in the isothermal case is 
heuristically explained by the presence of negative specific heat. That 
is, the co-existence of the regions with negative and positive specific heat 
leads to the violation of thermal balance and finally causes the 
catastrophe temperature growth. This can be achieved when the radius of the 
adiabatic wall or equivalently the density contrast becomes 
sufficiently larger so that the effect of self-gravity is 
important at the inner dense region 
and becomes negligible at the outer sparse region. 
Although this intuitive explanation heavily 
relies on the standard notion of extensive thermostatistics, it is emphasized 
that the  
same interpretation can be possible in the case of the stellar polytropic 
distributions \citep[][]{TS2003a}. 
Therefore, despite some distinctive features, the stellar polytropes 
are consistently characterized as equilibrium distribution 
by means of the non-extensive formalism and their thermodynamic 
properties are similar to those of isothermal distribution.

Finally,  the above arguments indicate that 
the stellar polytropic distribution might be regarded as 
a sort of thermal equilibrium, since it is the extrema  
of the Tsallis entropy. However, polytropic equation of state for 
the stellar polytropes clearly shows that 
the one-dimensional velocity dispersion defined by 
\begin{equation}
\sigma_{\rm v,1D}^2 (r) =\frac{P(r)}{\rho(r)}
\end{equation}
manifestly depends on the radius $r$ (see Eq.[\ref{eq: polytrope}]). 
Only in the isothermal case $n \rightarrow \infty$, 
$\sigma_{\rm v,1D}$ is kept spatially constant. 
Thus, it is expected that a gradient of the velocity dispersion is 
relaxed on timescales of two-body encounter. 
This means that the stellar polytrope is no longer the equilibrium but 
quasi-equilibrium state. At present, it might be also possible to give 
another interpretation that the 
stellar polytropic states with $n<5$ and with $n>5$ and $D<D_{\rm crit}$ 
are dynamical equilibrium states 
that are even stable with respect to a non-linear perturbation 
\citep[][]{Chavanis2003,CS2004}. 
In subsequent section, we report the results of  
the $N$-body simulations, which are carried 
out to investigate how the stellar polytrope actually evolves.
%
%
%
%
%
%
%
%
%
%
%
%
%
%
%
%
%
%
%
%
%
%
%
%
%
%
%
%
%
%
%
\section{Numerical simulations}
\label{sec: simulation}
%
%
%
%
%
%
%
We are in a position to discuss the long-term dynamical evolution from 
the non-equilibrium $N$-body system and to explore the reality of stellar 
polytropes as quasi-equilibrium state.  
The $N$-body experiment considered here is the same situation as 
investigated in classic papers. That is, we treat the self-gravitating 
$N$-body system confined in an adiabatic wall of sphere. 
Hereafter, we set the units to $G=M=r_e=1$ without loss of generality.  
According to the Antonov problem, all the particles are assumed to 
have the same mass $m=1/N$, where $N$ 
is the total number of the particles. 
The initial conditions for particle data 
were whole created by a random realization of the stellar models.  
To investigate the evolutionary states of the $N$-body system 
from the thermostatistical point-of-view,  
we deal with the three kinds of the stellar models: 
(i) isothermal distribution (ii) stellar polytropes and (iii) 
the stellar models with cuspy density profile. 
Table \ref{tab:model_iso}, \ref{tab:model_poly} and 
\ref{tab:model_tremaine} summarize the simulation runs, 
which are frequently referred in subsequent section. The 
generation of random initial data was basically followed by 
the rejection method described in \citet{AHW1974} 
\citep[see also Chap.8 of][]{Aarseth2003}. Using 
the analytic form of the mass profile $m(r)$ and the 
one-particle distribution function $f(\epsilon)$ as function of 
specific energy $\epsilon=v^2/2+\Phi(r)$, 
this method generates the random distribution that possesses the 
same phase space structure as in the initial stellar model.

We are specifically concerned with collisional aspects of $N$-body dynamics,   
the timescale of which is much longer than the two-body relaxation time. 
For this purpose, we utilized a special-purpose hardware, GRAPE-6, 
which is especially designed to accelerate the gravitational force 
calculations for collisional $N$-body system \citep[]{MFKN2003}. 
With this implementation, 
we use the fourth-order Hermite integration scheme with 
individual time-step algorithm, 
which is suitably efficient to combine with GRAPE-6 facility.

In our setup,  the adiabatic wall is implemented by the same 
procedure as used by \citet{EFM1997}.
The adiabatic wall reverses the radial components of the velocity 
for particles just located at the wall. Since we use an individual time-step 
algorithm, the positions of the particles are monitored at each fixed 
time interval $\Delta T$, which is chosen to be a synchronized time 
interval in our time-step algorithm. This guarantees that 
the penetration of particles into the wall can be ignored. 
At these times, the radial components of velocities for particles 
outside the wall are reversed if the radial velocity vector is 
directed outward. Namely, 
\begin{equation}
\mbox{\boldmath$v$} \longrightarrow 
\mbox{\boldmath$v$} - \frac{(\mbox{\boldmath$v$}\cdot \mbox{\boldmath$x$})}
{|\mbox{\boldmath$x$}|}\,\,\frac{\mbox{\boldmath$x$}}{|\mbox{\boldmath$x$}|}
\end{equation}
After particles were reversed, we recalculated the force and adjusted 
the time-step of particles.

Note that we did not use the regularization scheme to treat the 
close two-body or multiple encounter \citep[e.g.,][]{Aarseth2003}. 
Because of the adiabatic wall, the standard scheme 
of regularization method is not directly applicable 
and the implementation of the regularization 
keeping the high accuracy requires a considerable amount of programming. 
Since we are interested in a quasi-equilibrium evolution before 
the core-collapse stage, absence of the regularization scheme itself is not 
crucial. Rather, a serious problem might arise from an introduction of the 
potential softening in order to reduce the numerical error. 
In general, the potential softening 
diminishes the close-encounter of each 
particle and it makes the energy exchange 
of the particles inefficient. This would lead to the overestimation of 
the relaxation timescales even before the core-collapse stage. 
In appendix A, 
the significance of the potential softening to the time-scale of 
core-collapse and/or quasi-equilibrium sequences is examined. 
Based on these experiments, 
we adopt the Plummer softened potential ($\phi=1/\sqrt{r^2+\epsilon^2}$) 
and basically set the softening parameter to $\epsilon= 1/N$ 
when estimating the time-scale (Sec.\ref{subsec:polytrope}). Otherwise, 
we set $\epsilon=4/N$ (Sec.\ref{subsec:isothermal} and 
\ref{subsec:Tremaine}).\footnote{Hereafter, the quantity $\epsilon$ is 
interchangeably used to imply the specific energy of a particle 
and the softening length. Readers should not confuse the meanings 
of $\epsilon$. } 
%
%
%
%
%
%
%
%
%
%
%
%
%
%
%
%
%
\section{Results}
\label{sec: results}
%
%
%
In our present situation with units $G=M=r_e=1$, the dynamical time given by 
roughly corresponds to $t_{\rm dyn}=(3\pi/16G\rho_0)^{1/2}\simeq 1$ and thus 
the global relaxation time 
becomes $t_{\rm relax} = 0.1 (N/\ln{N})\,t_{\rm dyn}\sim N/\ln{N}$. 
While this relation gives a crude estimate of the relaxation timescales, 
a more useful convention might be the half-mass relaxation time 
\citep[e.g.,][]{Spitzer1987,BT1987}:
\begin{equation}
t_{\rm rh}= 0.138\,\,\frac{N}{\ln\Lambda}
	\left(\frac{r_{\rm h}^3}{GM}\right)^{1/2}, 
\end{equation}
where the Coulomb logarithm $\ln\Lambda$ is usually taken as 
$\ln\Lambda=\ln(0.4 N)$ or $\ln(0.1N)$ \citep[][]{Spitzer1987,GH1994}. 
In what follows, adopting the latter convention ($\Lambda=0.1N$), 
all the simulation results are presented by rescaling the timescale  with 
the half-mass relaxation time evaluated at an initial time, $t_{\rm rh,i}$. 
In tables  \ref{tab:model_iso}-\ref{tab:model_tremaine}, half-mass radii 
for each initial distribution are evaluated and their numerical values 
are summarized. 
%
%
%
%
%
\subsection{Isothermal distribution}
\label{subsec:isothermal}
%
%
%
%
%
Let us first check our numerical calculations by examining the 
cases with isothermal initial conditions and comparing those 
results with previous work by \citet{EFM1997}, who have investigated the 
gravothermal expansion of the isothermal distribution under the same setup 
as examined in our simulation. Following 
their paper, we examine the three kinds of initial conditions, 
summarized in table \ref{tab:model_iso}.

\begin{figure}
  \begin{center}
    \epsfxsize=5.5cm
    \epsfbox{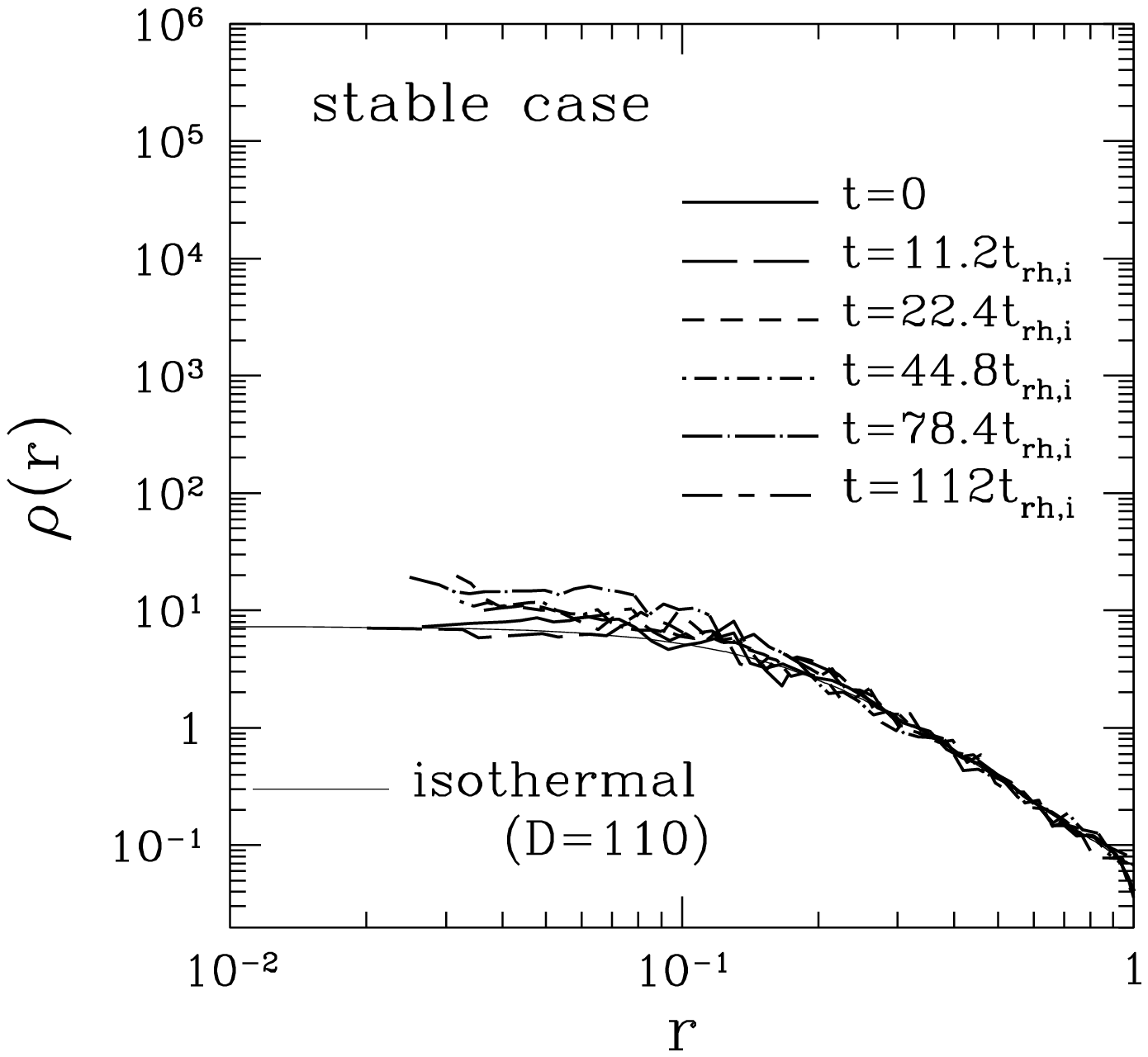}
\hspace*{0.2cm}
    \epsfxsize=5.5cm
   \epsfbox{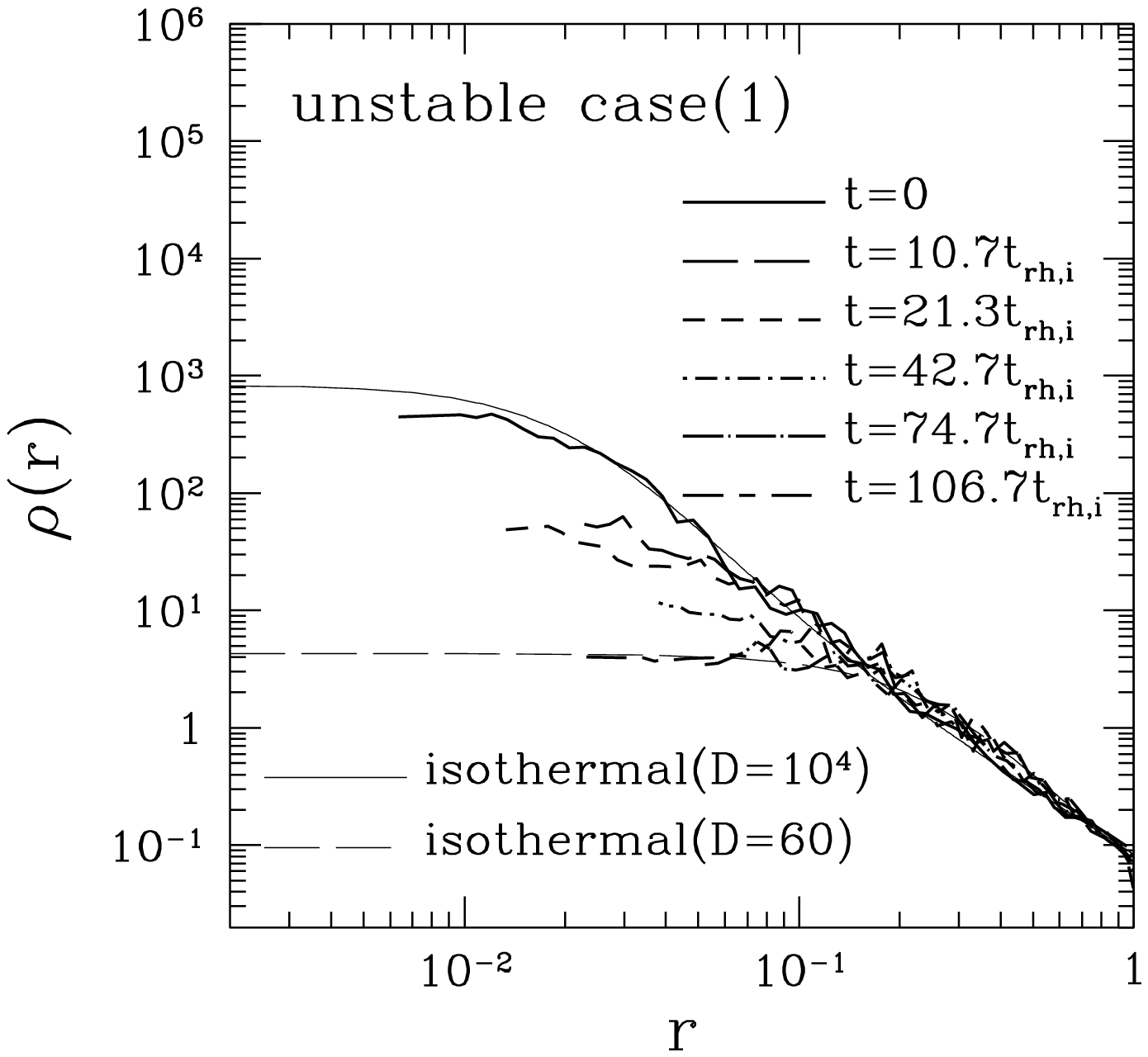}
\hspace*{0.2cm}
    \epsfxsize=5.5cm
    \epsfbox{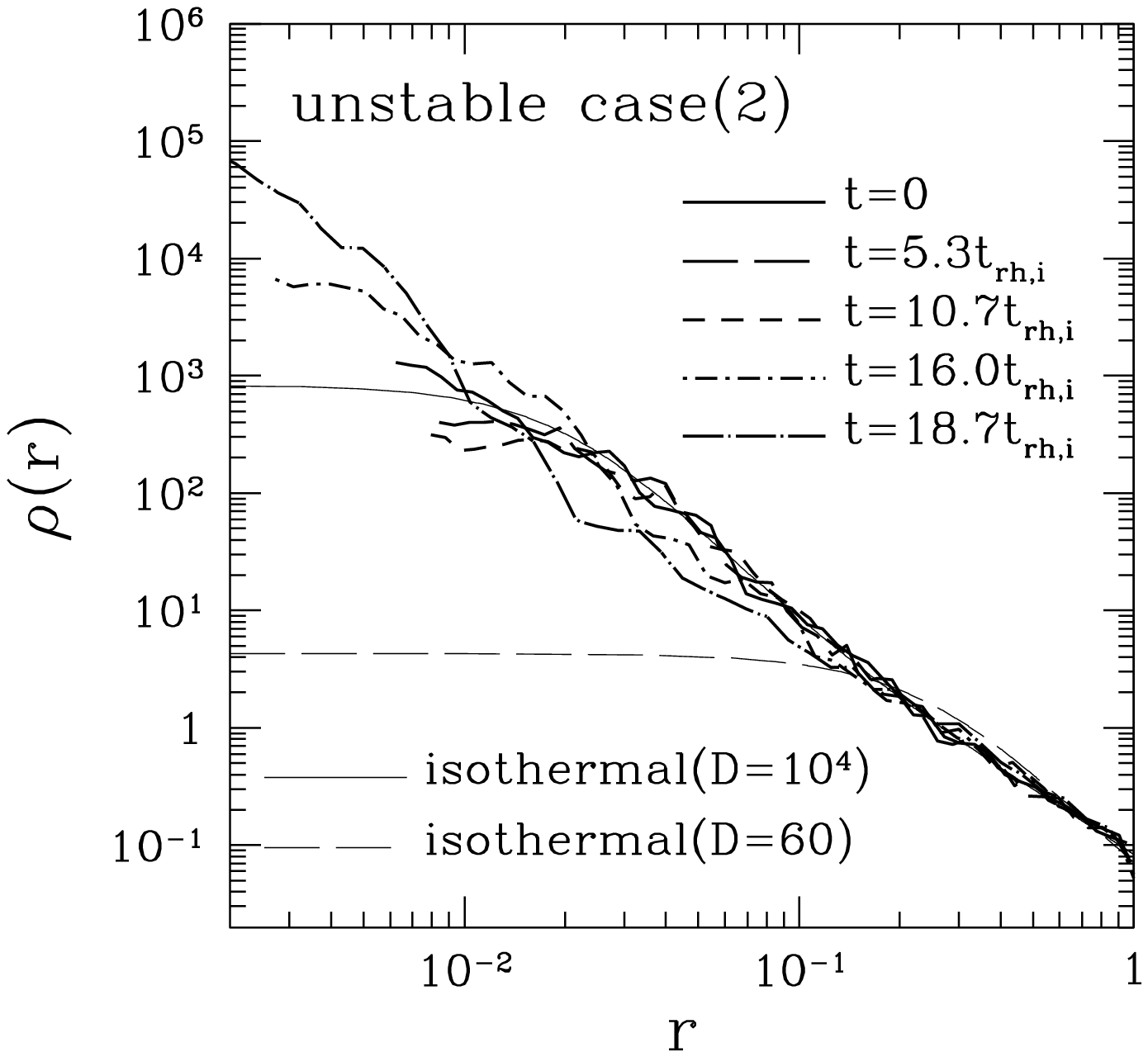}
  \end{center}
    \caption{Snapshots of density profile in the stable case({\it Left}) 
 and the unstable cases({\it Middle} and {\it Right}). In each panel, 
 the thick lines represent the snapshots of simulation results, while the 
 thin lines indicate the ones obtained from the Emden solutions 
 of isothermal distribution. 
    \label{fig:isothermal} }
  \begin{center}
    \epsfxsize=5.5cm
    \epsfbox{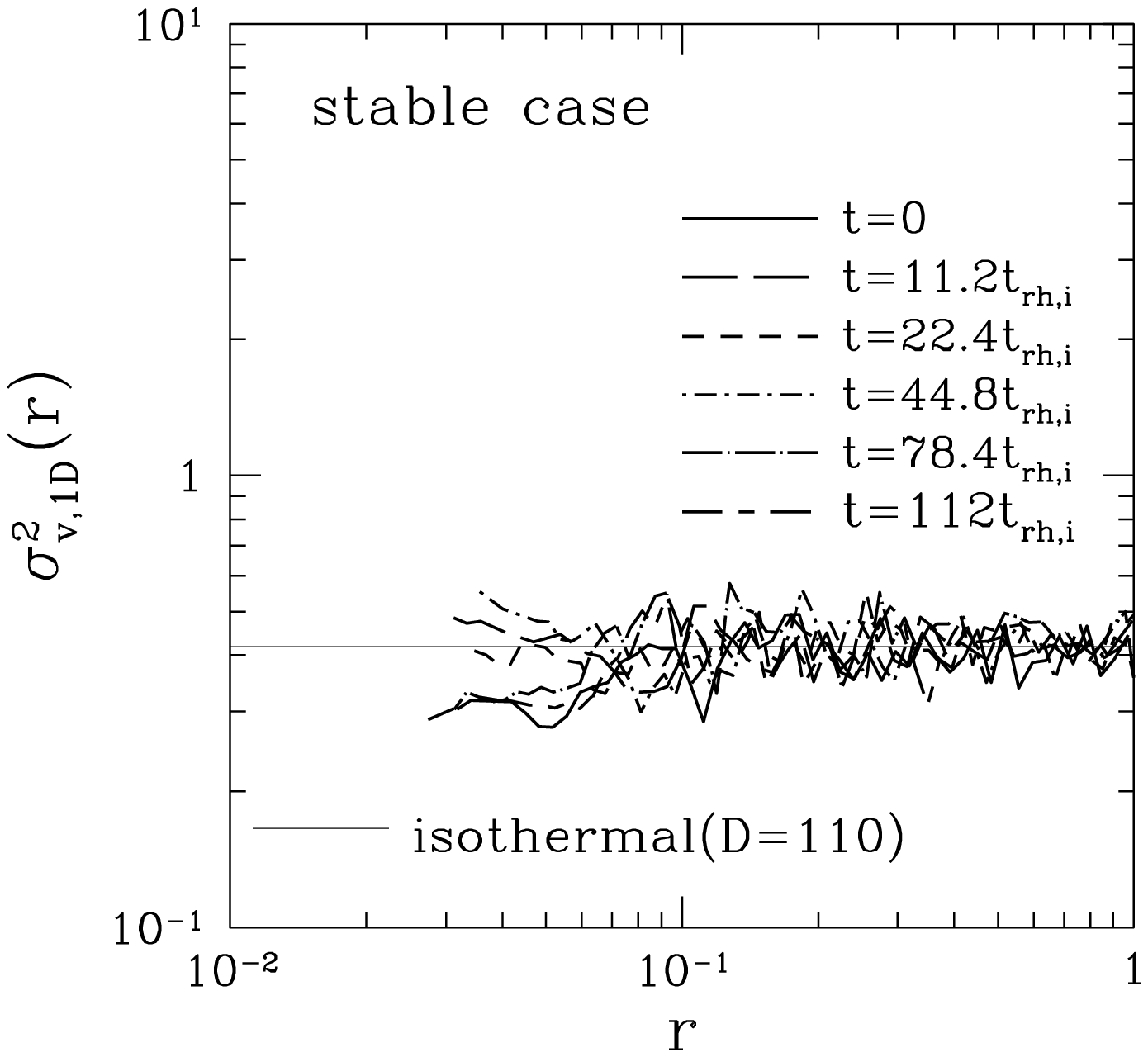}
\hspace*{0.2cm}
    \epsfxsize=5.5cm
    \epsfbox{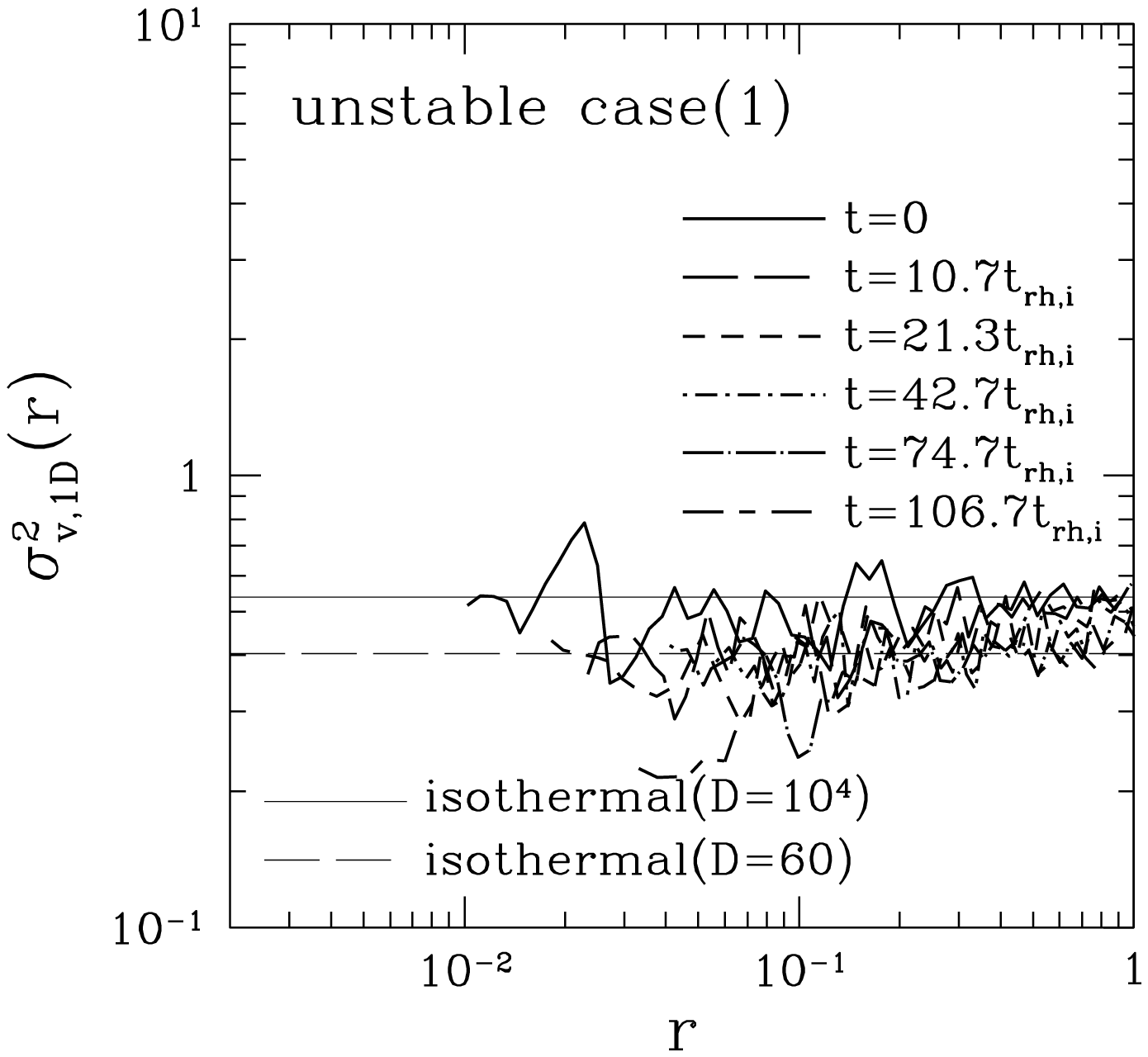}
\hspace*{0.2cm}
    \epsfxsize=5.5cm
    \epsfbox{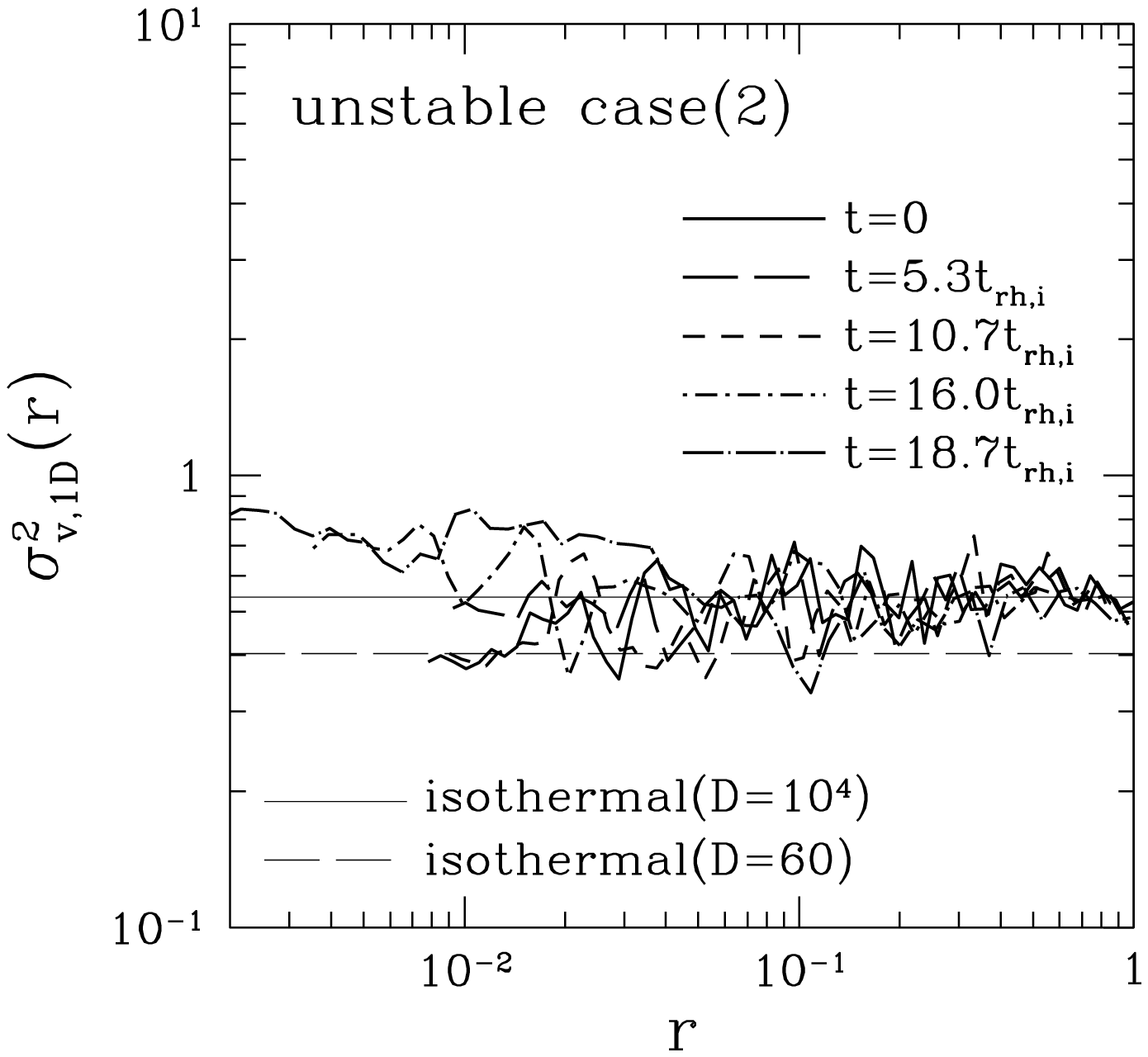}
  \end{center}
    \caption{Snapshots of one-dimensional velocity dispersion profile 
	in the stable case({\it Left}) and the unstable cases({\it
 Middle} and {\it Right}).
    \label{fig:isothermal_sigma} }
\end{figure}

Figure \ref{fig:isothermal} and \ref{fig:isothermal_sigma} respectively 
show the snapshots of evolved 
density profiles and one-dimensional velocity dispersion profiles 
starting from a stable initial configuration({\it left}) 
and unstable initial distributions({\it middle, right}). 
Here, the term, stable or unstable is used according to 
the initial density contrast $D$ smaller or larger than the 
critical value $D_{\rm crit}=709$, 
which is obtained from the thermostatistical 
prediction. The evolutionary path for each run is depicted in figure 
\ref{fig: lambda_poly}. 
In figure \ref{fig:isothermal}, while the evolved density profiles 
starting from the the stable initial condition is, by construction, stable 
and almost remain the same, the fate of the 
unstable cases sensitively depend on the randomness of the initial 
data, leading to the very different endpoints. 
The results depicted in the middle and the right panels 
of figure \ref{fig:isothermal}, both of which were started with the same 
initial parameters, are indeed such examples. 
Discriminating the final fate of the system from 
the initial random distributions seems generally difficult, however, 
the evolved density profiles seen in figure \ref{fig:isothermal} are 
intimately connected with the response of the velocity structure 
shown in figure \ref{fig:isothermal_sigma}. 
Suppose that the velocity dispersion at the central part 
is initially higher than that at the outer part. In this case, 
the energy exchange by the two-body encounter yields the kinetic energy 
transport from the core to the halo. In other words, 
the outward heat flows occur and due to the negative specific heat, 
the inner part gets hotter, leading to the catastrophic growth of 
velocity dispersion. As a consequence, 
the core-collapse takes place. On the other hand, 
when the velocity dispersion at the inner part is relatively lower than 
that at the outer 
part, the inward heat flow conversely occur, which makes the core expand. 
In contrast to the former case, the heat flow can stop 
after balancing the thermal inertia at inner and outer parts. As a result, 
the system finally reaches the stable isothermal configuration. 
Though not clearly seen its signature, the response of the velocity 
dispersion profiles seen in figure \ref{fig:isothermal_sigma} 
are consistent with the evolution of the density profiles. 
Compared the results in the unstable cases 
with the results obtained by \citet{EFM1997},  
the resultant density profiles are similar and  
the evolved timescales for the unstable cases almost agree with each 
other.

As for the stable case, 
a calculation was proceeded up to $t=112t_{\rm rh,i}$, corresponding 
to  $t=2000$ in unit of $N$-body time, which was 
longer calculation time than that of the other runs.   
A closer look at figure \ref{fig:isothermal} shows that 
the central density slightly increases and the system 
seems to become out of equilibrium. 
The total energy of the system was conserved to 0.05\% 
at the end of the calculation, which is relatively larger value 
than the errors in the other runs. Thus the destabilization of 
the core may be attributed to the numerical error in the 
energy conservation. Though the effect of destabilization yields 
a serious problem in estimating the time scale of core-collapse, 
we are specifically concerned with 
the evolutionary sequence before entering the core-collapse phase.  
Hence, at a level of this present accuracy, one can ignore the 
destabilization effect as long as the total energy of the system 
is well conserved. 
%
%
%
%
%
%
%
%
%
%
%
%
\begin{table}
\caption{\label{tab:model_iso}Model parameters of initial condition in
 the case of the isothermal distribution}

\vspace*{0.0cm}

  \begin{center}
{\large
\begin{tabular}{lcccc}
\hline
 run $\#$ & parameters & half-mass radius($r_{\rm h}/r_e$) &$\#$ of particles  & status\\
\hline
 stable    & $D=110$  & 0.4834 & 2k  &  stable \\
 unstable1 & $D=10^4$ & 0.4993 & 2k  &  expansion\\
 unstable2 & $D=10^4$ & 0.4993 & 2k  &  collapse\\
\hline
\end{tabular}
}
  \end{center}
\end{table}
%
%
%
%
%
%
%
%
\subsection{Stellar polytropes}
\label{subsec:polytrope}
%
%
%
%
%
%
Having checked the validity of the numerical calculations, 
we next investigate the non-equilibrium evolution away from 
the isothermal equilibrium state. In this subsection, we 
deal with a class of initial conditions 
characterized by the stellar polytropes. 
Table \ref{tab:model_poly} summarizes the 
model parameters of initial distributions. The 
evolutionary track for each run is depicted in 
the energy-density contrast relation in figure \ref{fig: lambda_poly}.  

Overall behaviors of the simulation results are as follows. 
Before entering the core-collapse stage, most of the system 
experiences the quasi-equilibrium stage, 
in which the distribution function slowly changes in time.  
The evolutionary sequence in the quasi-equilibrium stage  
can be well-approximated by the one-parameter family of stellar 
polytropes with the time-varying polytrope index. 
Apart from some fluctuations, the fitted value of 
the polytrope index, on average, increases monotonically with time, 
implying that the system tends to approach the exponential distribution. 

To see the quasi-equilibrium behavior quantitatively, 
we first show the representative result obtained from the run $n3A$, 
i.e., the stellar polytrope with index $n=3$ and with the initial density 
contrast $D=10^4$. Then we discuss the other runs, $n5A\sim n5D$ and 
$n6A,B$ in section \ref{subsubsec:others}. 
%
%
%
%
%
%
%
%
%
%
\begin{table}
\caption{\label{tab:model_poly}Model parameters and the 
evolutionary states in the cases starting from the stellar polytropes}
  \begin{center}
{\large
\begin{tabular}{lcccccc}
\hline
 run $\#$ & model parameters & half-mass radius$(r_{\rm h}/r_e$) & 
$\#$ of particles  & realizations & fitting to stellar polytrope \\
\hline
 n3A$^1$ & $n=3$, $D=10^4$ & 0.3282 & 2k, 4k, 8k & 2, 1 ,1 & successful \\
 n5A     & $n=5$, $D=10^2$ & 0.4611 & 2k         & 2 & successful \\
 n5B     & $n=5$, $D=10^3$ & 0.3114 & 2k         & 3 & successful \\
 n5C     & $n=5$, $D=10^4$ & 0.2025 & 2k         & 2 & successful \\
 n5D     & $n=5$, $D=10^6$ & 0.08196 & 2k        & 2 & failed \\
 n6A$^2$ & $n=6$, $D=110$  & 0.4531 & 2k         & 1 & successful \\
 n6B$^3$ & $n=6$, $D=10^4$ & 0.1794 & 2k         & 1 & successful \\
\hline
\end{tabular}
}

$^1$ initial distribution corresponding to {\it run A} in \citet{TS2003c} \\
$^2$ initial distribution corresponding to {\it run B1} in \citet{TS2003c} \\
$^3$ initial distribution corresponding to {\it run B2} in \citet{TS2003c} 
  \end{center}
\end{table}
%
%
%
%
%
%
%
%
\subsubsection{run $n3A$}
\label{subsubsec: run_n3A}
\begin{figure}
  \begin{center}
    \epsfxsize=7.3cm
    \epsfbox{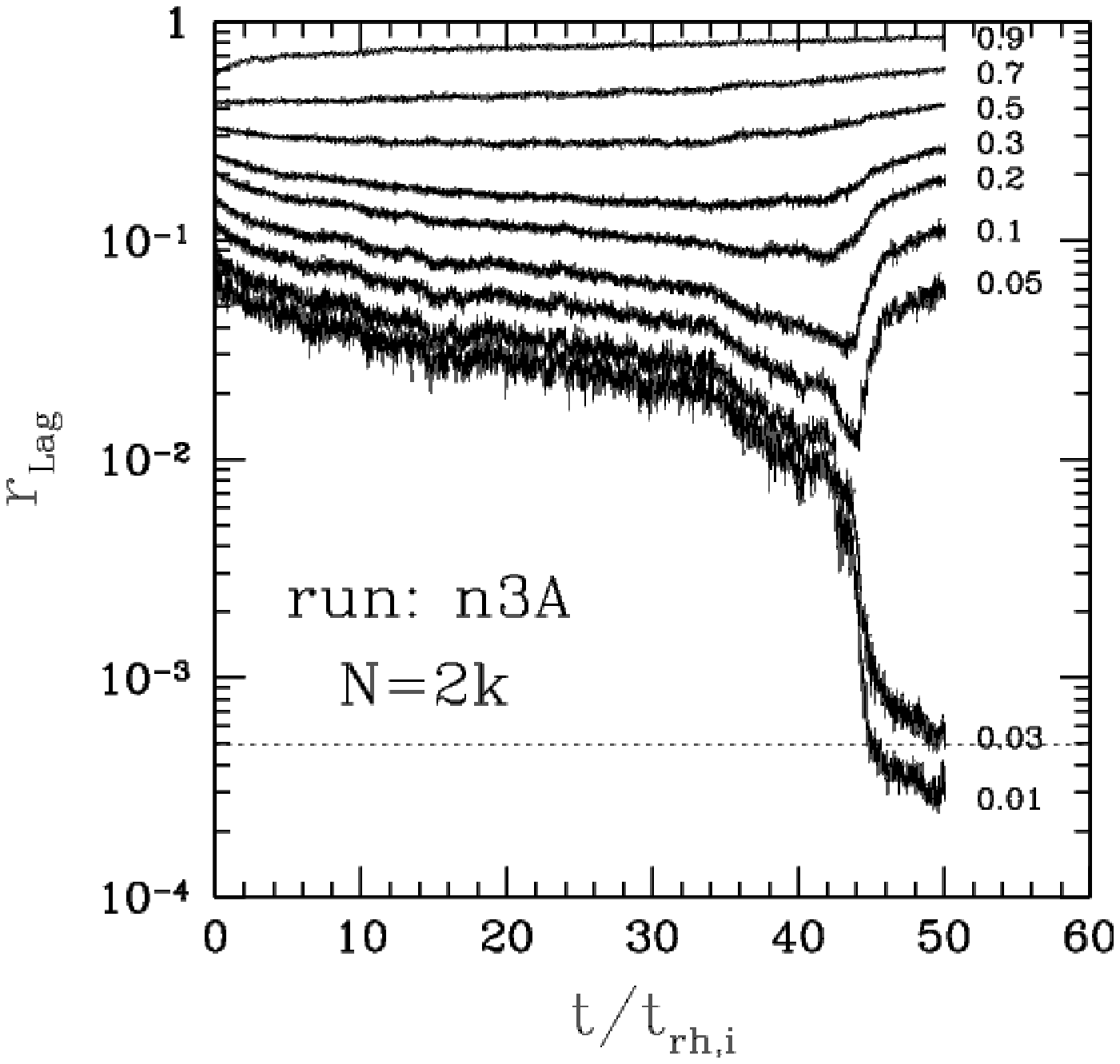}
  \end{center}

\vspace*{-0.2cm}

    \caption{Lagrangian radii as function of time obtained from the run $n3A$. 
      Each line represent 
      the radius of the shell which contains a constant mass fraction.    
      The numerical values of the mass fraction are indicated 
      in the right part of figure. The horizontal dotted line is the 
 potential softening scale, $\epsilon=1/N\simeq4.9\times10^{-4}$.
    \label{fig: poly_n3A_r_Lag} }
  \begin{center}
    \epsfxsize=5.5cm
    \epsfbox{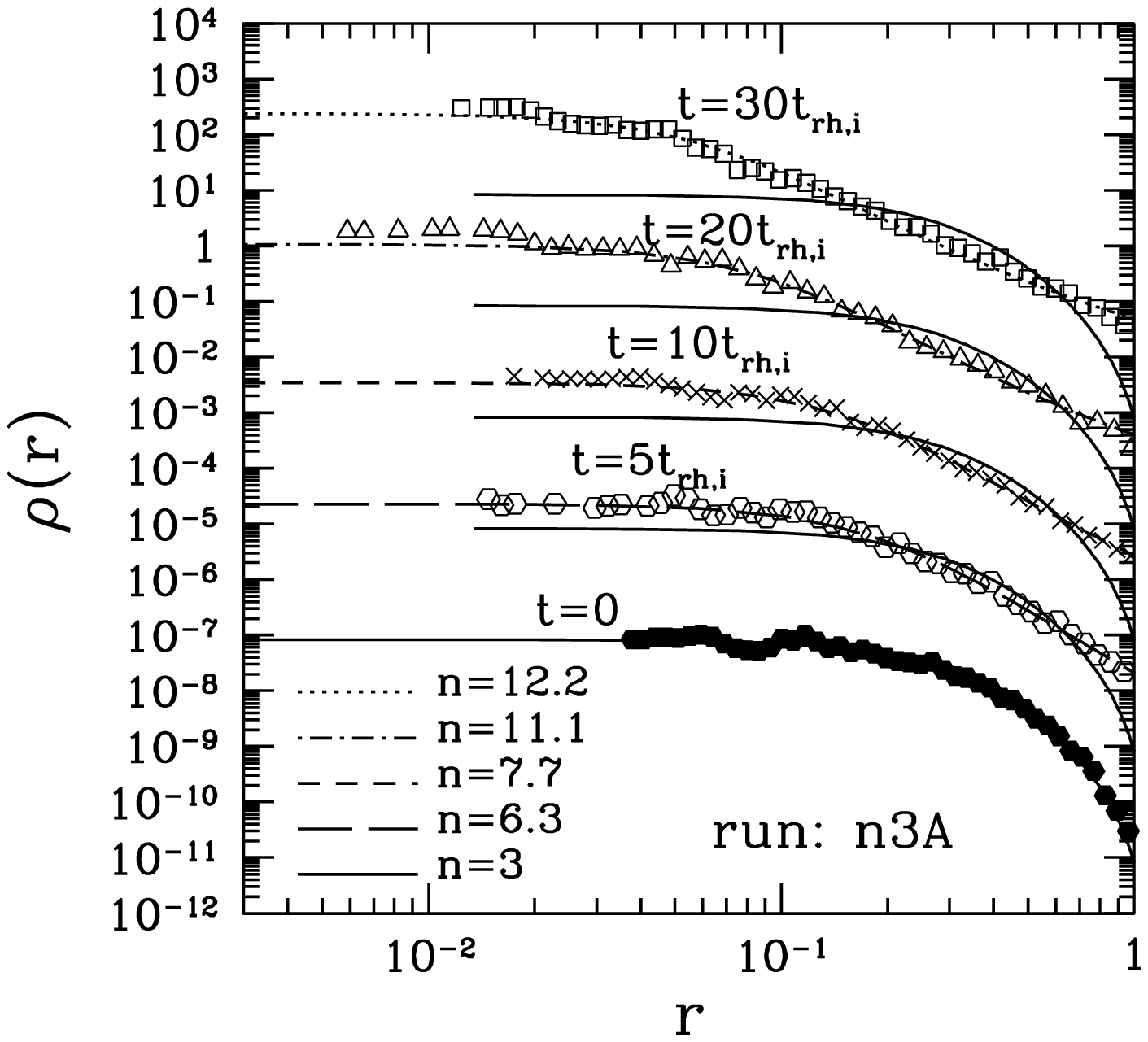}
    \epsfxsize=5.5cm
    \epsfbox{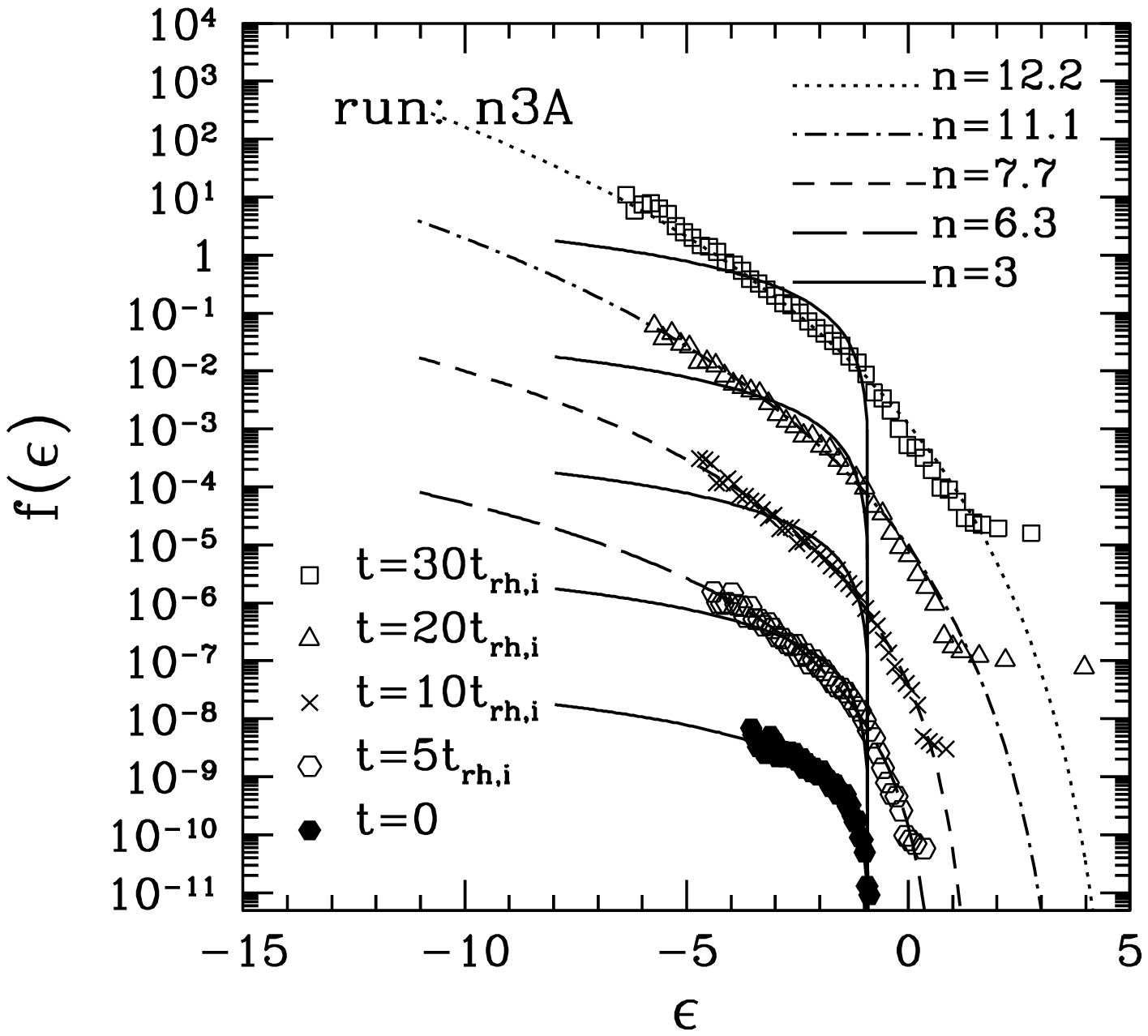}
    \epsfxsize=5.5cm
    \epsfbox{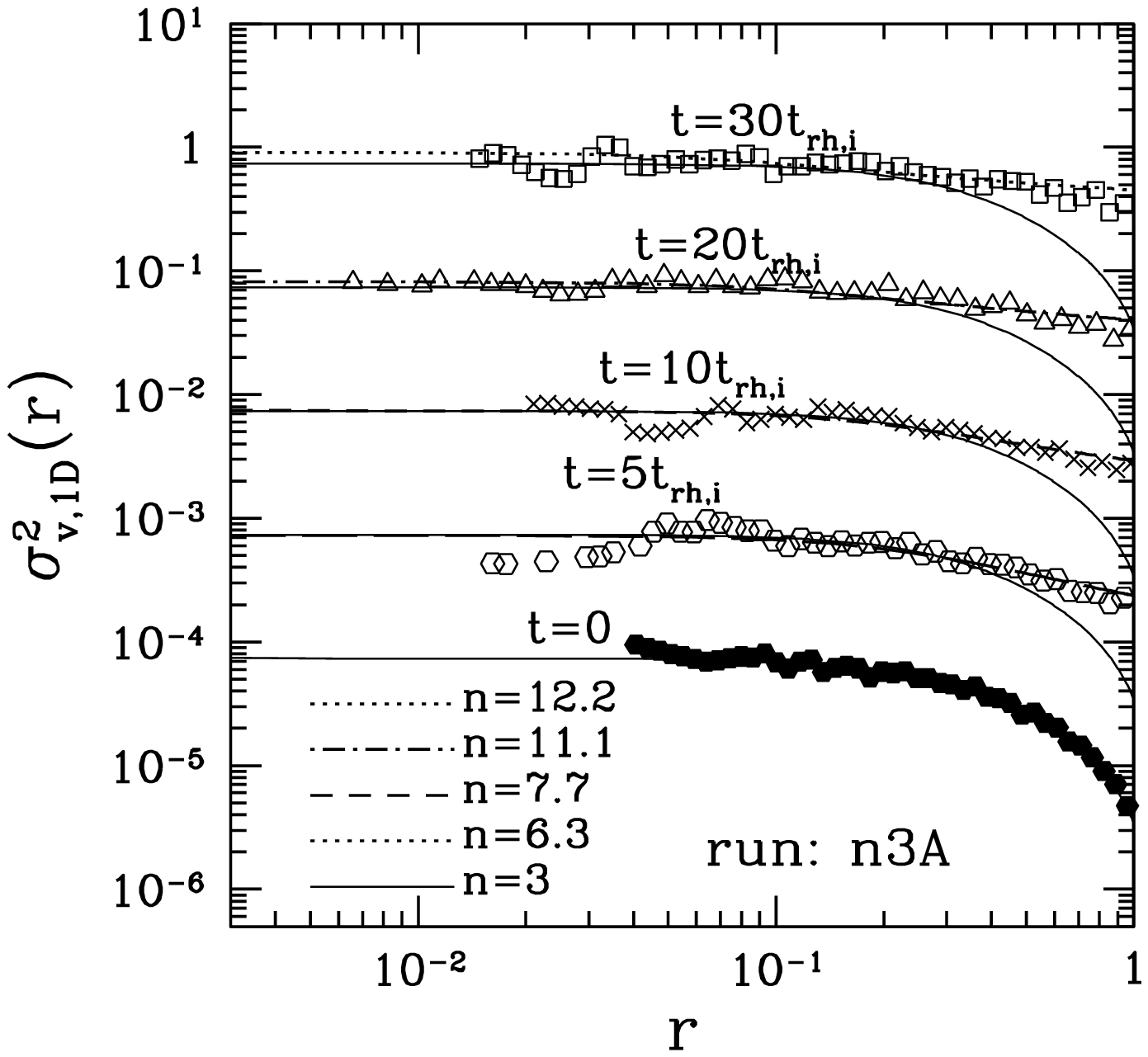}
  \end{center}

\vspace*{-0.2cm}

    \caption{Snapshots of density profile $\rho(r)$ ({\it Left}), 
      distribution function $f(\epsilon)$ as function of specific energy of 
      particles $\epsilon=v^2/2+\Phi(r)$ ({\it Middle}) 
      and one-dimensional velocity dispersion $\sigma^2_{\rm v,1D}(r)$ 
      ({\it Right}) from the run $n3A$. In each panel, symbols represent 
      the $N$-body results, while the solid lines show the Emden solutions of 
      initial distribution. The lines except for the solid lines are  
      the results fitted to the one-parameter sequence of stellar polytropes. 
      Note that for clarify, the results are offset vertically, 
      successively by two-digits below, except for the final output.  
    \label{fig: poly_n3A_fit_poly} }
  \begin{center}
    \epsfxsize=7cm
    \epsfbox{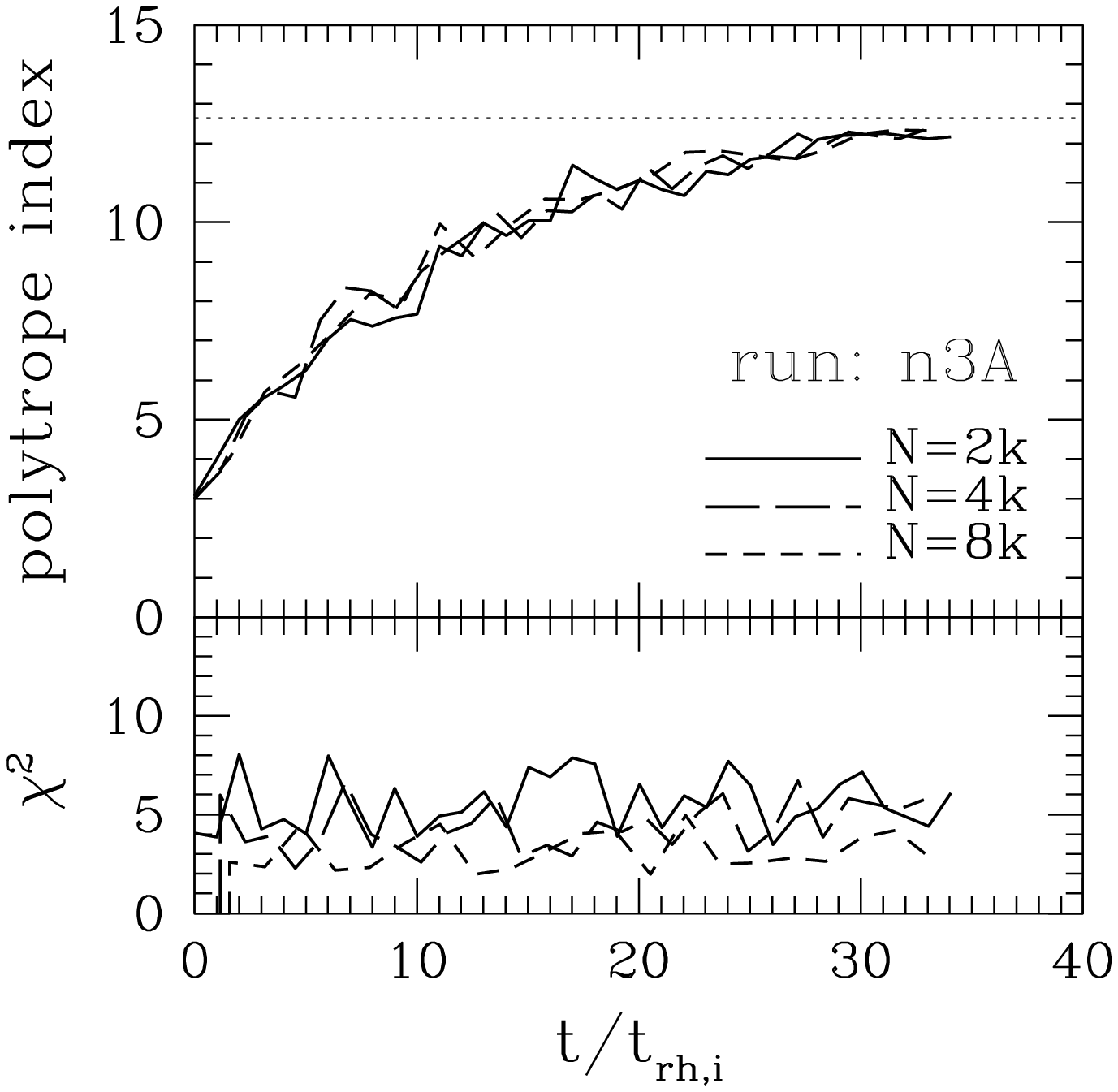}
  \end{center}

\vspace*{-0.2cm}

    \caption{Time evolution of polytrope index fitted to the $N$-body 
      simulations in the case of the run $n3A$. 
      The upper panel shows the fitting results for the runs with particles 
      $N=2,048$({\it solid}), $4,096$({\it long-dashed}) and 
      $8,192$({\it short-dashed}). The horizontal dotted line represents  
      the critical index $n_{\rm crit}$, corresponding to the marginal
      state of the entropy $S_q$ satisfying the condition 
      $d\lambda/dD=0$ in the case of run $n3A$. 
      The lower panel represents the goodness 
      of fit by plotting the function $\chi^2$ defined in equation 
      (\ref{eq:def_of_chi^2}). 
    \label{fig: poly_n3A_evolve_n} }
\end{figure}

%
%
%
%
Figure \ref{fig: poly_n3A_r_Lag} shows the time evolution of the 
Lagrange radii taken from the run with $N=2K$, plotted 
as function of time in unit of the half-mass relaxation time. With the 
softening parameter $\epsilon=1/N$, the core-collapse takes place at 
$t\sim 44 t_{\rm rh,i}$ and the core-halo structure was developed at 
the end of the calculation. Looking at an earlier phase, 
the Lagrangian radii evolve very slowly and one can clearly 
distinguish the timescales between 
the early relaxation phase and 
the late-time core-collapse phase. 
Since the early stage of the time evolution seems nearly equilibrium,  
we will especially call it quasi-equilibrium evolution. 
In a quasi-equilibrium regime, while decreasing the inner 
Lagrangian radii, the outer Lagrangian radii slightly expands to compensate 
the slow contraction of the core.

To see the quasi-equilibrium structure in more detail, 
in figure \ref{fig: poly_n3A_fit_poly}, we plot the snapshots of the density 
profile as function of radius ({\it left}), 
the distribution function as function of specific energy of the particles  
$\epsilon=v^2/2+\Phi(r)$ ({\it middle}) and the velocity 
dispersion profile as function of radius ({\it right}) 
during the quasi-equilibrium regime. 
 In each panel, the symbols denote the simulation results. 
Note that for clarify, the results are offset vertically, successively by 
two-digits below, except for the final output at $t=30t_{\rm rh,i}$.

In figure \ref{fig: poly_n3A_fit_poly}, the solid lines show the 
initial stellar polytrope with $n=3$ evaluated from the Emden solutions.   
Comparing those curves with simulation results, one deduces that the system 
at an earlier time is slightly out of equilibrium and it gradually 
deviates from the initial state. While the velocity dispersion profile 
monotonically increases at an outer part, the density profile first 
increases at both the inner and the outer 
parts ($t\simlt 5t_{\rm rh,i}$), leading to a slight decrease of the density 
contrast $D=\rho_c/\rho_e$ (see the arrow labeled by {\it n3A} in figure 
\ref{fig: lambda_poly}). Later, the increase of the core density surpasses 
that of the edge density and thereby the density contrast turns to 
increase ($t\simgt 10t_{\rm rh,i}$).

In figure \ref{fig: poly_n3A_fit_poly}, in order to characterize the 
evolutionary sequence of the quasi-equilibrium state, 
the simulation results are compared with a sequence of the 
stellar polytropes. The fitting results are then plotted 
as long-dashed, short-dashed, dot-dashed and 
dotted lines from the data at $t=5t_{\rm rh,i}$ to that at $t=30t_{\rm rh,i}$. 
In fitting the simulation data to the stellar polytropes, we first 
quantify the radial density profile $\rho(r)$ from each snapshot data.
Selecting the $100$ points from it 
at regular intervals in logarithmic scale of radius $r$, the results are 
then compared with the Emden solutions.
\footnote{Reason why we adopted the density profile instead of 
the distribution function is that the density profile can be very 
sensitive to the dynamical evolution process. 
Further, in spherically symmetric and isotropic cases, the density profile is  
uniquely determined from the distribution function through the 
Eddington formula \citep[][]{BT1987}. Thus, in the present 
situation, it seems better to perform the fitting  by using the 
radial density profile.} 
Note that in our present situation, the total energy and the mass are 
conserved. Thus, the only fitting parameter is the polytrope index. 
We determine the index $n$ so as to 
minimize the function $\chi^2$:\footnote{Strictly speaking, this is not the 
$\chi^2$ function usually used in the likelihood analysis.} 
\begin{equation}
\chi^2(n)=\sum_{i=1}^{100} 
\left\{ \frac{\rho_{\rm sim}(r_i)-\rho_{\rm Emden}(r_i;n)}
{\rho_{\rm sim}(r_i)} \right\}^2,  
\label{eq:def_of_chi^2}
\end{equation} 

Clearly from figure \ref{fig: poly_n3A_fit_poly}, 
the stellar polytropes quantitatively characterize the 
evolutionary sequence of the simulation results. Note that 
we obtained $\chi^2 \simeq 4$--$8$ for each time-step,  
indicating that the fitting result is satisfactory. 
A closer look at the low-energy part of the distribution 
function $f(\epsilon)$ 
reveals that the simulation results partly resemble the exponential form 
rather than the power-law function. Nevertheless, the most remarkable 
fact is that the stellar polytropes as simple power-law 
distribution globally 
approximate the simulation results in a quite good accuracy. Moreover, 
the fitting results in figure \ref{fig: poly_n3A_fit_poly} 
indicate that the fitted values of the polytrope index monotonically 
increase in time, in contrast to the non-monotonic behavior of the 
density contrast.  In order to quantify the evolution of polytrope index, 
we collect the new snapshot data at each time interval $\Delta t=10$ in 
$N$-body units for the run with $N=2K$. 
Repeating the same fitting procedure as in figure 
\ref{fig: poly_n3A_fit_poly}, the polytrope indices are estimated at 
each time step and the resultant values are plotted in 
figure \ref{fig: poly_n3A_evolve_n} together with the $\chi^2$ value of 
the fitting results. Here we also plot 
the results obtained from the run with $N=4K$ and $N=8K$, in which the time 
intervals are respectively chosen as $\Delta t=20$ and $50$.

Clearly, the fitted values of the polytrope index monotonically increase 
apart from the fluctuations during the short time interval. 
The growth rates of the polytrope index normalized by half-mass relaxation 
time almost coincide with each other, consistent with the fact that 
the quasi-equilibrium sequence is evolved via two-body relaxation. 
In upper panel of figure \ref{fig: poly_n3A_evolve_n}, 
the horizontal dotted line denote the critical index $n_{\rm crit}$, 
corresponding to the point at which $d\lambda/dD=0$ for a given 
$\lambda$ in $(\lambda,~D)$-plane.  
According to the prediction from the non-extensive thermostatistics with 
Tsallis entropy, after reaching the critical index $n_{\rm crit}$,  
the system is expected to enter the gravothermally unstable regime. 
That is,  
the relaxation timescales between the inner and the outer parts are 
decoupled and the system 
finally undergoes the core-collapse. 
In a rigorous sense, 
the thermodynamic prediction itself cannot be applicable to the 
out-of-equilibrium  state, however, 
it turns out that 
the predicted value $n_{\rm crit}$ or $D_{\rm crit}$ is 
consistent with the $N$-body results. In fact, 
the successful fit is obtained until $t\simeq 34 t_{\rm rh,i}$, 
while the core-collapse takes place lately at 
$t\sim 44 t_{\rm rh,i}$ (see Fig.\ref{fig: poly_n3A_r_Lag}). 
Thus, the prediction  based on the non-extensive thermostatistics 
may provide a crude estimate of the boundary 
between the stability 
and the instability in 
the general non-isothermal states. 
This point will be further discussed in other runs. 
%
%
%
%
%
%
%
%
%
%
%
%
%
%
%
\subsubsection{Other runs}
\label{subsubsec:others}
\begin{figure}
  \begin{center}
    \epsfxsize=5.5cm
    \epsfbox{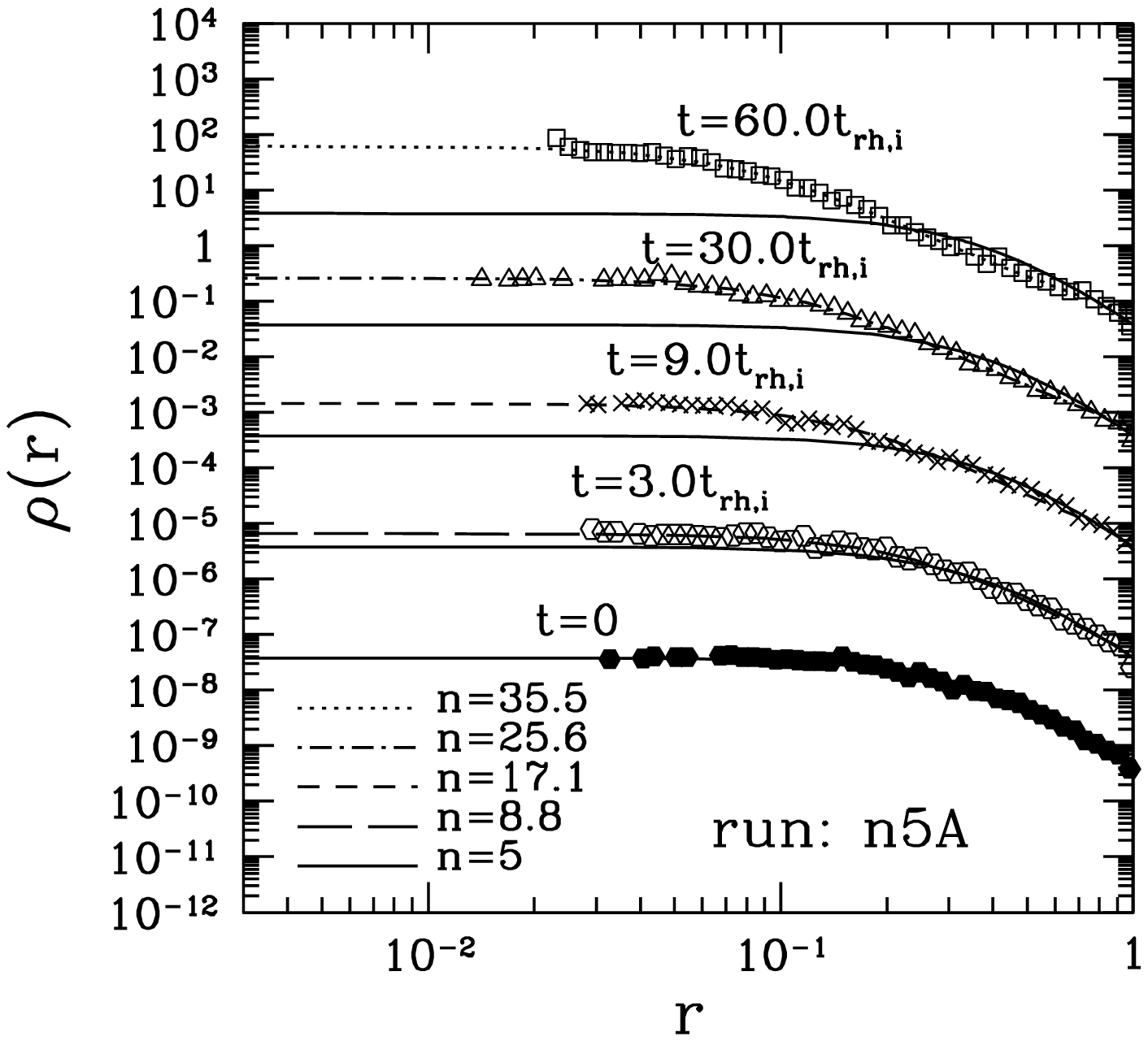}
    \epsfxsize=5.5cm
    \epsfbox{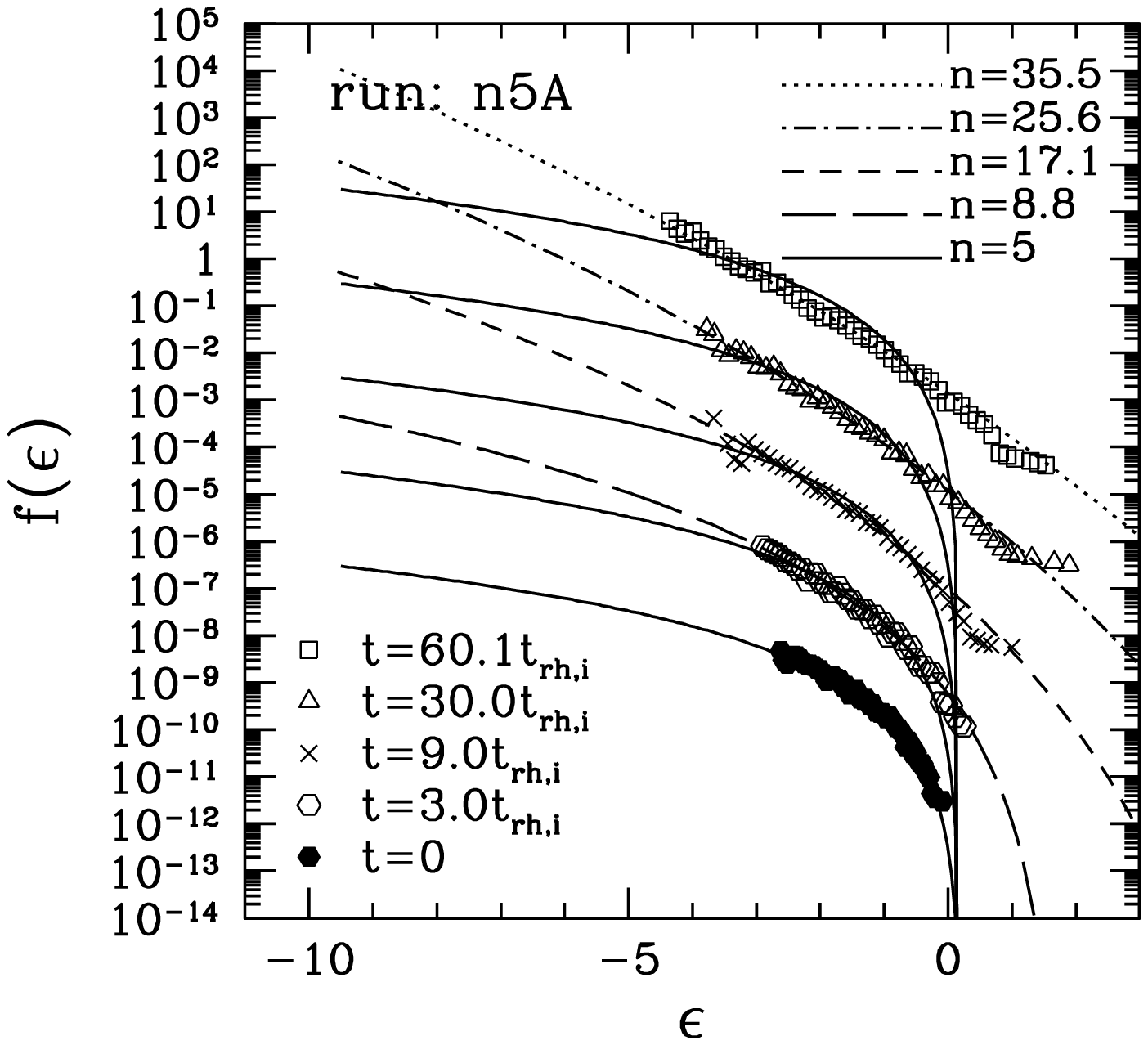}
    \epsfxsize=5.5cm
    \epsfbox{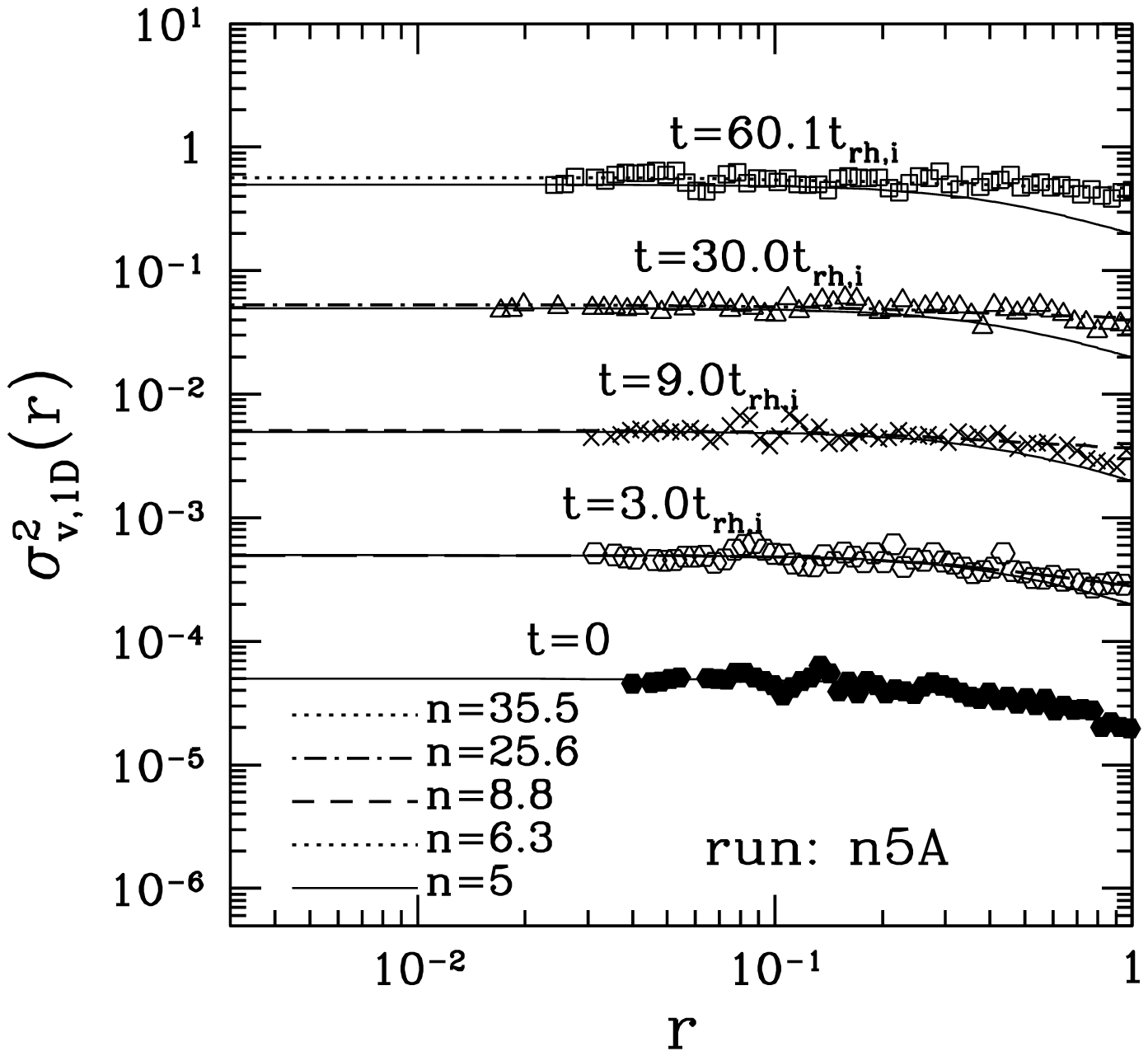}
  \end{center}

\vspace*{-0.2cm}

    \caption{Snapshots of the density profile, distribution function 
	and the velocity dispersion profile from the run $n5A$
    \label{fig: snapshot_n5A} }
  \begin{center}
    \epsfxsize=5.5cm
    \epsfbox{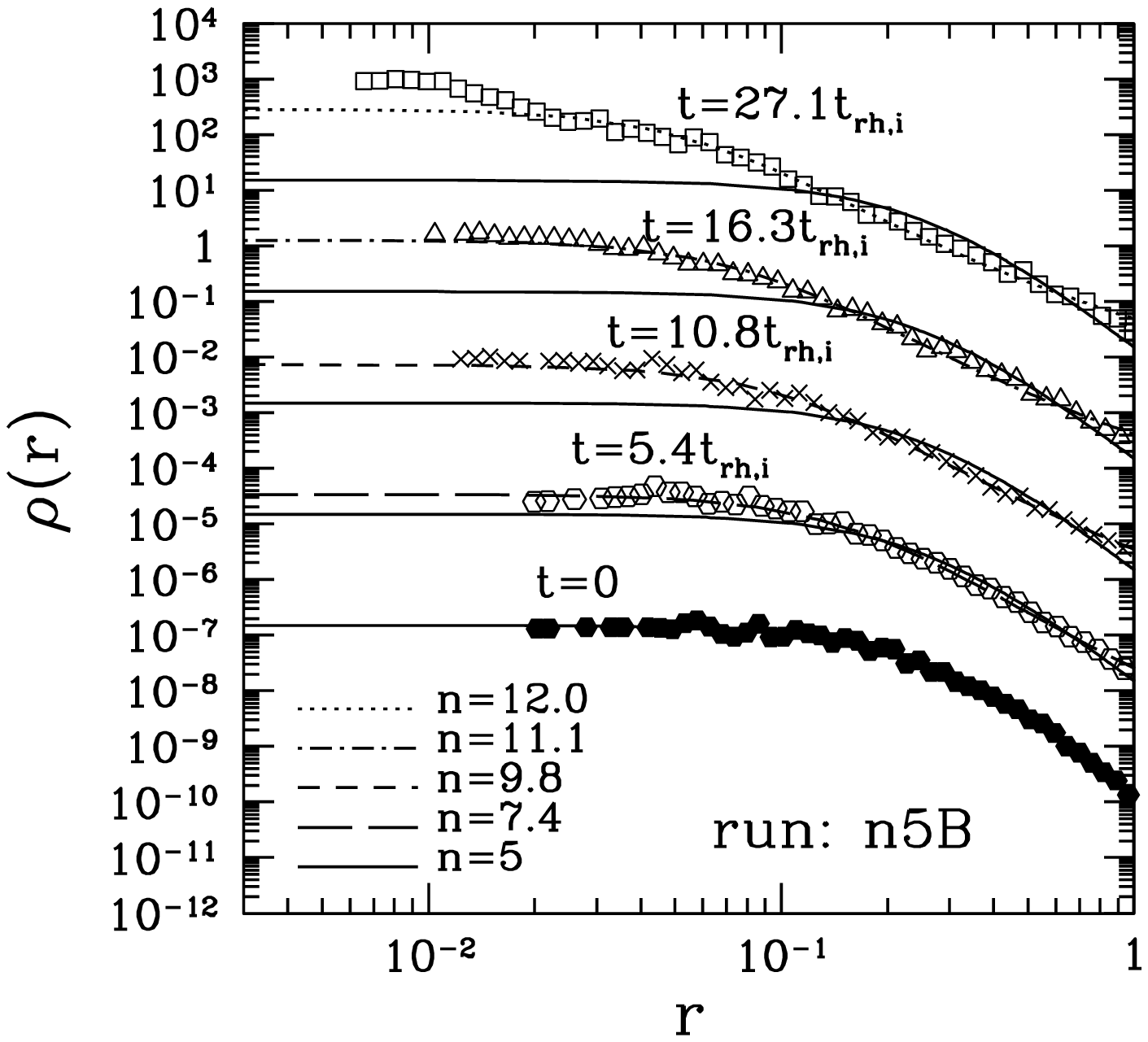}
    \epsfxsize=5.5cm
    \epsfbox{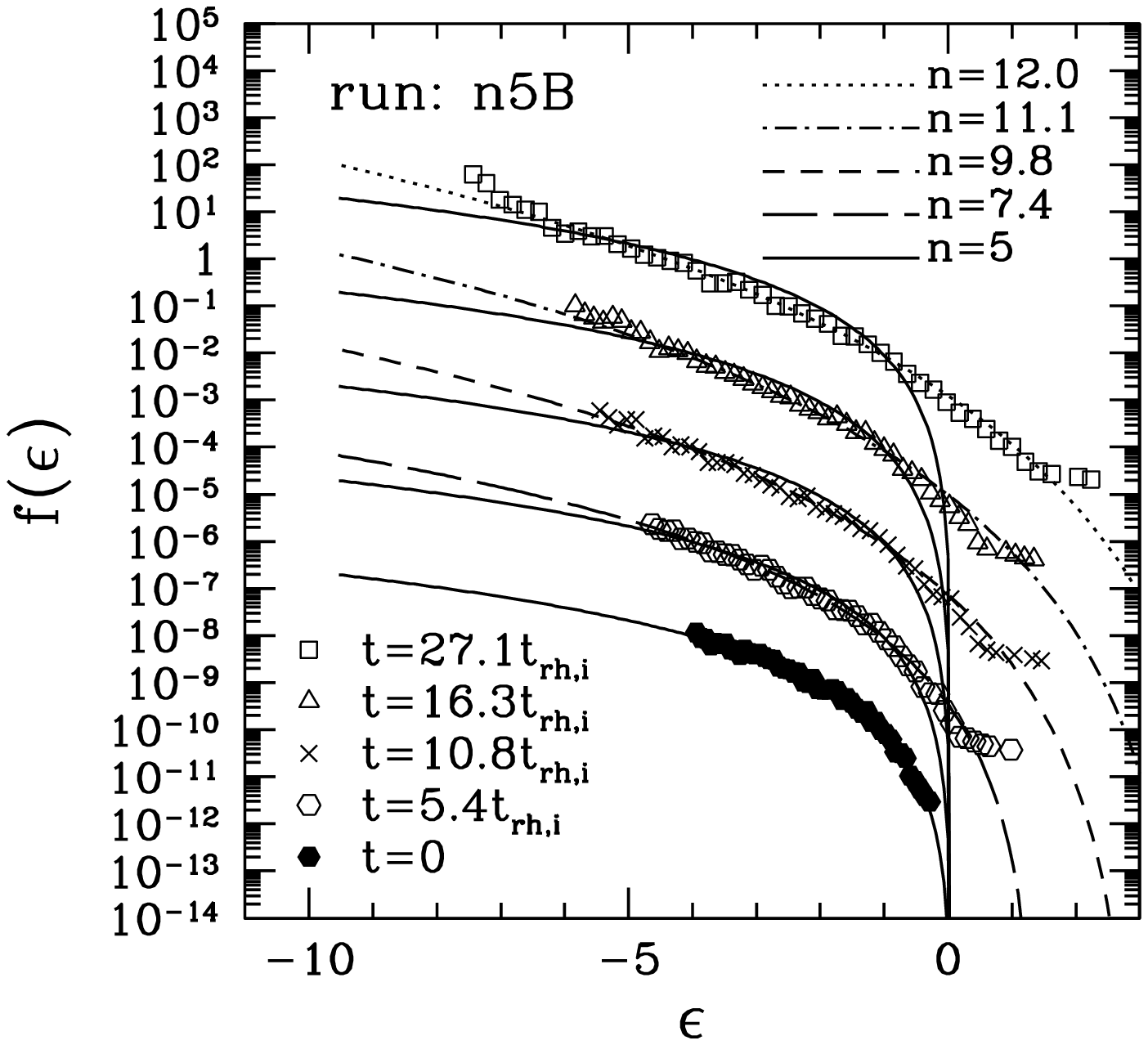}
    \epsfxsize=5.5cm
    \epsfbox{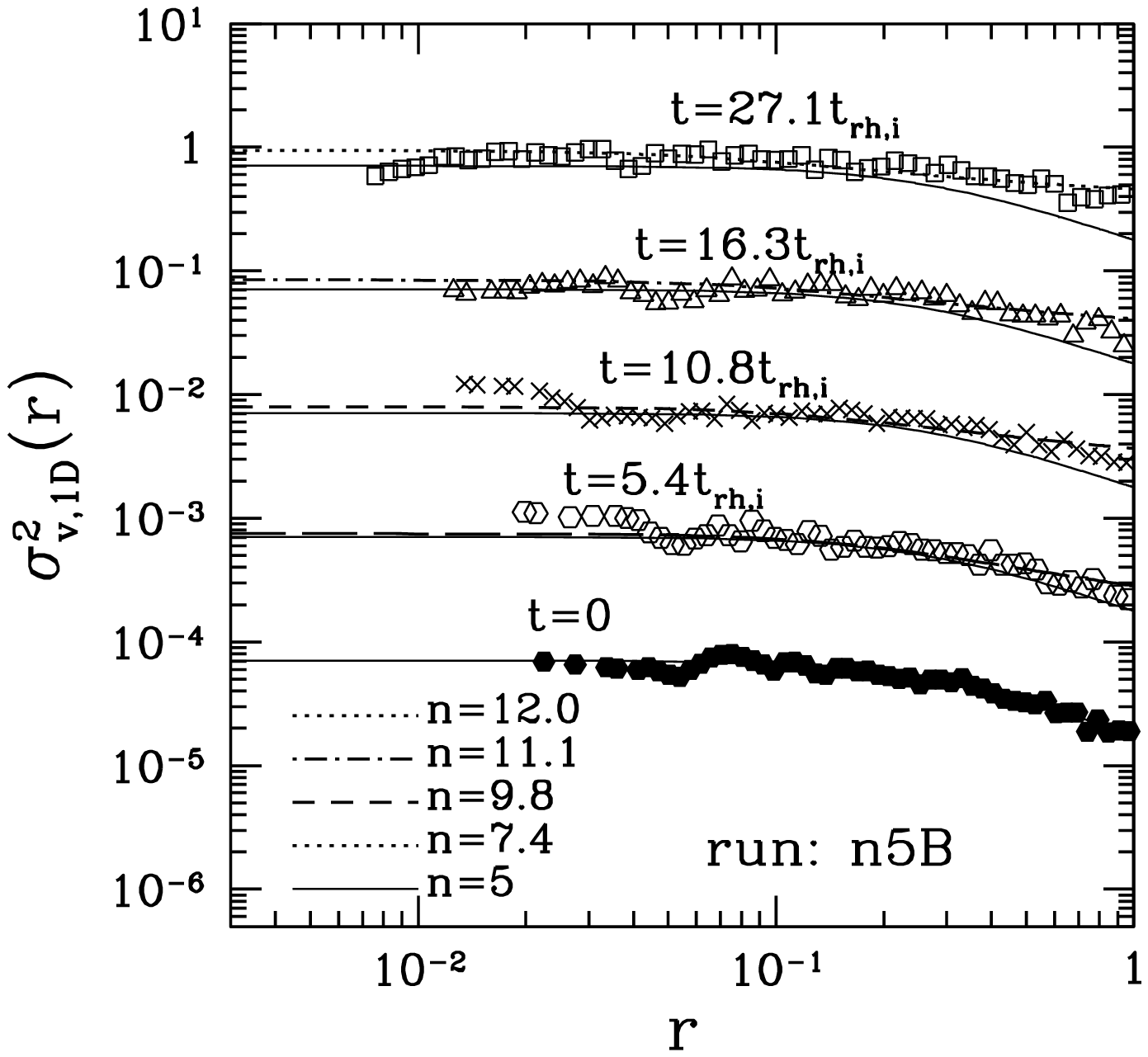}
  \end{center}

\vspace*{-0.2cm}

    \caption{Same as Fig.\ref{fig: snapshot_n5A}, 
	but in case of the run $n5B$. 
    \label{fig: snapshot_n5B} }
  \begin{center}
    \epsfxsize=5.5cm
    \epsfbox{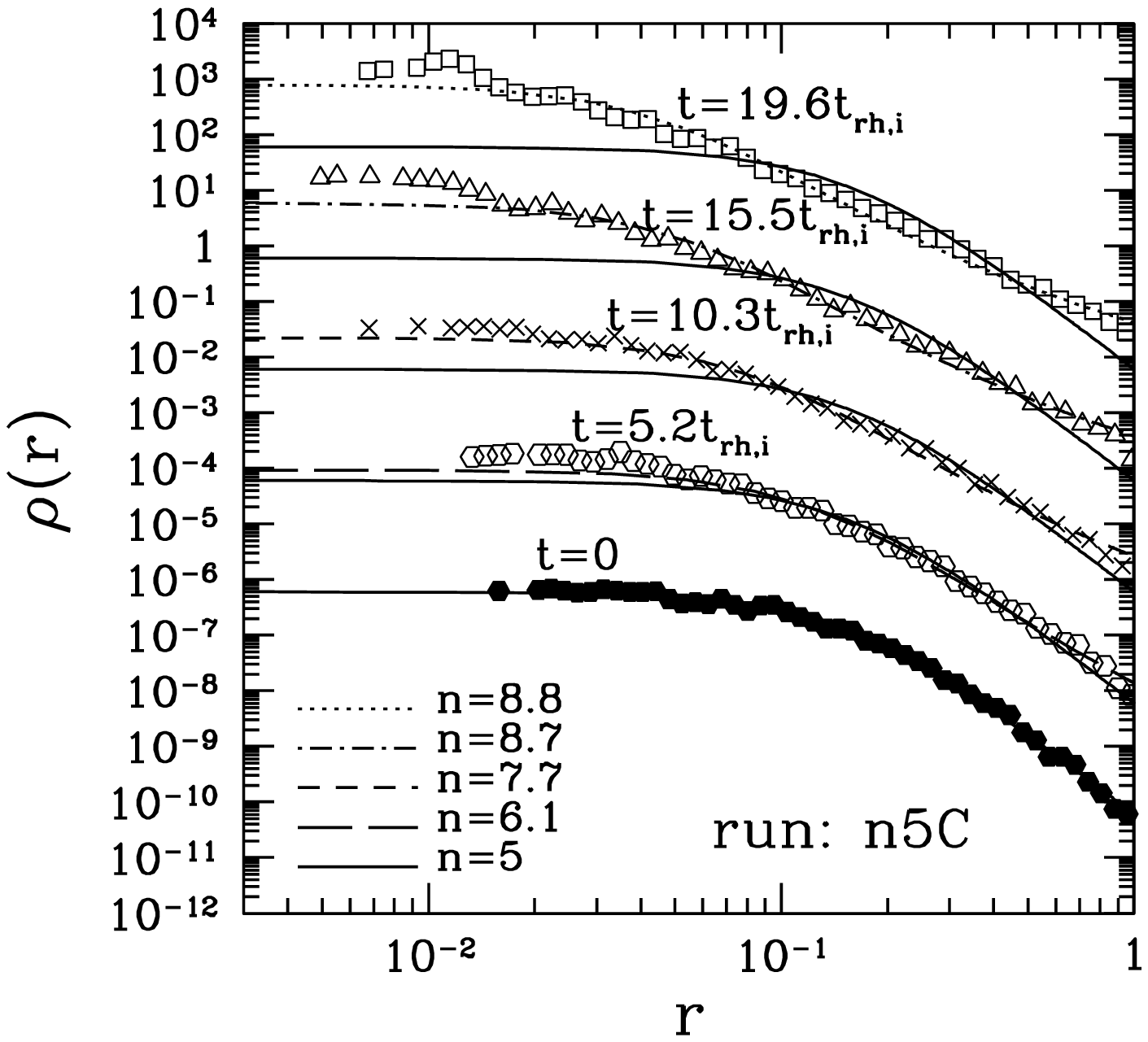}
    \epsfxsize=5.5cm
    \epsfbox{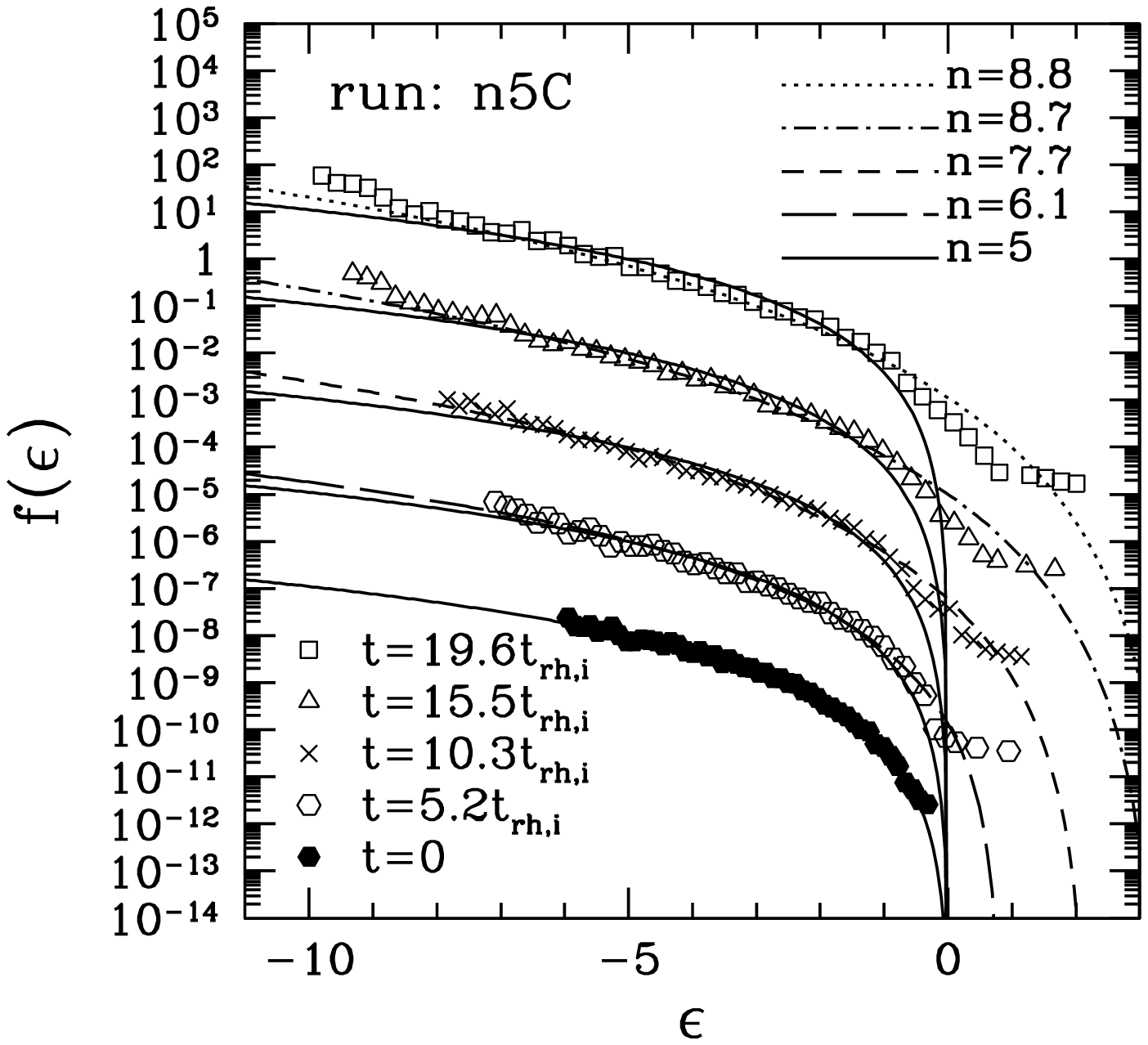}
    \epsfxsize=5.5cm
    \epsfbox{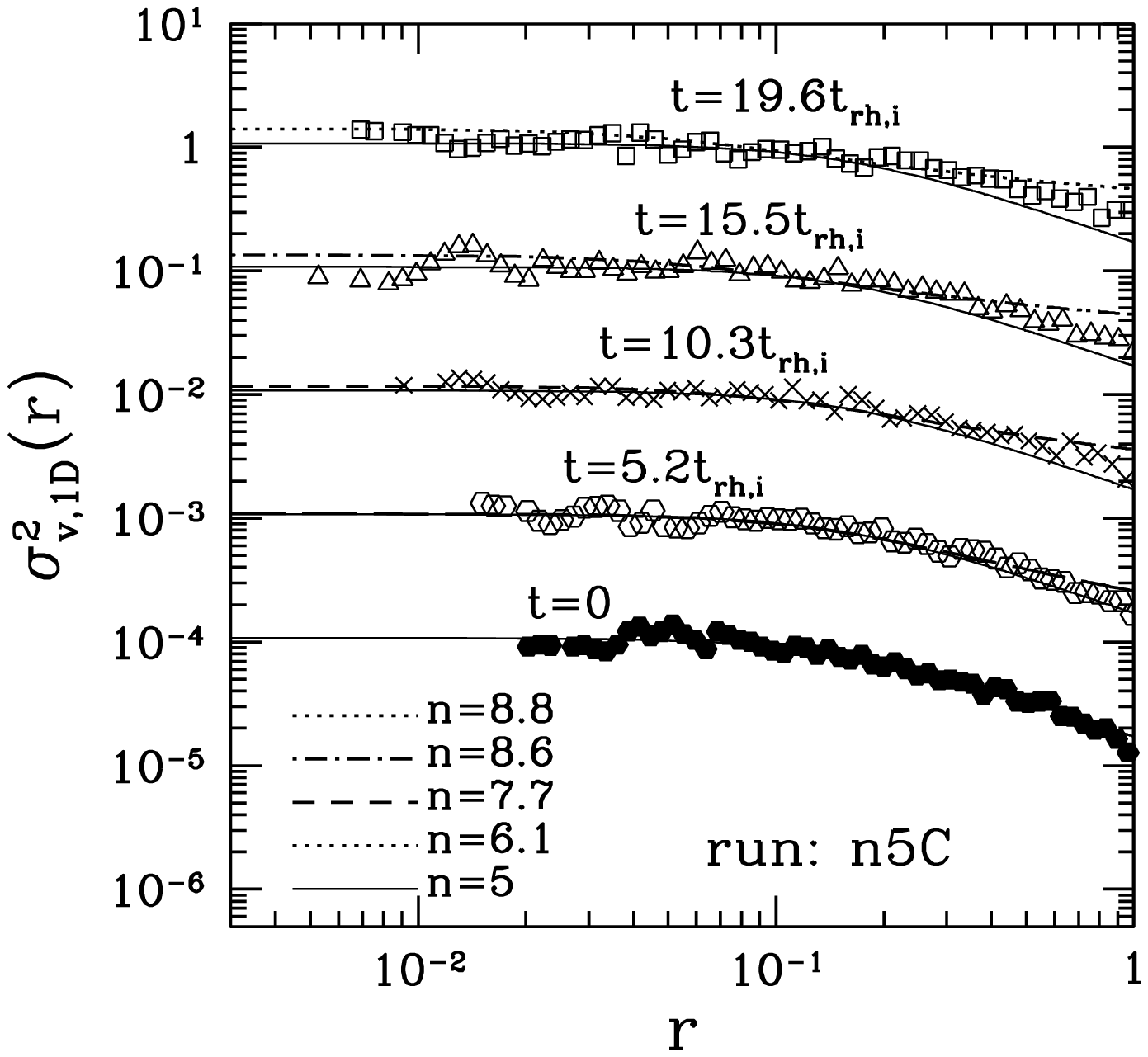}
  \end{center}

\vspace*{-0.2cm}

    \caption{Same as Fig.\ref{fig: snapshot_n5A}, 
	but in case of the run $n5C$. 
    \label{fig: snapshot_n5C} }
  \begin{center}
    \epsfxsize=5.5cm
    \epsfbox{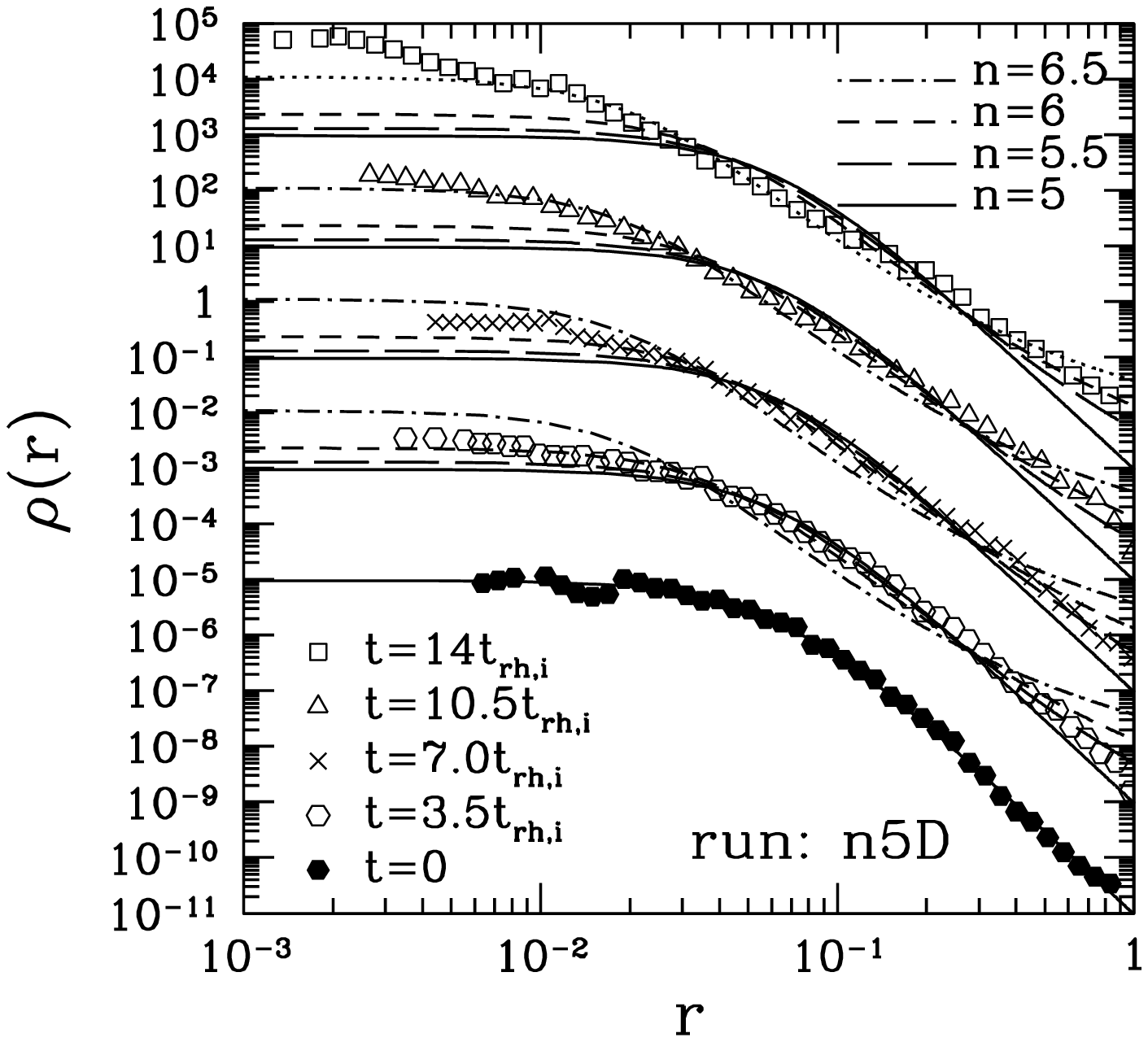}
    \epsfxsize=5.5cm
    \epsfbox{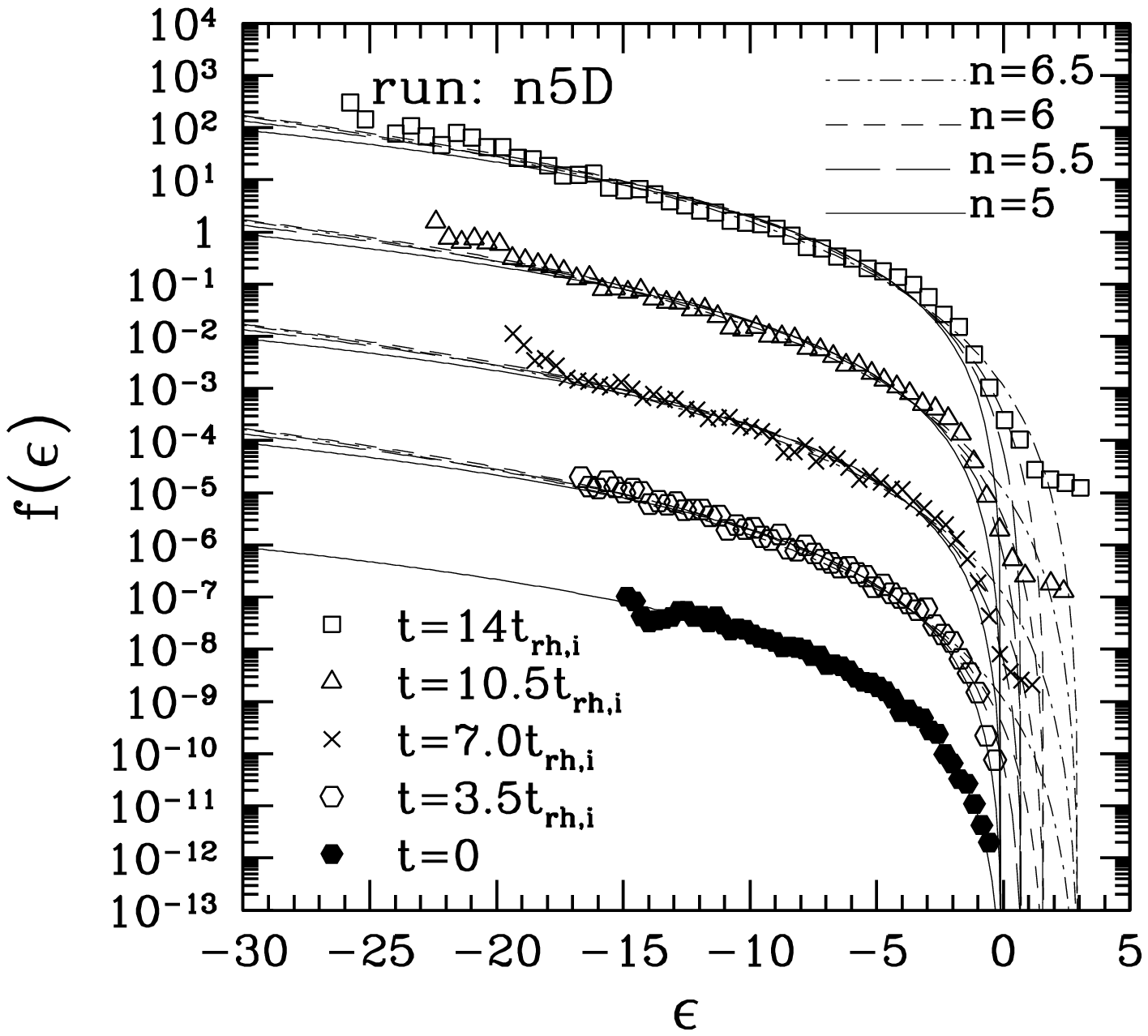}
    \epsfxsize=5.5cm
    \epsfbox{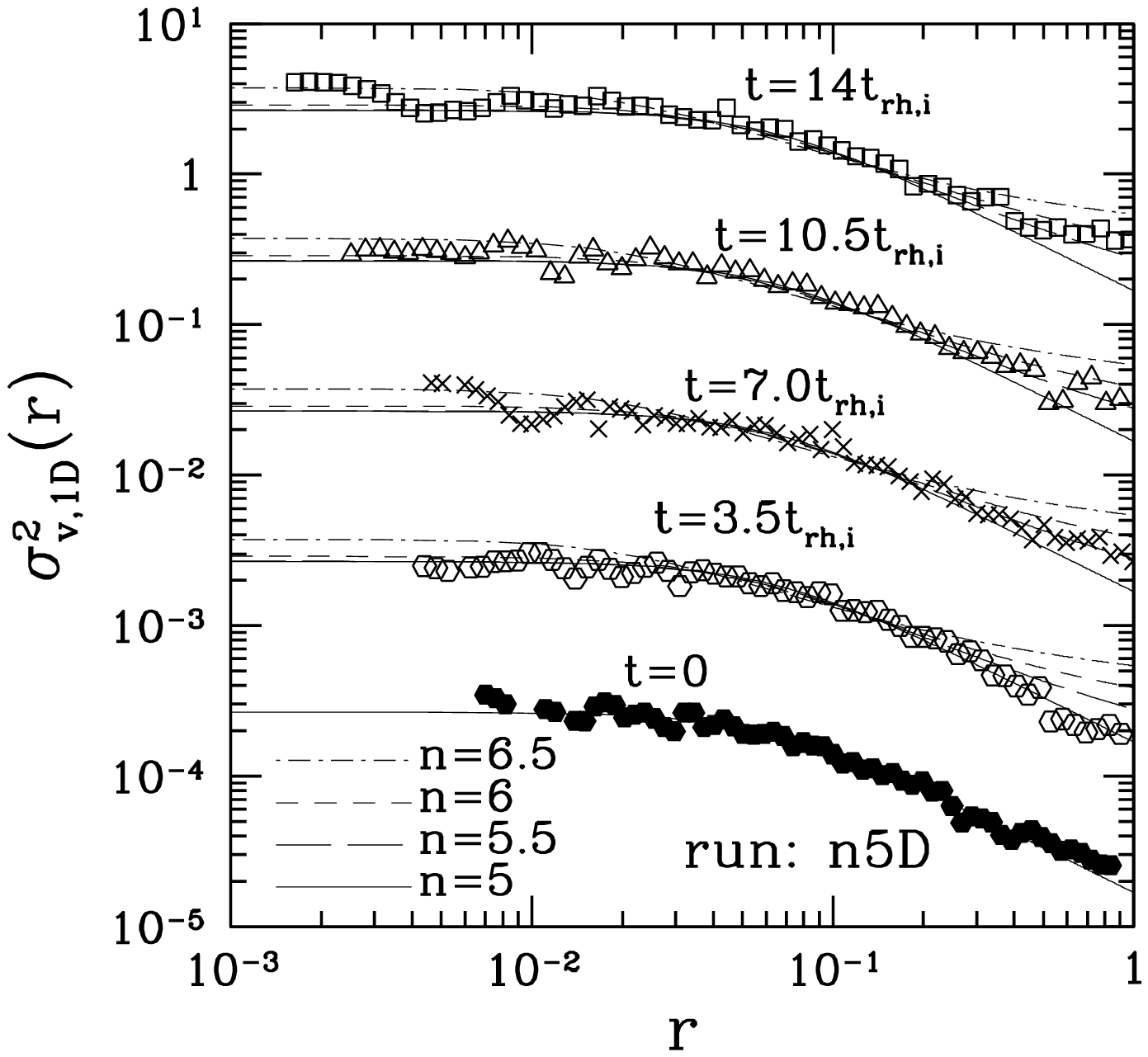}
  \end{center}

\vspace*{-0.2cm}

    \caption{Same as Fig.\ref{fig: snapshot_n5A}, 
	but in the case of the run $n5D$. 
    \label{fig: snapshot_n5D} }
\end{figure}

\begin{figure}
  \begin{center}
    \epsfxsize=5.5cm
   \epsfbox{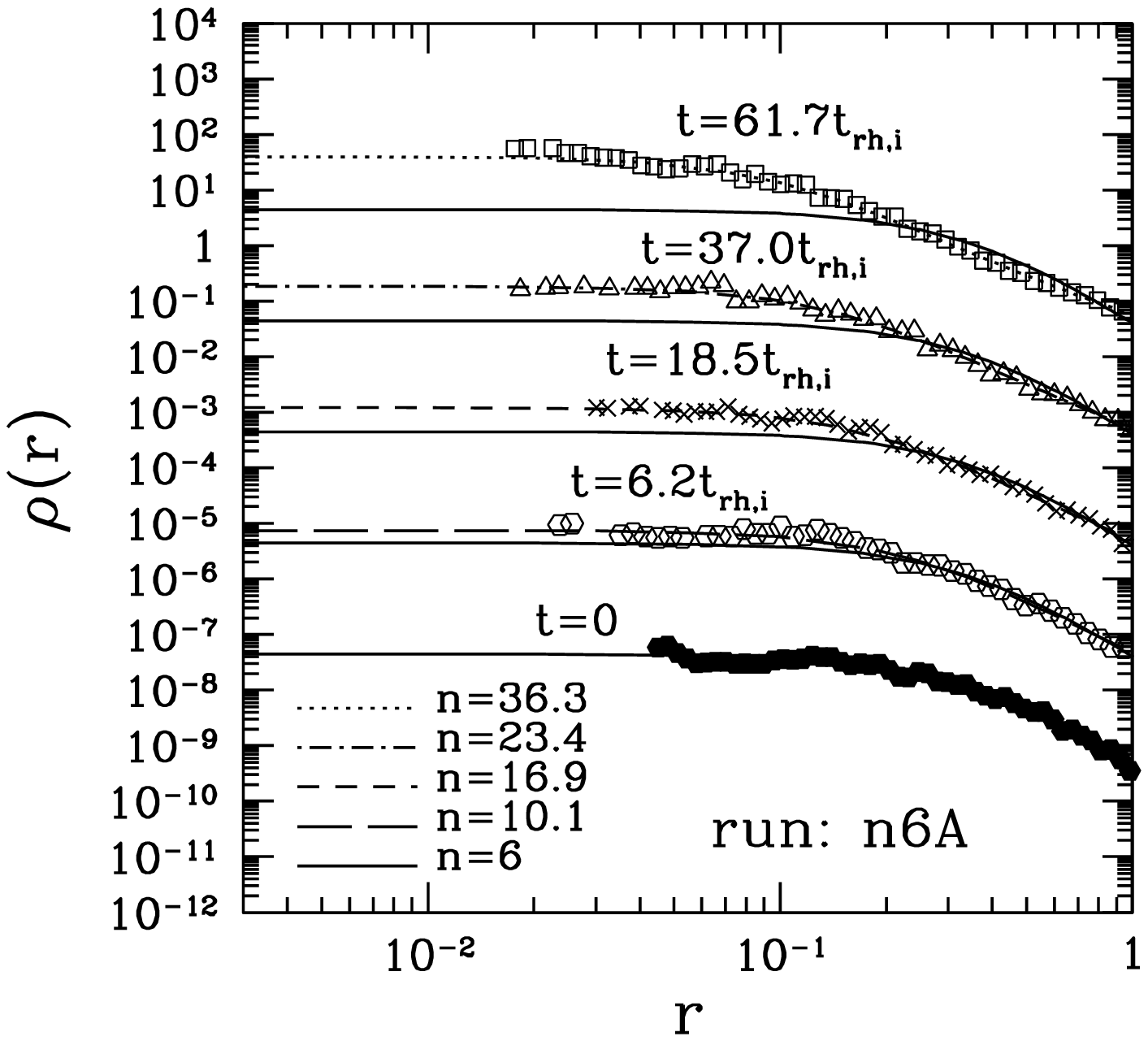}
    \epsfxsize=5.5cm
    \epsfbox{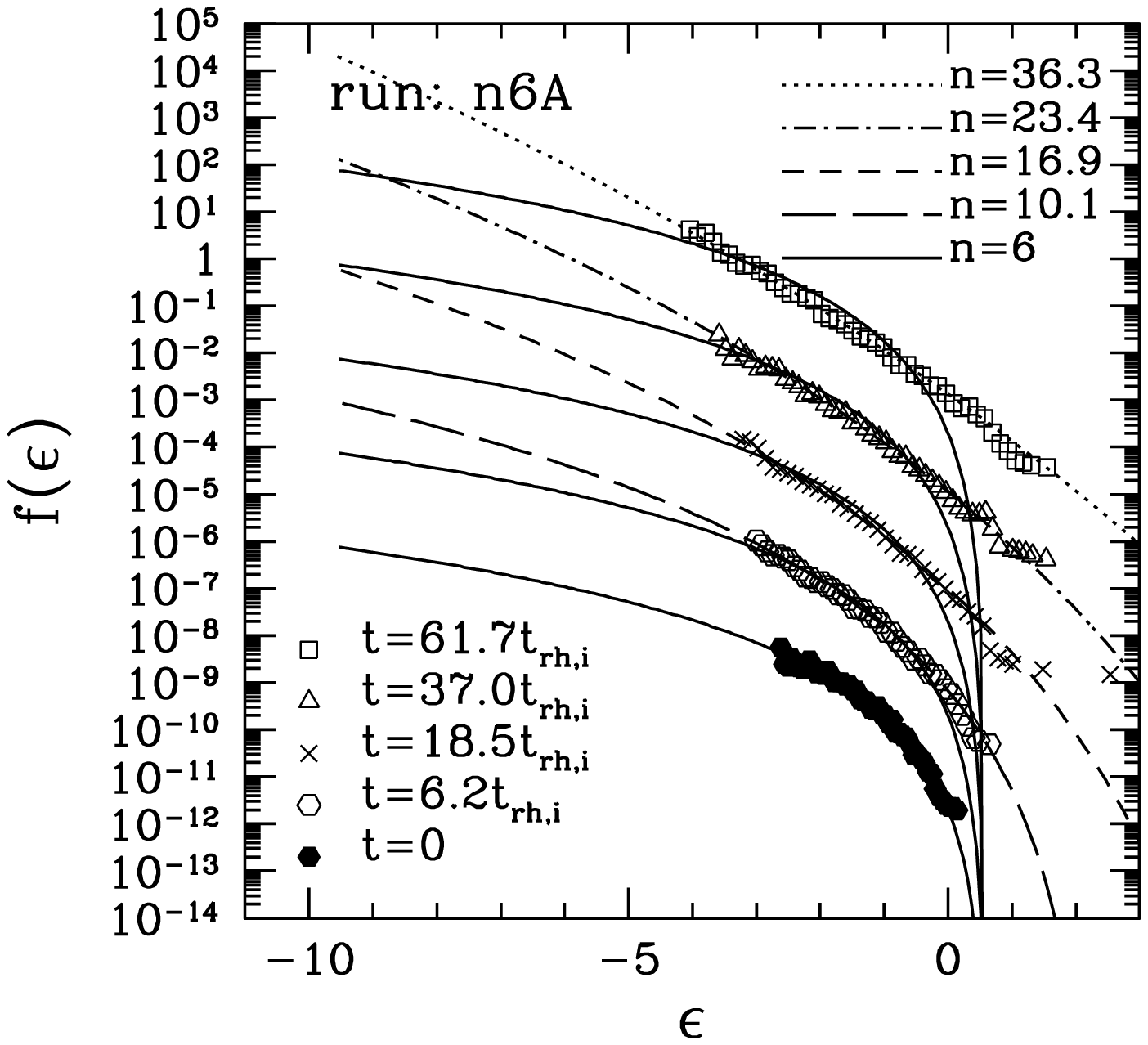}
    \epsfxsize=5.5cm
    \epsfbox{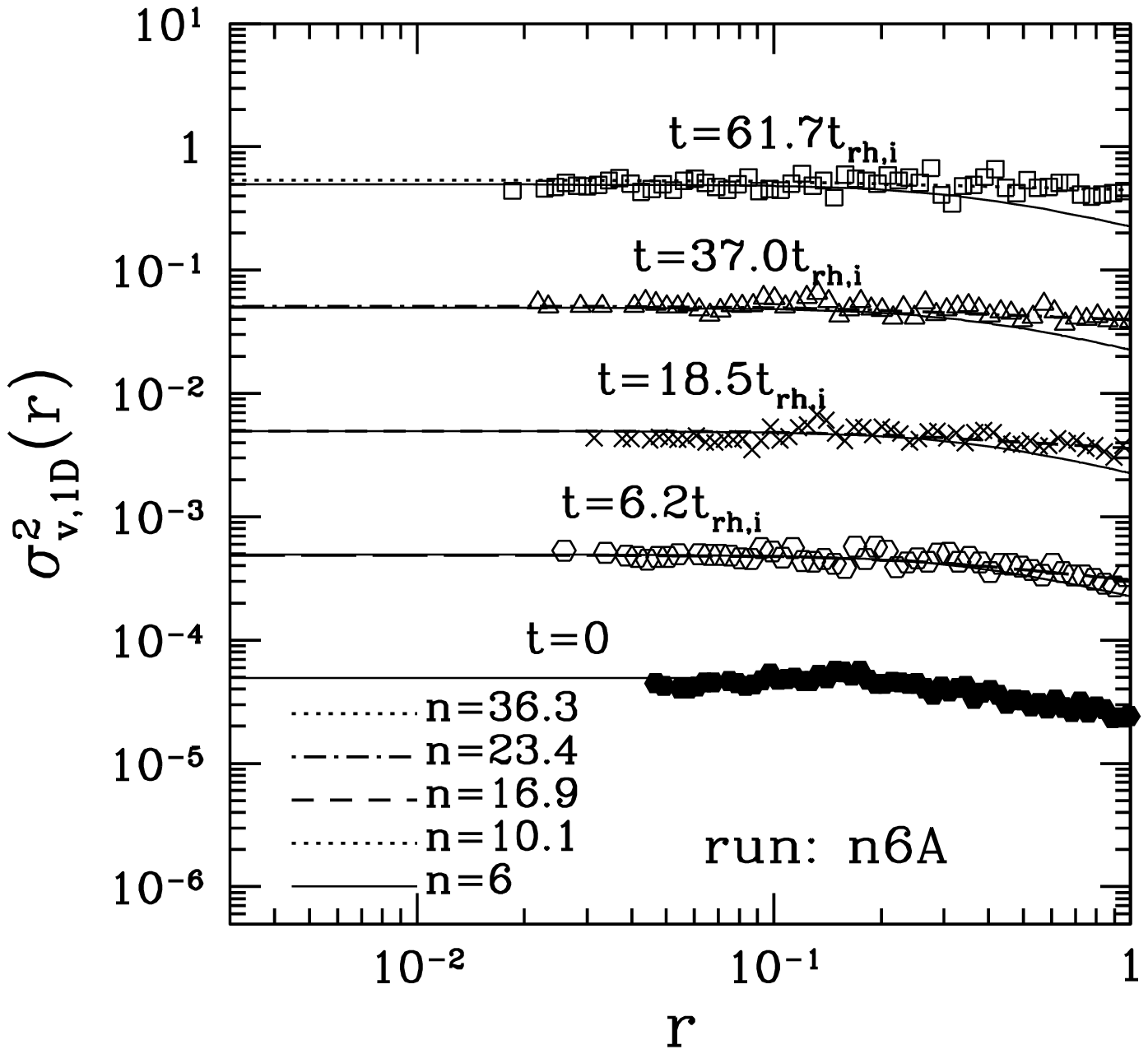}
  \end{center}

\vspace*{-0.2cm}

    \caption{Same as Fig.\ref{fig: snapshot_n5A}, 
	but in the case of the run $n6A$.
    \label{fig: snapshot_n6A} }
  \begin{center}
    \epsfxsize=5.5cm
    \epsfbox{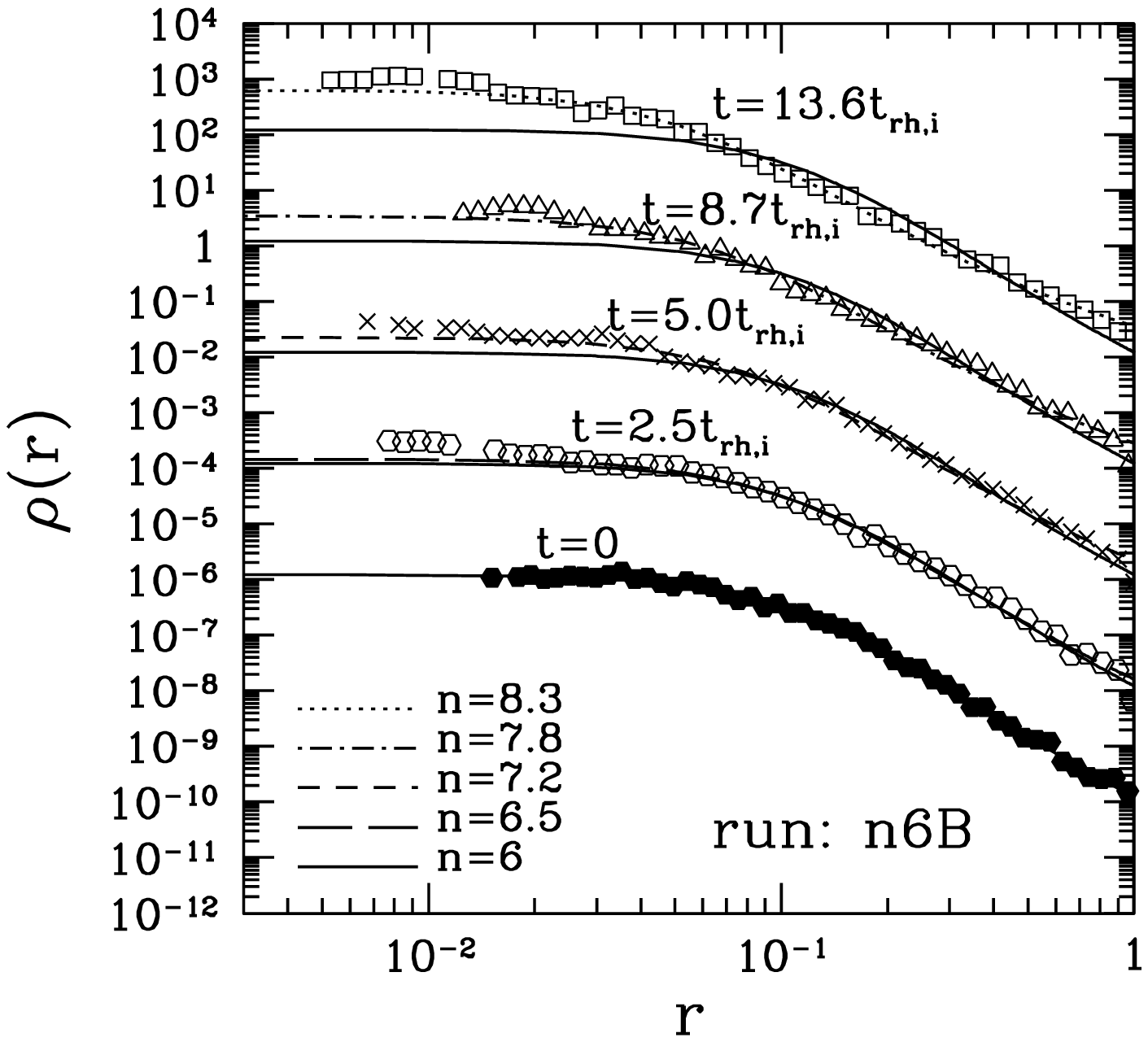}
    \epsfxsize=5.5cm
    \epsfbox{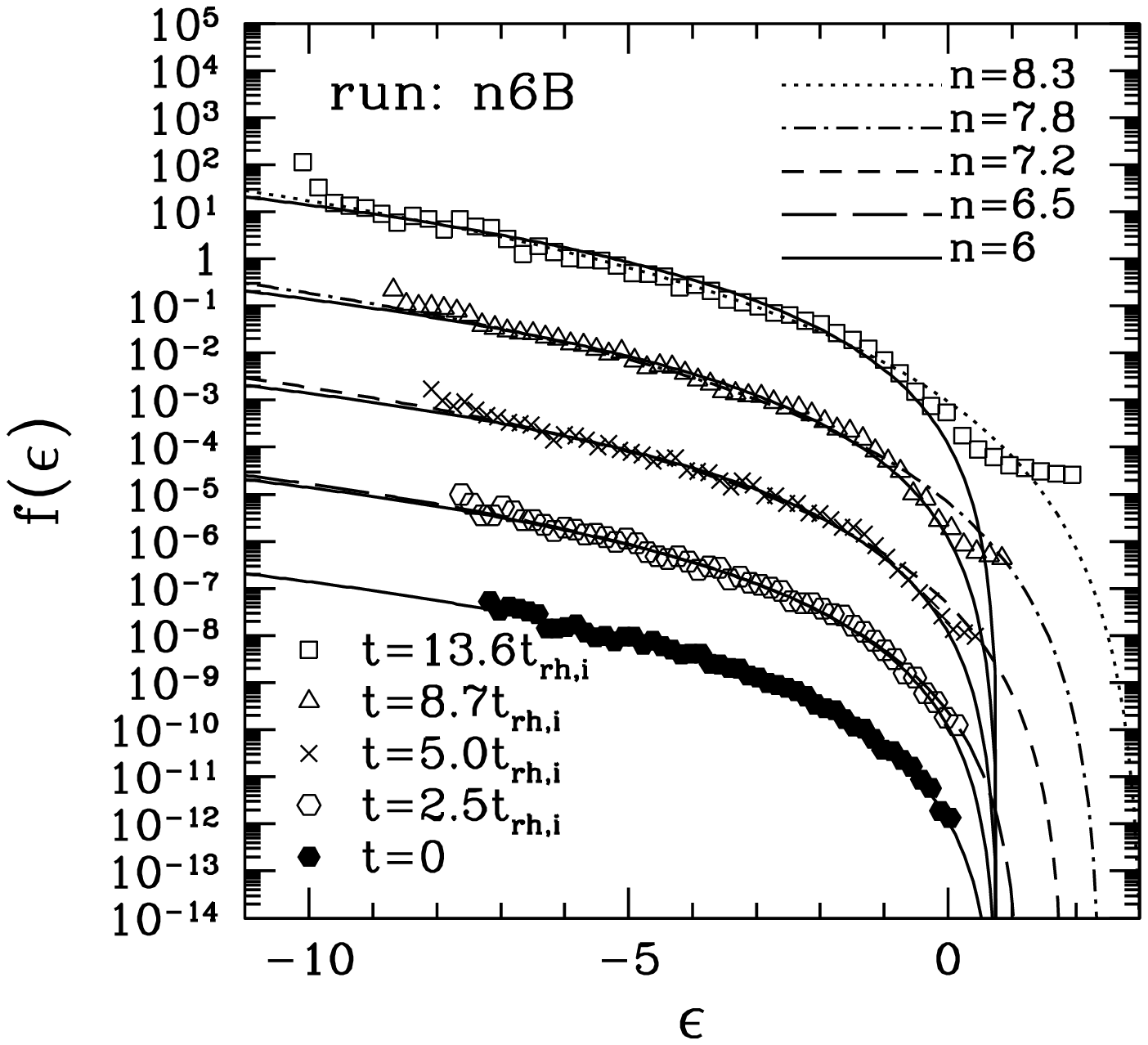}
    \epsfxsize=5.5cm
    \epsfbox{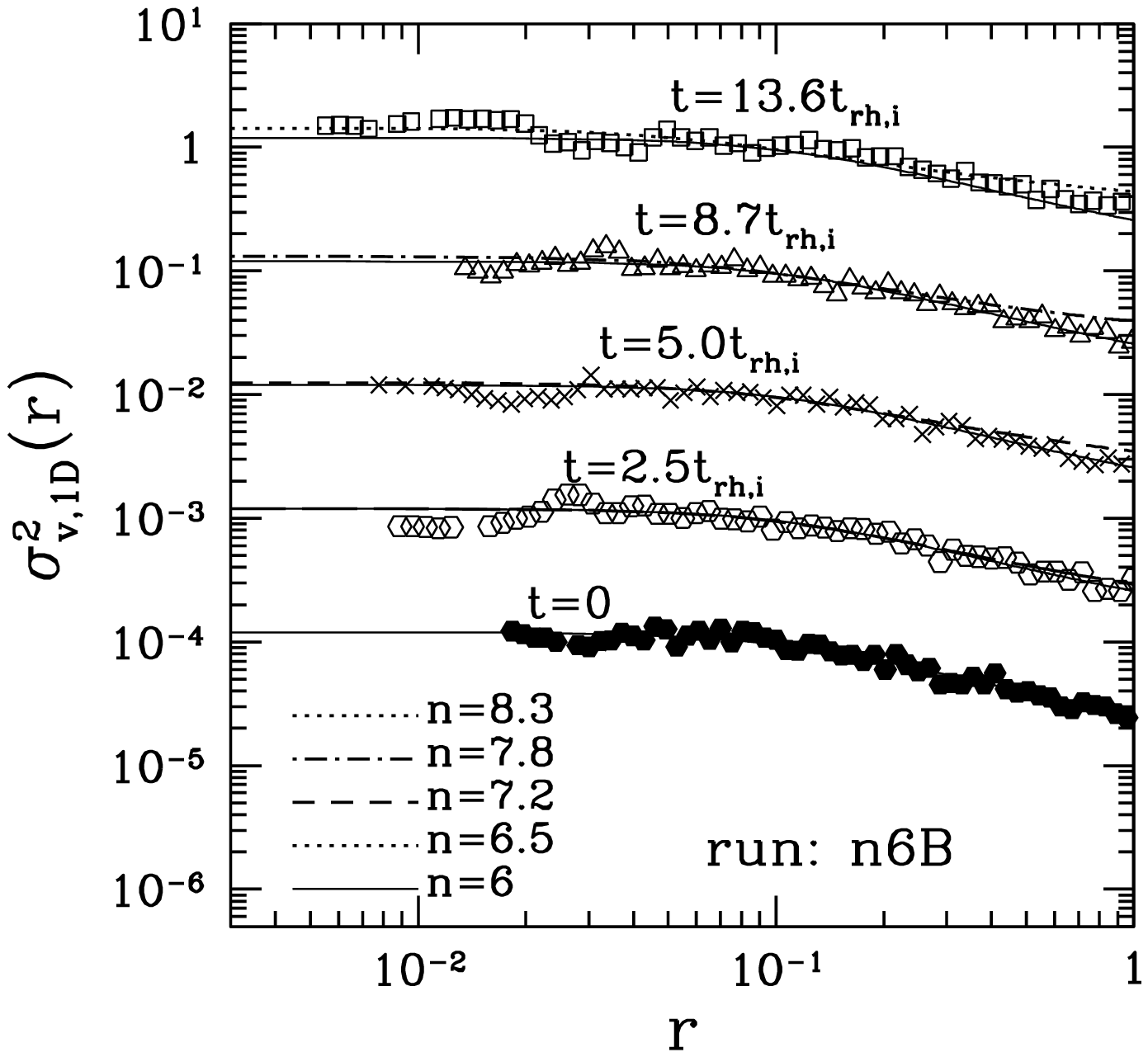}
  \end{center}

\vspace*{-0.2cm}

    \caption{Same as Fig.\ref{fig: snapshot_n5A}, 
	but in the case of the run $n6B$. 
    \label{fig: snapshot_n6B} }
  \begin{center}
    \epsfxsize=6.5cm
    \epsfbox{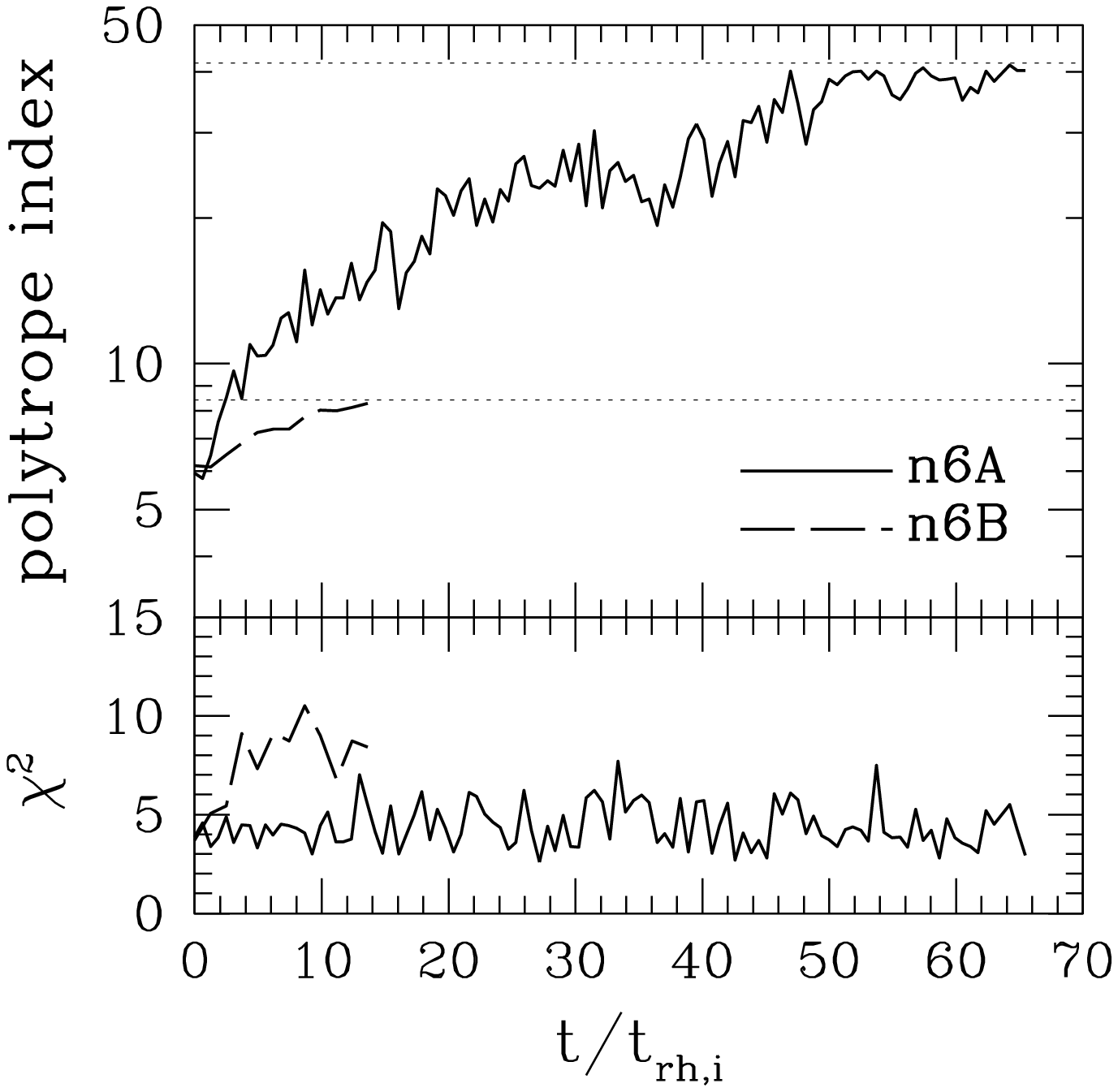}
  \end{center}

\vspace*{-0.2cm}

    \caption{Evolution of polytrope index in the cases of 
 runs $n6$A and $n6$B \label{fig: poly_n6AB_evolve_n} }
\end{figure}

%
%
%
%
%
%
Figures \ref{fig: snapshot_n5A}--\ref{fig: snapshot_n5D}  
show the snapshots of the 
density profile, the distribution function and the velocity dispersion profile 
obtained from the runs $n5A$--$n5D$, which were all started with the
same polytrope index $n=5$, but with the different initial density
contrasts; $D=100$(run $n5A$, Fig.\ref{fig: snapshot_n5A}): 
$D=10^3$(run $n5B$, Fig.\ref{fig: snapshot_n5B}), 
$D=10^4$(run $n5C$, Fig.\ref{fig: snapshot_n5C}) and 
$D=10^6$(run $n5D$, Fig.\ref{fig: snapshot_n5D}). 
Also in figures \ref{fig: snapshot_n6A} and  
\ref{fig: snapshot_n6B}, the results taken from 
the runs $n6A$ and $n6B$ are depicted, whose initial density contrasts are 
$D=110$ and $10^4$, respectively.

\begin{figure}
  \begin{center}
    \epsfxsize=6.5cm
    \epsfbox{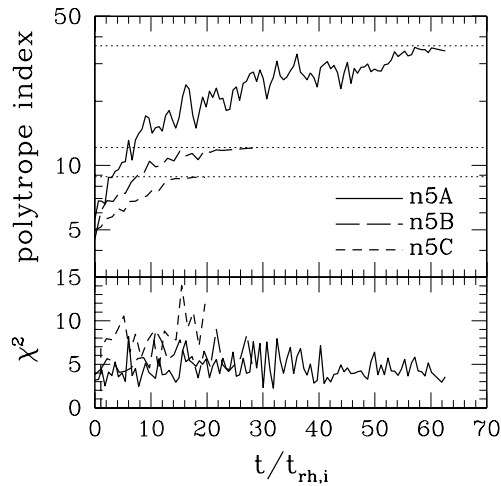}
  \end{center}

\vspace*{-0.2cm}

    \caption{Evolution of polytrope index in the runs $n5$A, $n5$B and $n5$C. 
    \label{fig: poly_n5ABC_evolve_n} }
\end{figure}

Similarly to the run $n3A$, the results from the runs $n5A$--$n5C$,   
$n6A$ and $n6B$ 
exhibit the quasi-equilibrium regime in the early phase of the evolution. 
The evolutionary sequences in the quasi-equilibrium regime can be 
quantitatively characterized by a family of stellar polytropes 
with time-dependent polytrope index. Figures  
\ref{fig: poly_n5ABC_evolve_n} and \ref{fig: poly_n6AB_evolve_n} 
summarize the fitting results for polytrope index as function of time 
in logarithmic scales. It seems apparently that for the cases starting with 
smaller value of $D$ (i.e., runs $n5A$ and $n6A$), 
time variation of polytrope index is 
systematically large and  
statistical fluctuation become noticeable, however,  
it turns out that this fact simply comes from the geometrical 
reason for the equilibrium sequence plotted in figure \ref{fig: lambda_poly}. 
That is, in $(\lambda,~D)$-plane, a number of trajectories of the stellar 
polytropes with different 
polytrope index $n$ are assembled at 
the narrow region with smaller value, $D$ or $\lambda$, causing a large 
variation and/or fluctuation in the time evolution of polytrope index.

Apart from this behavior, figures \ref{fig: poly_n5ABC_evolve_n} and 
\ref{fig: poly_n6AB_evolve_n} show that 
the timescales of quasi-equilibrium state crucially depend on 
the initial conditions, which are basically 
characterized by the initial density contrast $D$ or the dimensionless 
energy $\lambda$. 
Qualitatively, the timescale of quasi-equilibrium state is understood from 
the local estimates of the relaxation time, which are 
inversely proportional to the local density 
\citep[e.g.,][]{Spitzer1987,BT1987}:
\begin{equation}
t_{r}=0.065\,\,\frac{v^3}{G^2m\,\rho\,\,\ln\Lambda}.
\label{eq:local_t_r}
\end{equation}
For a system with small initial density contrast, the depth of the 
potential becomes shallower and 
equation (\ref{eq:local_t_r}) implies that the relaxation 
proceeds slowly enough in both core and halo. The quasi-equilibrium state 
is thus expected to be long-lived. As anticipated in figures 
\ref{fig: snapshot_n5A} and \ref{fig: snapshot_n6A}, 
fitting to the stellar polytropes is successful until $t\simeq63t_{\rm rh,i}$ 
for run $n5A$ and $66t_{\rm rh,i}$ for run $n6A$, much longer than the 
other cases.

On the other hand, for a system with large initial density contrast, 
equation (\ref{eq:local_t_r}) suggest that the decoupling of the 
timescales between core and halo would occur earlier and the 
quasi-equilibrium phase would be short-lived. 
Indeed, the distribution functions for runs $n5C$ and $n6B$ 
show that while the high-energy part of the function $f(\epsilon)$ almost 
remains the same,   
the low-energy part of the distribution is rapidly developed and stretches 
toward $\epsilon\to-\infty$. Also,  
the inner part of the density profile grows rapidly and shows 
an intermittent behavior with large fluctuation. 
As a result, fitting to the stellar polytropes 
failed earlier at $t\sim 20t_{\rm rh,i}$ for run $n5C$ and 
$12t_{\rm rh,i}$ for run 
$n6B$. After that, the system soon becomes unstable state, 
finally undergoing the core-collapse, 
consistent with the thermodynamic prediction based on the Tsallis entropy. 
Note that the fitting results in the runs $n5C$ and $n6B$ are slightly 
worse. From bottom panels of figures \ref{fig: poly_n5ABC_evolve_n} 
and \ref{fig: poly_n6AB_evolve_n}, the 
estimated $\chi^2$ values given by (\ref{eq:def_of_chi^2}) become 
$5\sim14$ in both runs, which is larger than the typical 
value $3-7$ in cases with the smaller initial density contrast.

The observation in both fitting results and the 
timescales for quasi-equilibrium states indicates 
that for some large values of the initial density contrast, 
the fitting to stellar polytropes would fail from the beginning and 
the quasi-equilibrium state ceases to exist. 
Indeed, such an example was obtained from the run $n5D$ (see 
Fig.\ref{fig: snapshot_n5D}). The initial density contrast 
of this run is $D=10^6$ and the dimensionless energy is $\lambda=2.345$.  
That is, the location of the initial state 
in the $(\lambda, D)$-plane is outside the region depicted 
in figure \ref{fig: lambda_poly}. In this case, the central part of the 
system is highly 
concentrated and the core-halo structure is developed from the beginning. 
The resultant relaxation timescale is 
much shorter than that of the other cases and the system soon 
becomes unstable, leading to the earlier core-collapse. 
Compared to the stable stellar polytropes with the same initial energy 
$\lambda=2.345$ and with $D<D_{\rm crit}$, 
none of the model parameters successfully reproduce 
the simulated density and/or velocity dispersion profiles 
(long-dashed, short-dashed and dot-dashed lines in 
Fig.\ref{fig: snapshot_n5D}). 
Though the collapse time crucially depends on the softening parameter, 
the convergence test as examined 
in appendix A 
suggests that the collapse time converges to $18t_{\rm rh,i}$, close to the 
standard result without the adiabatic wall, i.e., 
$t_{\rm coll}\simeq 16 t_{\rm rh,i}$ 
\citep[e.g., Table13.2 of][]{Aarseth2003}.

Therefore, for a general initial condition with 
large $D$ or $\lambda$, 
quasi-equilibrium behavior generally ceases to exist. 
In other words, long-lived quasi-equilibrium states appear at the smaller 
value of $\lambda$ or $D$ and the system is quantitatively characterized by 
the stellar polytropes. Their lifetime is expected to become  
much longer as approaching down to $\lambda= 0.335$ and/or $D= 709$, 
i.e., the critical values for the marginal stability in isothermal 
distribution. 
%
%
%
%
%
%
%
%
%
%
%
%
%
%
%
%
%
%
%
%
%
%
%
%
%
%
%
%
%
%
%
\subsubsection{Discussion}
\label{subsubsec:discussion}
%
%
%
%
So far, we have focused on the characterization of the quasi-equilibrium 
state using the one-parameter family of the stellar polytropes. While most of 
the transient  
state is well-approximated by the stellar polytropes with varying polytrope 
index, one may criticize that the use of the stellar polytropes is not the 
best characterization for the quasi-equilibrium sequence. Indeed, even 
restricting the stellar distribution to the stationary solutions 
of the Vlasov equation, one can, in principle, construct the infinite set of 
one-parameter family of stellar models. In this sense, the stellar polytropes 
should be regarded as a particular set of stellar models. 
Further, there is no rigorous proof for 
the uniqueness to characterize the quasi-equilibrium states.

In real astronomical systems such as globular clusters,  
the stellar polytropes had been known to poorly fit to the observed 
structure of stellar distribution. 
Rather, the majority of the Galactic globular 
clusters is quantitatively characterized by the King model 
\citep[][]{King1966,BT1987,Spitzer1987,MH1997}. 
In contrast to the stellar polytropes as {\it $q$-exponential} 
distribution, King model is represented by the {\it truncated}  
exponential distribution:  
\begin{eqnarray}
	f(\epsilon) \propto  
\left\{
\begin{array}{lcl}
e^{-\beta \epsilon'}-1 &;& \epsilon' <0
\\
0 &;& \epsilon'\geq0
\end{array}
\right.,
\label{eq:King_model}
\end{eqnarray}
where we define $\epsilon' = \epsilon-\phi(r_B)$ with radius $r_B$ being 
the truncation radius and $\epsilon$ is the specific energy of a particle. 
Provided the dimensionless energy $\lambda$, 
the equilibrium sequence of the King model is then 
characterized by the one-parameter $W_0=2\beta[\phi(r_B)-\phi(0)]$, 
which represents the depth of the gravitational potential.

\begin{figure}
  \begin{center}
    \epsfxsize=5.5cm
    \epsfbox{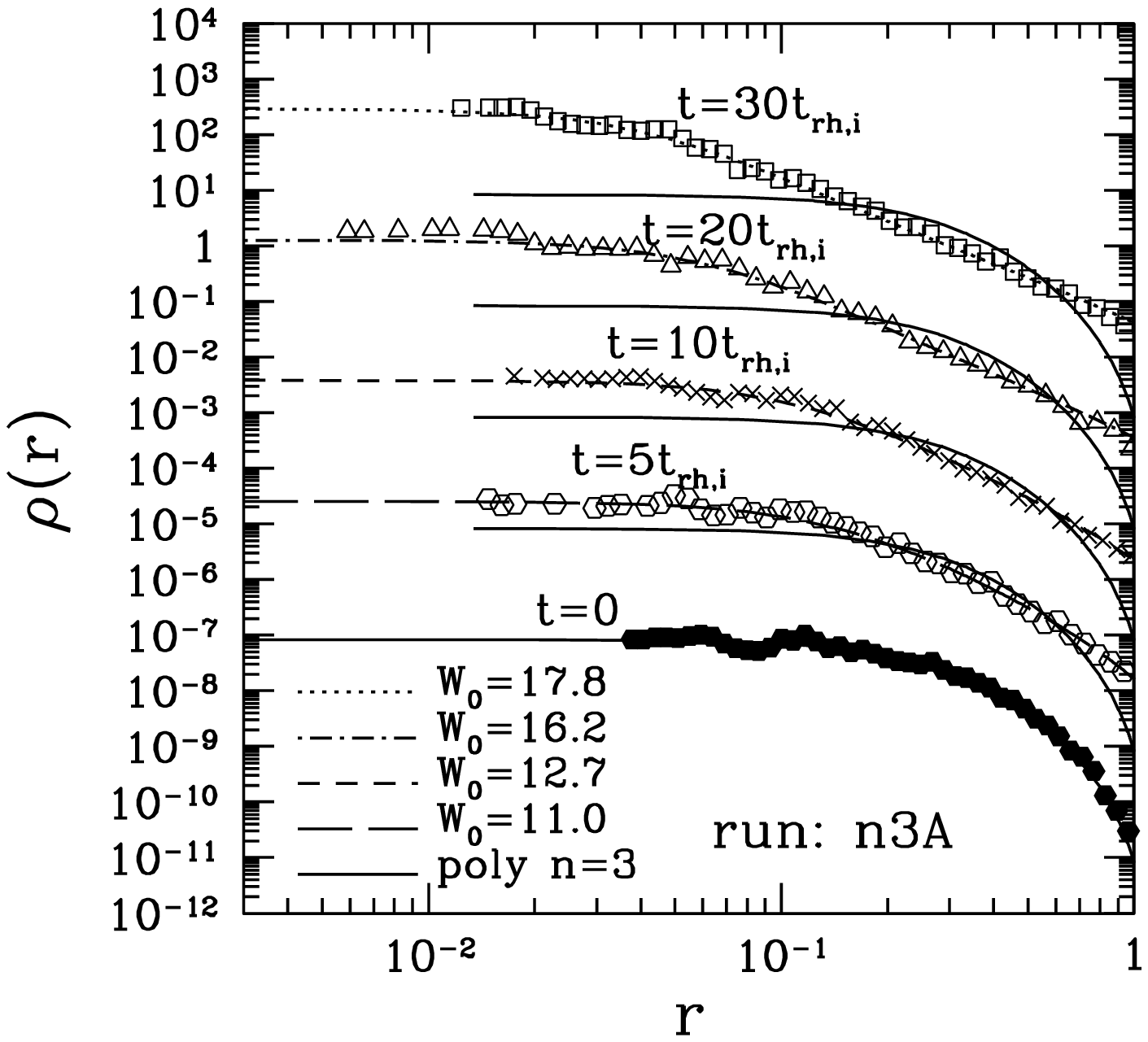}
    \epsfxsize=5.5cm
    \epsfbox{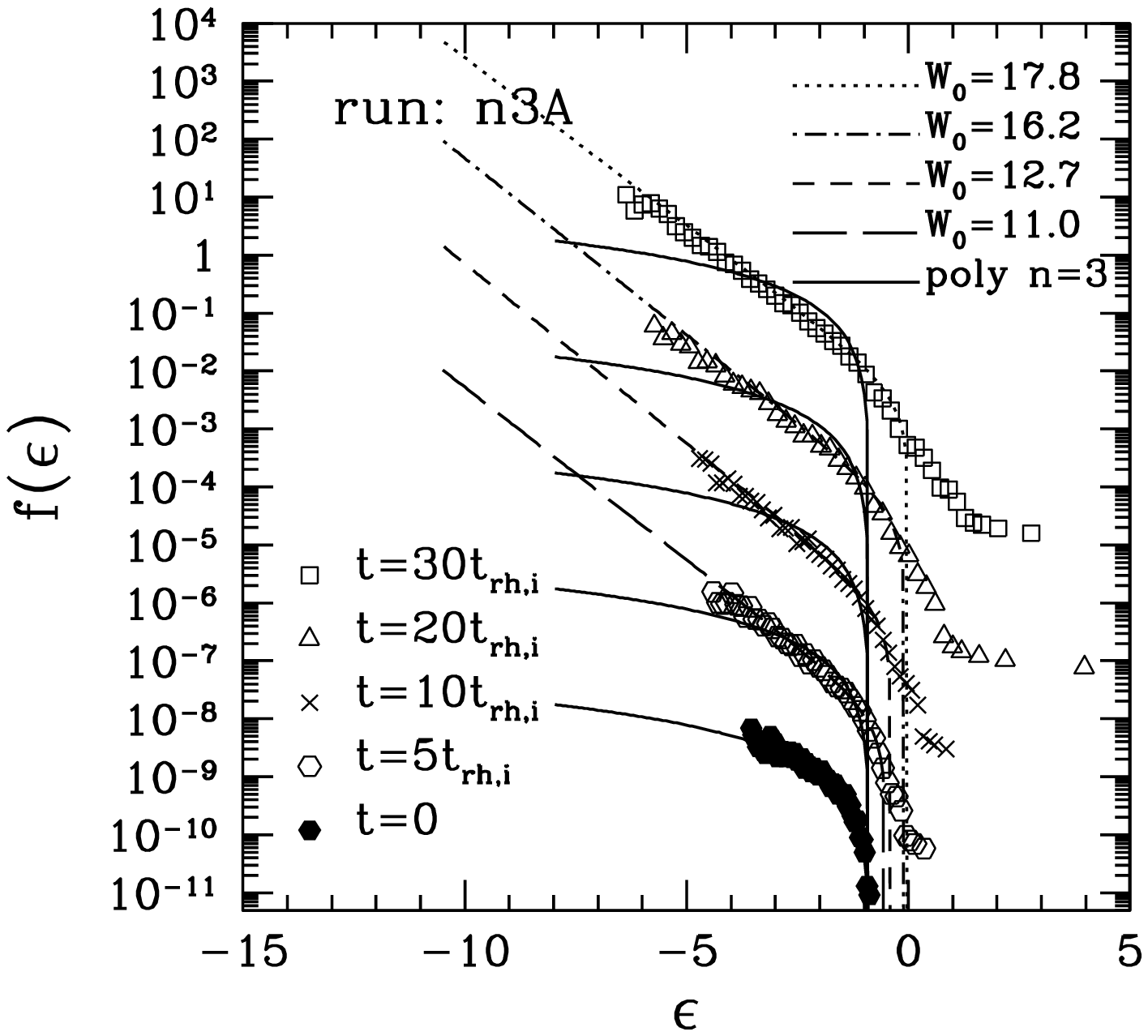}
    \epsfxsize=5.5cm
    \epsfbox{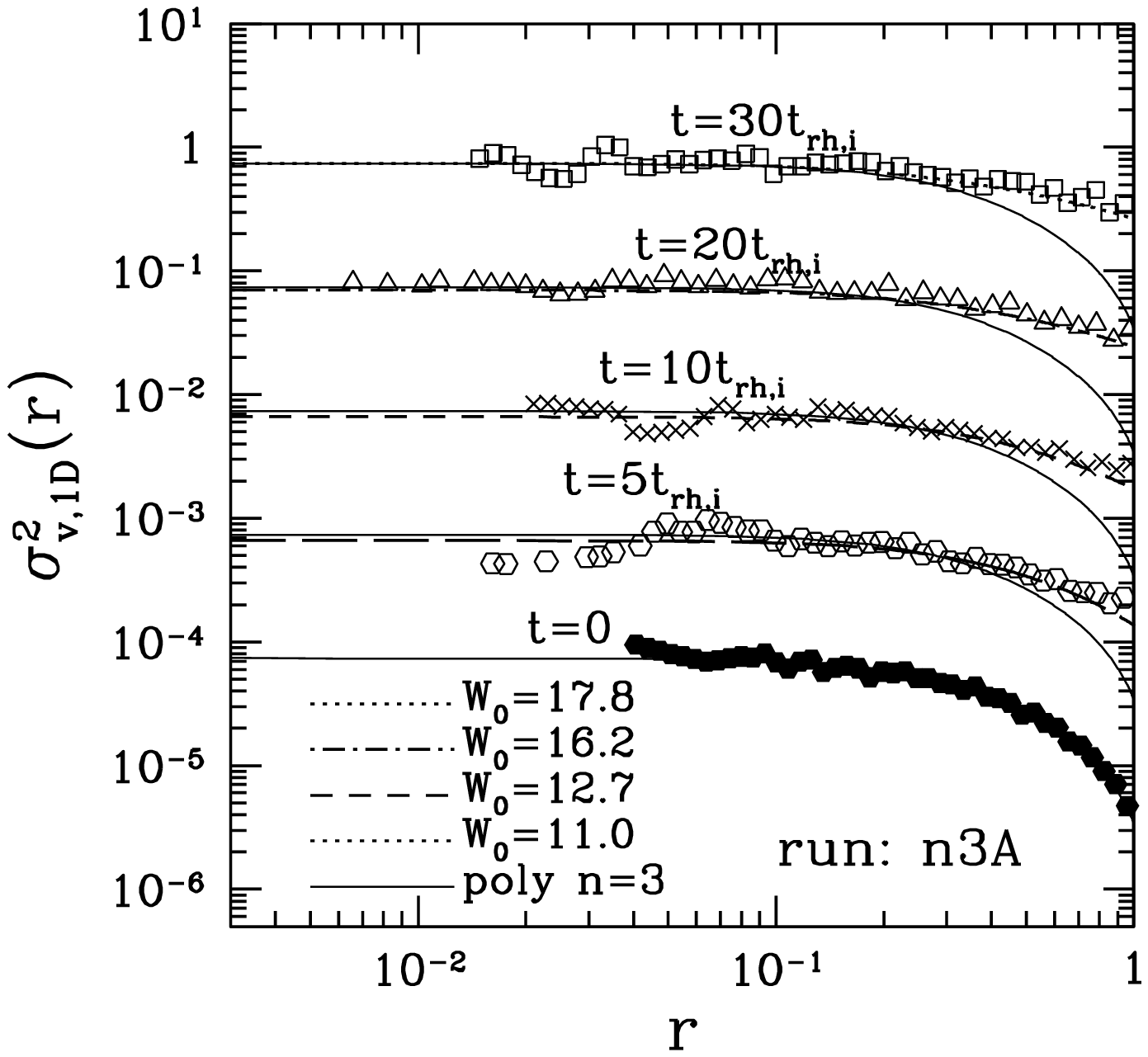}
  \end{center}

\vspace*{-0.2cm}

    \caption{Same as Fig.\ref{fig: poly_n3A_fit_poly}, but results are 
	compared with the King models. The fitting results shown in 
 the lines except for solid lines are obtained by varying the
 dimensionless parameter, $W_0=2\beta[\phi(r_B)-\phi(0)]$ under 
 keeping the mass and the total energy fixed. 
    \label{fig: poly_n3A_fit_king} }
\end{figure}

In figure \ref{fig: poly_n3A_fit_king}, the $N$-body data taken from the 
run $n3A$ is used to compare with the King model. Note that the fitted values 
of the parameter $W_0$ indicated by figure \ref{fig: poly_n3A_fit_king} 
were obtained under the suitable restriction, $r_B>r_e$. 
Similar to the stellar polytropes, the density and the velocity 
dispersion profiles reasonably fit to the King model. 
The fitted value of the parameter $W_0$ gradually increases as time goes on,  
indicating that the depth of the potential becomes deeper.  
Compared with the observed Galactic globular clusters with typical 
range $W_0=4$--$10$ \cite[][]{TKD1995},    
fitting results for $W_0$ are somewhat large. 

On the other hand, turn to focus on the distribution function, 
deviation from the King model 
becomes manifest at the high-energy tails $\epsilon>0$. While the 
distribution function for King model falls off at the truncation 
energy $\epsilon =\phi(r_B)$ which takes the negative value, 
number of high-energy particles with 
$\epsilon>0$ gradually increases in our $N$-body calculation, 
which contributes to a high-energy tail of the distribution function. 
Although the low-energy part of the distribution function 
resembles the exponential form of the King model, the discrepancy at 
the high-energy part implies that the boundary condition 
in our idealistic situation is very different from that of the Galactic 
globular clusters.
In fact, the influence of external tidal field 
is significant for the Galactic globular cluster and the 
resultant distribution function sharply falls off at the tidal 
boundary \citep[]{BM1998}. The stellar particles which are 
usually bounded to the system tend to escape from the 
globular cluster system \citep[e.g.,][]{FH1995,BM2003,TF2005}.
To mimic this effect, 
truncation radius $r_B$ is artificially introduced by hand in the 
King model. On the other hand, in presence of the adiabatic wall, 
the high-energy particles with $\epsilon>0$, which are 
usually unbounded, cannot freely escape 
outward from the system and thereby no specific energy cutoff appears.

 As a result, the stellar polytropic distribution which has no 
energy cutoff successfully reproduces the quasi-equilibrium states in  
our $N$-body setup. This means that, in presence of the adiabatic wall,
the simple power-law distribution 
provides a better characterization than the truncated exponential 
distribution. Since the adiabatic wall is an artificial but the 
simplest boundary condition,  the stellar polytropic distribution 
might be regarded as a fundamental stellar model theoretically, 
though not practically useful in characterizing  
the observed structure of Galactic globular clusters.  
%
%
%
%
%
%
%
%
%
%
%
%
%
%
%
%
%
%
%
%
%
%
%
%
%
\subsection{A family of stellar models with cusped density profile}
\label{subsec:Tremaine}
%
%
%
%
%
The analyses in previous subsection have revealed that 
the quasi-equilibrium evolution characterized by the 
stellar polytropic distribution can appear at the  
energy $\lambda$ close to the critical value of the isothermal distribution. 
We then attempt to clarify the generality and/or the physical conditions for 
the quasi-equilibrium state in more general initial conditions. 
In this subsection, as a special class of initial 
conditions that contain the non power-law features,  
we treat a family of stellar models with cusped density profile 
\citep[]{TRBDFGKL1994}. The models contain two parameters,  
one is the scale-radius $a$, and the other is related to the slope of the 
inner density profile $\eta$. The density profiles of these models are 
expressed as: 
\begin{equation}
\rho(r) \propto \frac{1}{(r/a)^{3-\eta}(1+r/a)^{1+\eta}}.   
\label{eq:cusped_profile}
\end{equation}
Note that the above expression includes 
the models considered by \citet{Hernquist1990} for $\eta=2$ and 
by \citet{Jaffe1983} for  $\eta=1$  as special cases. 
%
%
%
%
%
%
%
%
\subsubsection{$N$-body setup and overview}
\label{subsubsec:setup}
%
%
%
%
%
%
%
For a spherically symmetric configuration with isotropic velocity 
distribution, using the Eddington formula, 
the one-particle distribution function can be uniquely reconstructed from 
the density profiles (\ref{eq:cusped_profile}) \citep[][]{BT1987}. 
\citet{TRBDFGKL1994} gave  
useful analytic formulas for various physical quantities such as 
the distribution function, the potential and the velocity dispersion 
profiles for a system extending over the infinite radius, $r\to\infty$. 
Based on their formulas, \citet{Quinlan1996} numerically estimated the 
dependence of the slope $\eta$ on the timescale of core-collapse using 
the Fokker-Planck code. 
For present purpose, however, a direct application of their formulas 
is inadequate in presence of the adiabatic wall.

\begin{figure}
  \begin{center}
    \epsfxsize=7.5cm
    \epsfbox{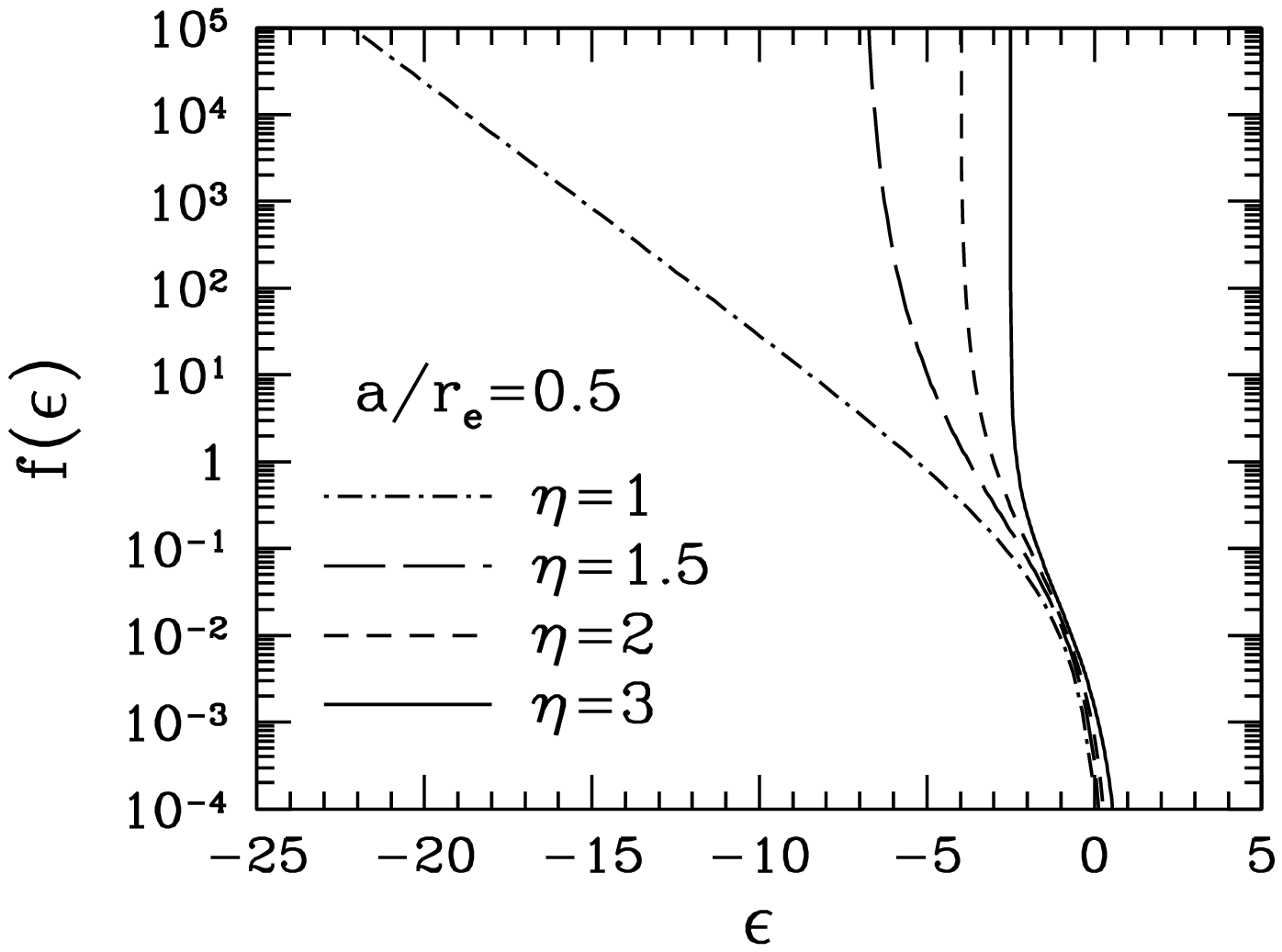}
    \epsfxsize=7.5cm
    \epsfbox{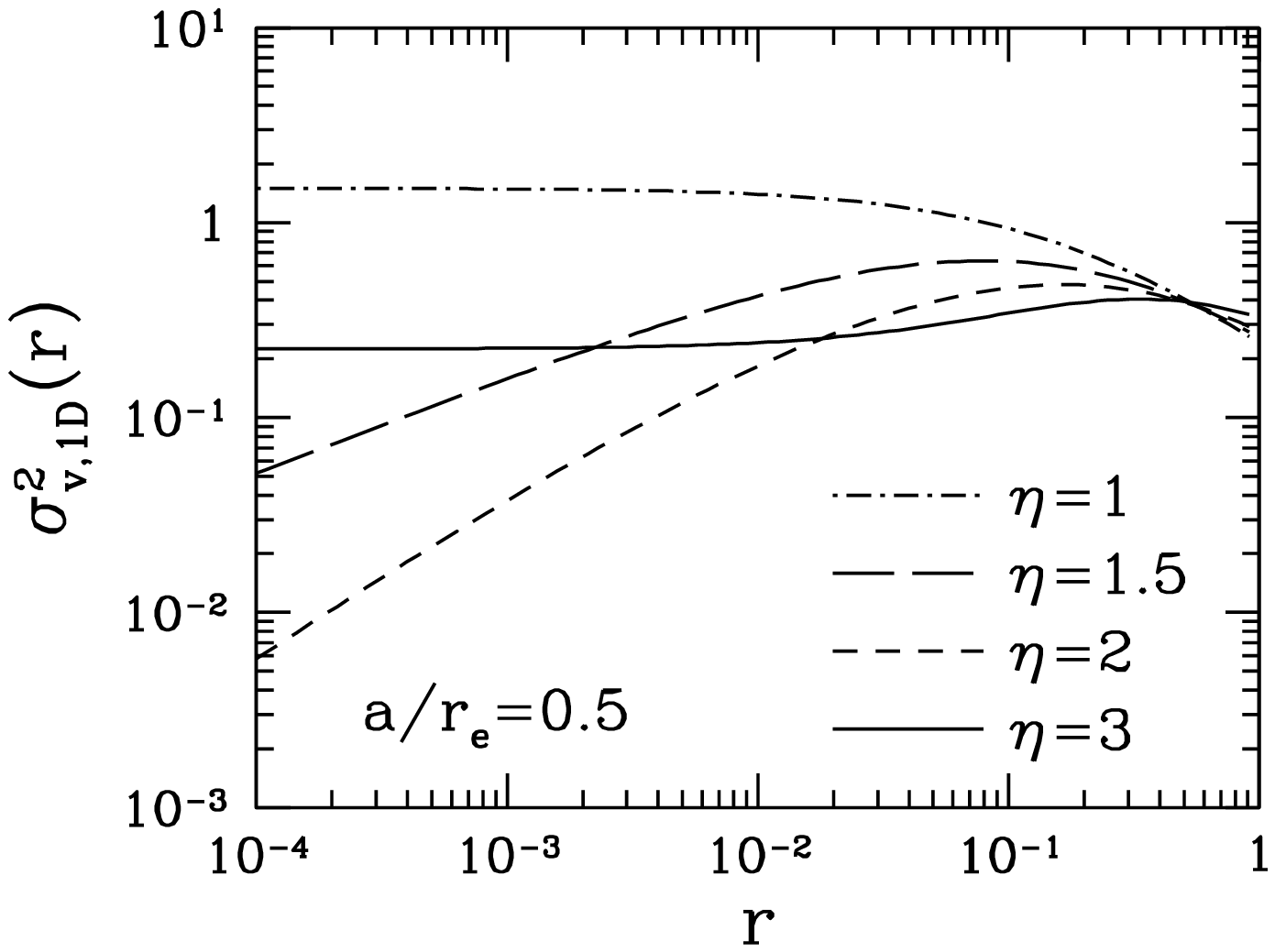}
  \end{center}

\vspace*{-0.2cm}

    \caption{Distribution function({\it Left}) and velocity 
 dispersion profile({\it Right}) for a family of stellar model 
 by Tremaine et al. (1994) in presence of an adiabatic wall. 
 For a specific value of the scale radius $a/r_e=0.5$, 
 the analytic results with various slopes $\eta$ are plotted 
 based on the formulae in Appendix B. 
    \label{fig: tremaine_theor} }
\end{figure}

In appendix B, 
taking account of the truncation 
radius $r_e$, we re-derive 
the analytic formulas for distribution functions 
as well as the other physical quantities. 
Based on this, figure \ref{fig: tremaine_theor} plots the 
theoretical curves for the distribution function({\it left}) and the 
velocity distribution profile({\it right}) 
with a specific choice of the parameters,  
$\eta=1$, $1.5$, $2$ and $3$, fixing the scale-radius to $a/r_e=0.5$. 
For models with $\eta>1$,  
the distribution function exhibits a divergent behavior at a finite energy 
$\epsilon=\epsilon_{\rm min}<0$, $f(\epsilon_{\rm min})=+\infty$. Further, the 
velocity dispersion profile shows non-monotonic behavior; it 
first increases and eventually turns to decrease as approaching the center. 
These peculiar features imply that the inward heat flow occurs along a 
course of the relaxation, causing the central part of the system expand, 
which  is the same phenomenon as observed in the 
unstable case of the isothermal distribution 
(see Sec.\ref{subsec:isothermal}).

To perform a simulation, we increased the number of particles 
 to $N=8K$ in order to resolve the central part of the cuspy density 
profiles.  The softening parameter of gravitational potential is set 
to $\epsilon=4/N$. Note that the convergence test 
in appendix A 
suggests that much smaller value of the softening parameter 
should be used for a system with highly concentrated core. 
However, decreasing the softening parameter requires a much longer  
calculation time and the probability of binary-formation via 
three-body interaction increases around the core. Since we 
use a simple individual time-step algorithm without any regularization 
schemes, the $N$-body integration becomes heavily time-consuming 
once a tight binary is formed. 
Hence, we do not discuss here the timescale of the quasi-equilibrium 
state and focus only on the condition for quasi-equilibrium states. 
The quantitative estimates of the timescale will be presented in future task.

\begin{figure}
  \begin{center}
    \epsfxsize=8.5cm
   \epsfbox{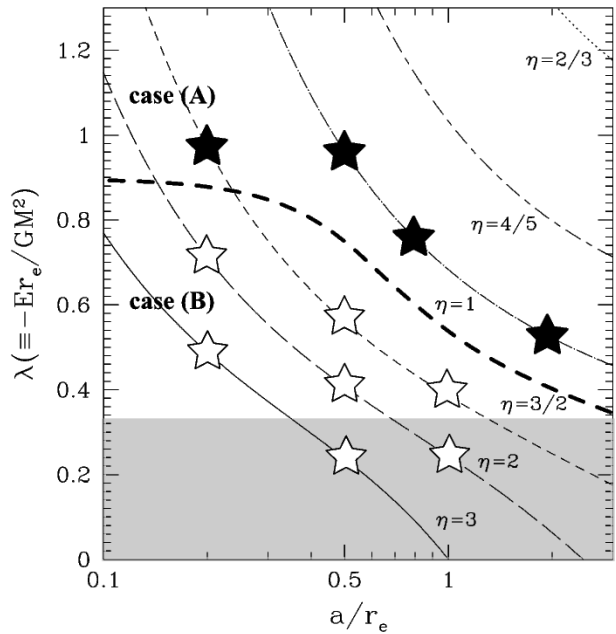}
  \end{center}

\vspace*{-0.2cm}

    \caption{Energy-scale radius relationship in stellar models 
	by Tremaine et al. (1994) in presence of an adiabatic wall. 
 The thin lines represent the equilibrium sequences of stellar model 
 with different values of index $\eta$. Along each line, we plot the 
 star symbols, which indicate the result of 
 $N$-body simulations fitted to the stellar polytropes. The fitting 
results shown in the figure are categorized as 
case (A) and case (B), which are roughly separated by thick dashed
line. The case (A) denoted by filled symbols 
 means that fitting to the (stable) stellar polytropes 
 was failed, while the case (B) denoted by open symbols indicates that the 
 transient states approximated by the stellar polytropic distribution 
 appeared as quasi-equilibrium state.    
     \label{fig: lambda_tremaine} }
\end{figure}

In table \ref{tab:model_tremaine}, the model parameters of 
the initial conditions examined here are summarized, together with the 
evolutionary status. Also in figure \ref{fig: lambda_tremaine},  
the result for each run is represented by the symbols 
in energy-scale radius relation. Roughly speaking, figure 
\ref{fig: lambda_tremaine} says that the quasi-equilibrium 
sequence characterized by the stellar polytrope appears({\it open
stars}, case(B)) when the total energy of the system is not 
far from the critical energy of isothermal distribution,  
$\lambda_{\rm crit}=0.335$. However, 
this is not a sufficient criterion   
for the presence of quasi-equilibrium state.  
For models with $\eta=1$, whose density profile 
resembles a singular isothermal sphere at the center, 
the system cannot be fitted by the stellar polytrope ({\it filled
stars}, case(A)). On the other hand, for the initial conditions with 
$\lambda<\lambda_{\rm crit}$ ({\it stars} in shaded-region), the system 
finally approaches the isothermal state. In this case, 
the transient state of the system could be also 
fitted by the stellar polytropes,  although  
the fitted value of the polytrope index is so large that 
one cannot be easily discriminate it from the isothermal distribution.
%
%
%
%
%
%
%
%
\subsubsection{quasi-attractive behaviors and condition for 
quasi-equilibrium state}
\label{subsubsec:quasi-attractor}
%
%
%
%
%
%
%
Let us focus on the characteristic behaviors of the 
long-term evolution  by picking up some typical examples. 
Figures \ref{fig: tremaine_eta2C}, \ref{fig: tremaine_eta1.5B} and 
\ref{fig: tremaine_eta1C} show the results obtained from the 
runs $\eta2C$, $\eta1.5B$ and $\eta1C$.  
Figure \ref{fig: tremaine_eta2C} shows the typical example 
in which the system finally approaches the stable isothermal state. 
Due to the small value of $\lambda$, the effect of self-gravity is 
small and the system evolves very slowly.  
While we tried to fit the transient states of the system to 
the stellar polytropes,  
the resultant value of the polytrope index $n$ is quite large, 
$n\simeq 20$--$60$, indicating the system being in nearly isothermal 
state. For comparison, 
we also plot the theoretical curves for isothermal distribution.  
It seems difficult to discriminate which models 
are fitted to the simulation results better.

By contrast, in figure \ref{fig: tremaine_eta1.5B}, 
the quasi-equilibrium state approximated by the stellar polytropes 
appears. In the run $\eta1.5B$, because of the dimensionless 
energy $\lambda>0.335$, the system finally undergoes core-collapse.  
Looking at an early phase, however, the core expansion first 
takes place and the flatter core is formed. 
Then the core density turns to increase gradually and 
the transient state can be approximately described by the stellar 
polytropes for a long time ($t\simlt30t_{\rm tr,i}$). 
These behaviors, which may be regarded as quasi-attractive behavior, 
can be explained from the inner structure of the initial velocity 
dispersion profile. 
That is, due to the small value of the local relaxation time 
$t_r$ at the core, the inward heat flow first occurs toward the 
equipartition of the kinetic energy, 
leading to the uniform velocity dispersion at the inner part. Then, 
the outward heat 
flow next occurs and the inhomogeneity in the velocity dispersion 
is gradually erased. Although this slow relaxation does not stop and 
finally leads to the 
catastrophic heat flow, the system is remarkably long-lived.

On the other hand, the run $\eta1C$ has slightly smaller value of $\lambda$ 
than the run $\eta1.5B$ and one naively expects that the system is stable. 
However, the results shown in figure \ref{fig: tremaine_eta1C} is completely 
opposite. In contrast to the run $\eta1.5B$,  the initial condition of the 
run $\eta1C$ has uniform velocity dispersion at the inner part. This implies 
that the inward heat flow is only supplied by the randomness 
of the initial perturbation, as seen in the unstable isothermal case 
(see Sec.\ref{subsec:isothermal}). 
Thus, the amount of the heat flow is insufficient and the resultant core 
radius is rather small, whose density profile cannot be approximated by 
the stable stellar 
polytropic distribution. For comparison, in figure \ref{fig: tremaine_eta1C}, 
we plot the stellar polytrope with index $n\simeq17.6$, which is the 
marginal stable 
state that has the same total energy $\lambda$ as in the run $\eta1C$. 
While the 
distribution function and the velocity dispersion profile resemble the 
marginal stable state of the stellar polytrope, the discrepancy is manifest 
in the density profile.  
As a result, the system is short-lived and could not reach the 
quasi-equilibrium state.

Note that the lifetime in the model $\eta1C$ sensitively 
depends on the randomness of the initial perturbation. Figure 
\ref{fig: tremaine_eta1C} shows the 
run-by-run variation of the time evolution of the core radii, where the core 
radius was estimated according to the procedure given by \citet{CH1985} 
\citep[see also][]{FH1995}. Compared to the runs $\eta2C$ and 
$\eta1.5B$, the run-by-run variation in $\eta1C$ is significant and it 
seems to originate from the first stage of the core expansion. 
This behavior also holds for the other runs $\eta1A$  and 
$\eta1B$. Therefore, the lifetime of the system starting from the initial 
conditions with cusped density profile $\rho\propto r^{-2}$ 
would be generally stochastic. Although the present calculation 
with non-zero softening parameter 
is inadequate to estimate the precise core-collapse time, 
the uncertainty of the collapse 
time would remain true even if the appropriate regularization scheme is 
implemented in our code. 
In other words, the condition for quasi-equilibrium state 
as well as the lifetime of the system are much sensitive to the velocity 
structure of the initial condition. Though the present surveys do not 
give a conclusive statement for  
generality of quasi-equilibrium state, the out-of-equilibrium 
state with the cusped profile $\rho\propto r^{-\alpha},~(\alpha<2)$ 
possibly exhibits the quasi-equilibrium behaviors if the dimensionless 
energy $\lambda$ is close to $0.335$. 
%
%
%
%
%
%
%
%
\subsubsection{Entropy growth and quasi-equilibrium state}
\label{subsubsec:entropy_growth}
%
%
%
%
%
%
%
In order to elucidate the sequence of the quasi-equilibrium 
evolution from a thermodynamic point-of-view, 
we quantify the entropy growth for the $N$-body data, 
$\eta2C$, $\eta1.5B$ and $\eta1C$. 
Though the quasi-equilibrium behavior seen in 
the simulations may imply that the entropy growth can be better 
characterized by the non-extensive Tsallis entropy, 
the entropy measure in quantifying $S_q$ is not, strictly speaking, 
unique when comparing with different values of the polytrope indices $n$, 
or equivalently,  
the $q$-parameters. Therefore, as long as the cases with 
time-varying polytrope indices are concerned, it would be 
natural to quantify the entropy growth with the Boltzmann-Gibbs entropy.

\begin{figure}
  \begin{center}
    \epsfxsize=7.5cm
    \epsfbox{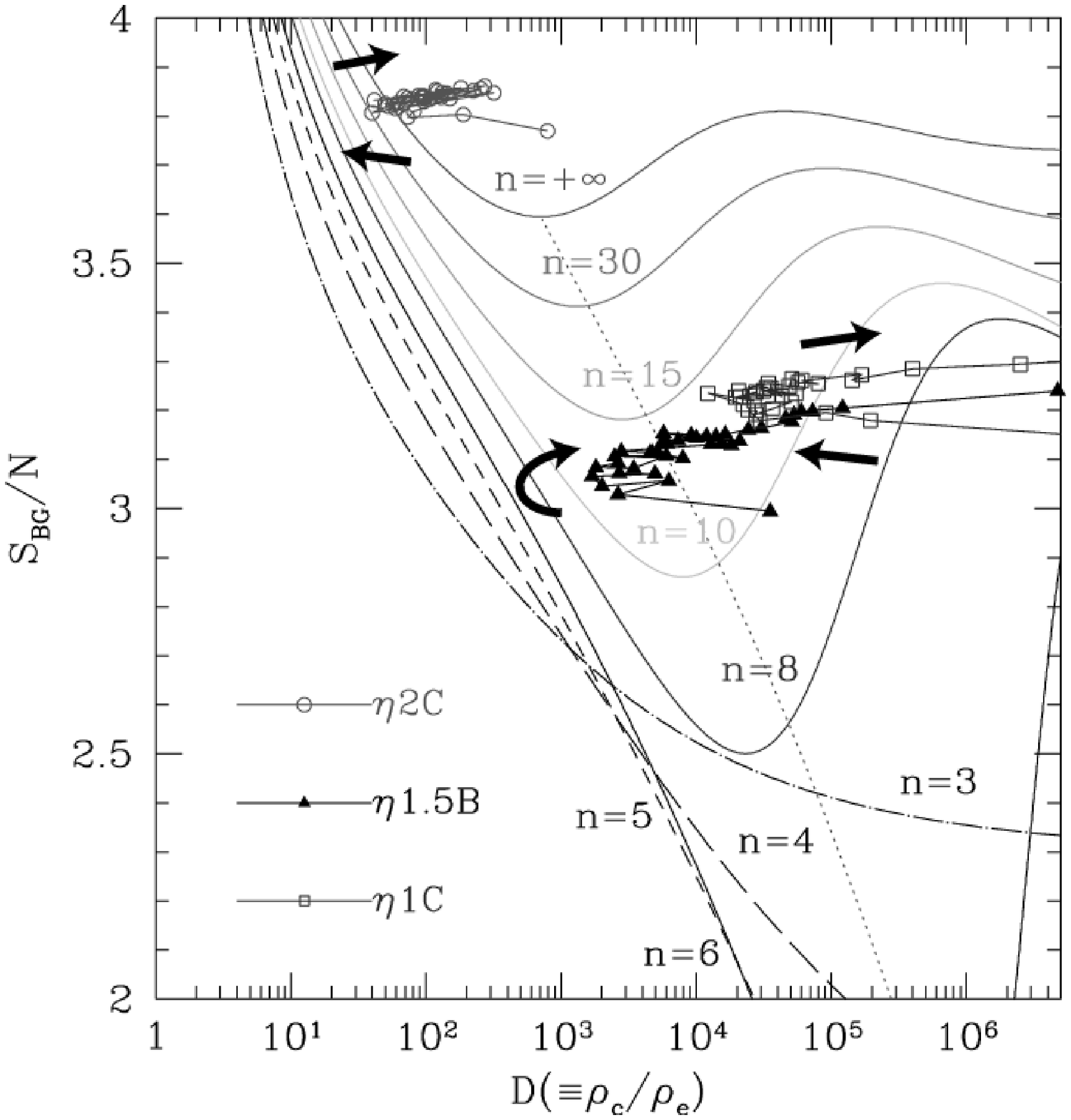}
\hspace*{0.5cm}
    \epsfxsize=7.3cm
    \epsfbox{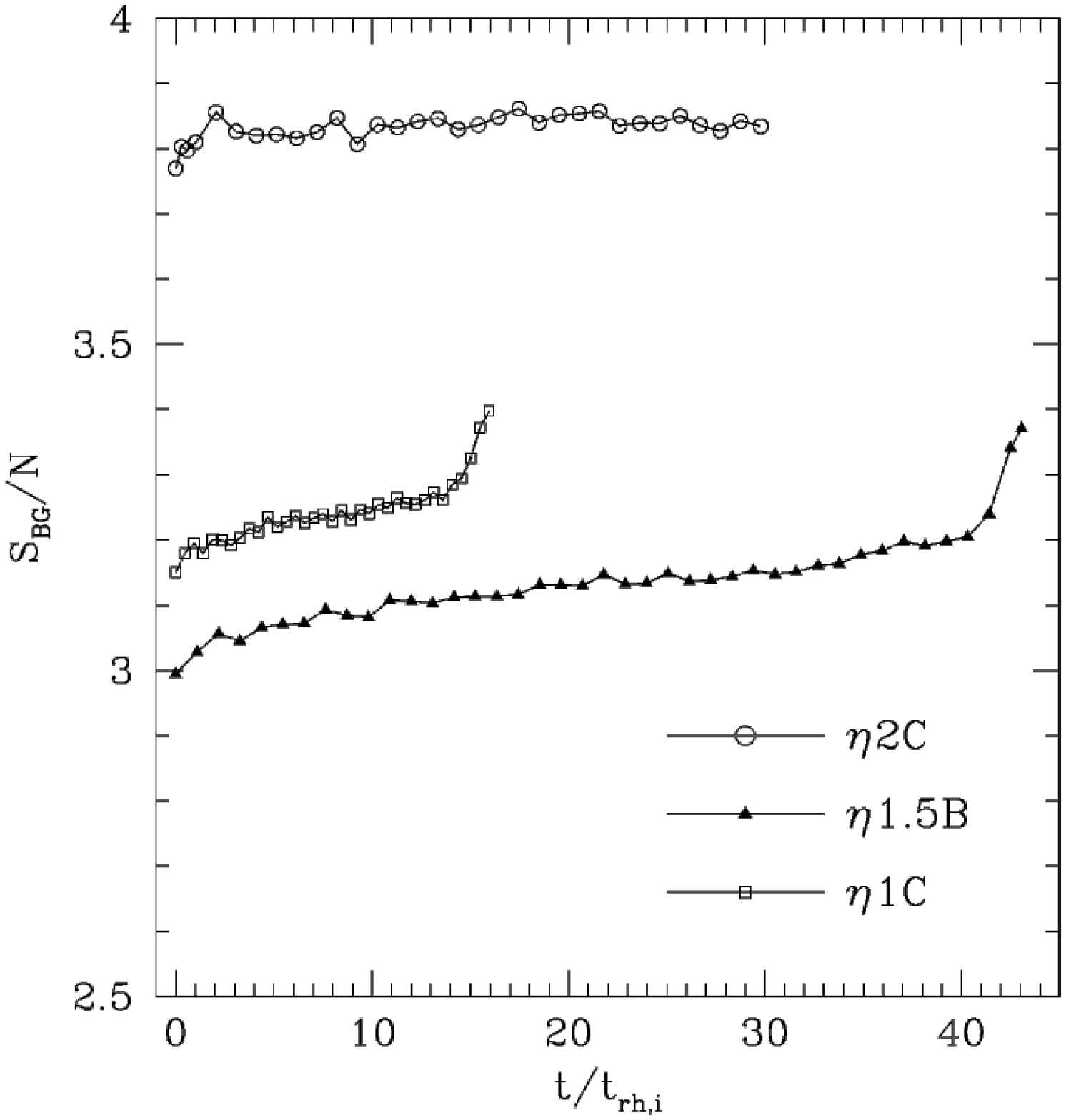}
  \end{center}

\vspace*{-0.2cm}

 \caption{{\it left}: Trajectories in $(S_{\rm BG},D)$-plane obtained 
 from the runs 
$\eta1C$({\it open-squares}), $\eta1.5B$({\it filled triangles}) 
	and $\eta2C$({\it open-circles}). The lines with symbols indicate 
 the simulation results, while the solid, the short-dashed, the 
 long-dashed and the dot-dashed lines are the equilibrium sequences 
of stellar polytropes with different polytrope index $n$, 
which are obtained from the 
 analytic results in Appendix C. On the other hand, 
 the dotted curve represents the 
 stability boundary of the stellar polytropes, inferred from the 
 energy-density contrast relation (dotted curve in 
 Fig.\ref{fig: lambda_poly}). 
	{\it Right}: Time evolution of Boltzmann-Gibbs entropy from the
 runs $\eta1C$, $\eta1.5B$ and $\eta2C$.
    \label{fig: s_BG_tremaine} }
\end{figure}

In left panel of figure \ref{fig: s_BG_tremaine}, the results are 
plotted as the trajectories in $(D,\,S_{\rm\scriptscriptstyle BG}/N)$-plane, 
together with the equilibrium sequence for the stellar polytropes 
denoted by continuous lines. The derivation of theoretical curves for 
isothermal and the stellar polytropes is described 
in appendix C. Also, in right panel, the entropy growth 
is quantified and is plotted as function of time. 
Note that the time interval between the symbols marked along each trajectory 
roughly corresponds to a half-mass relaxation time for the initial 
distribution.

\begin{figure}
  \begin{center}
    \epsfxsize=5.5cm
    \epsfbox{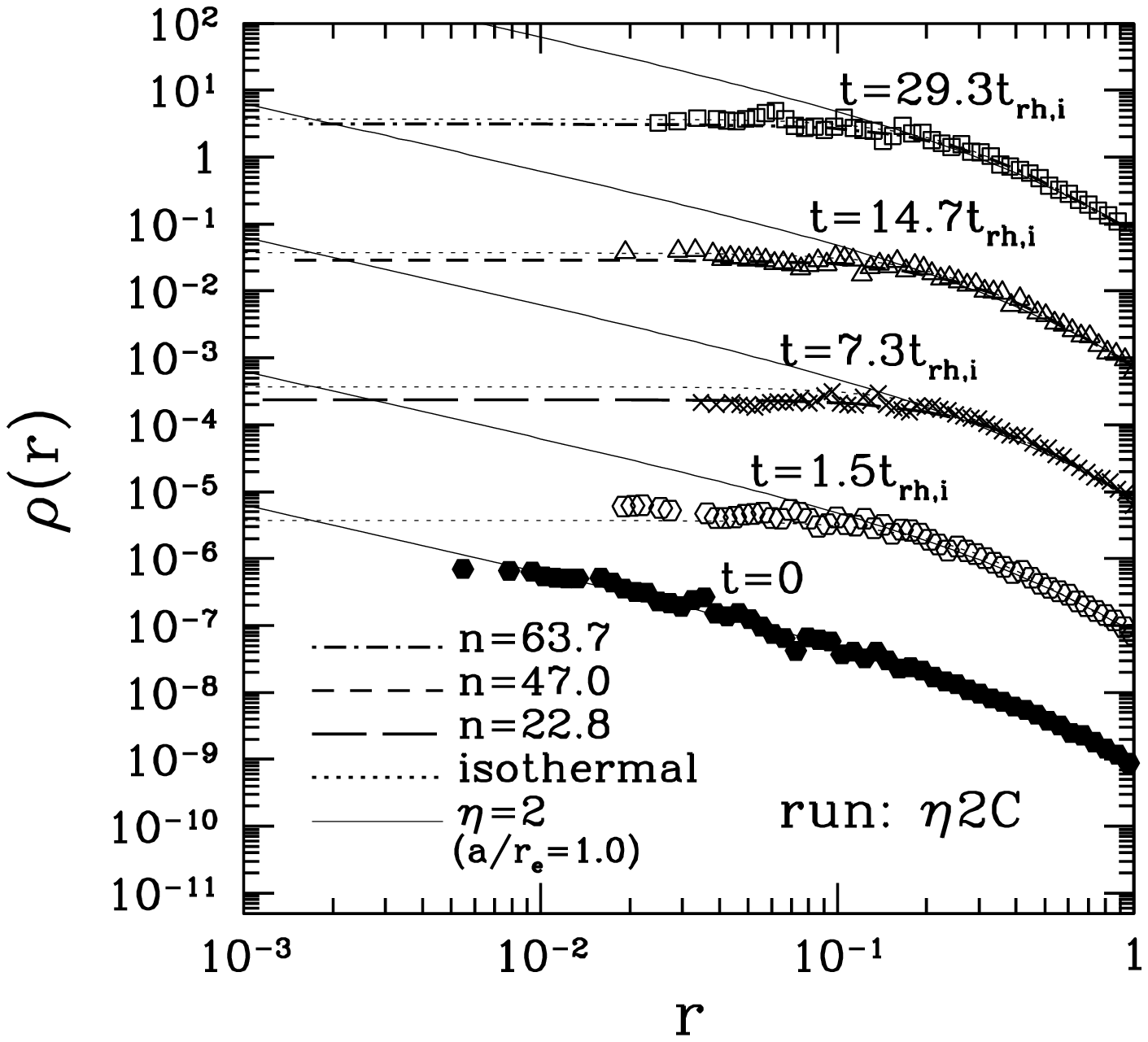}
    \epsfxsize=5.5cm
    \epsfbox{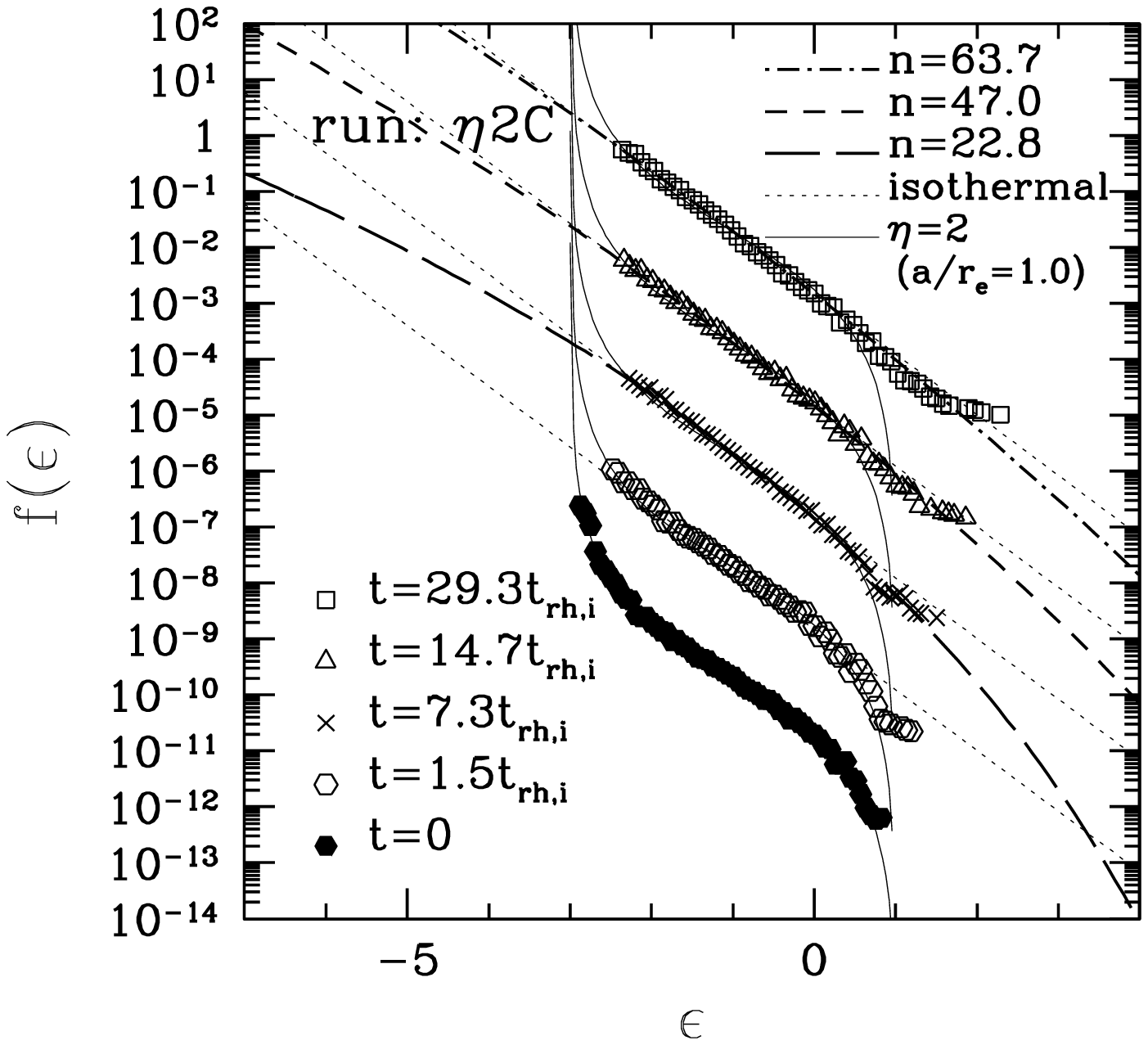}
    \epsfxsize=5.5cm
    \epsfbox{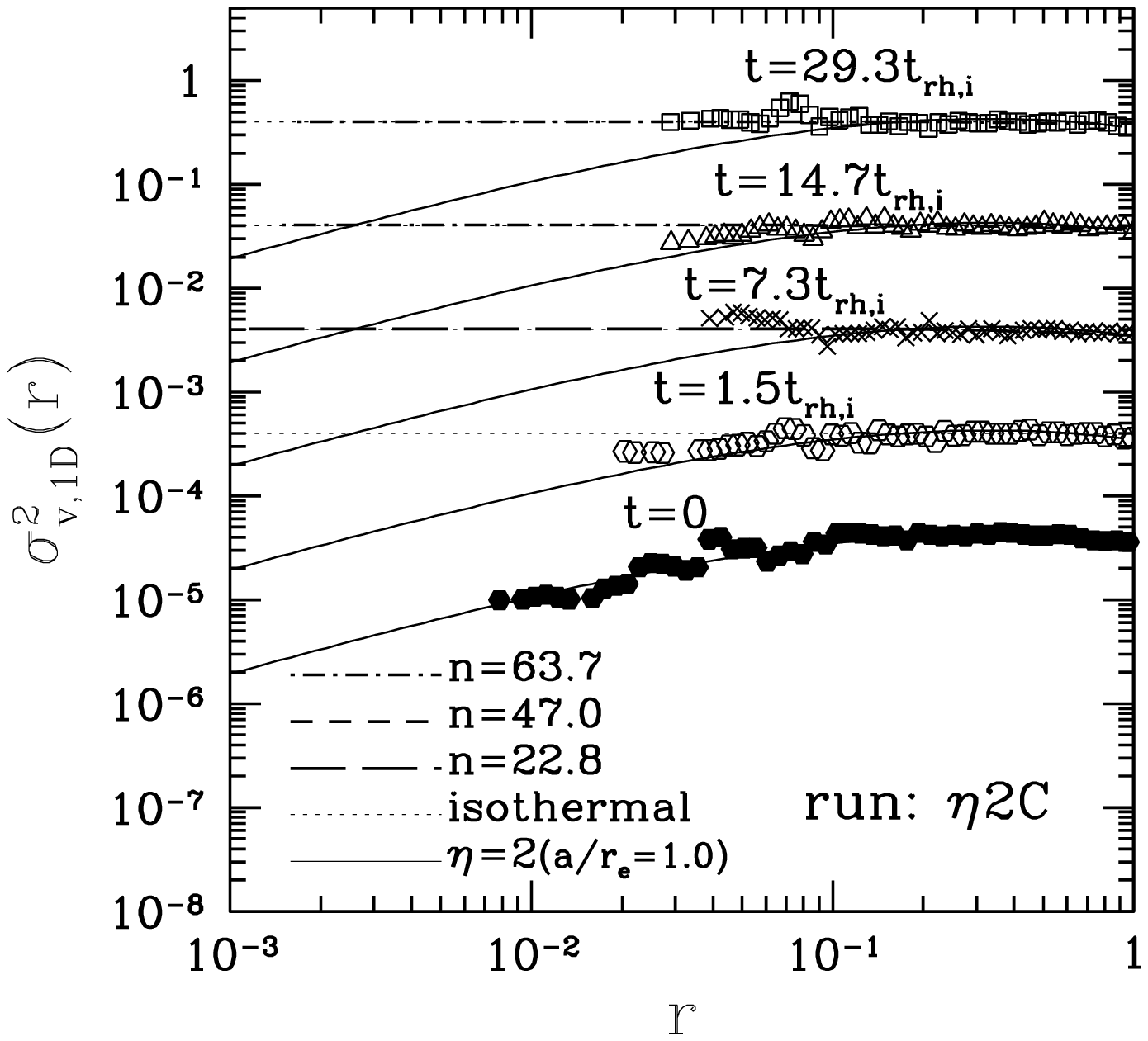}
  \end{center}

\vspace*{-0.2cm}

    \caption{Snapshots of density profile, distribution function and 
	velocity dispersion for run $\eta2C$. 
    \label{fig: tremaine_eta2C} }
  \begin{center}
    \epsfxsize=5.5cm
    \epsfbox{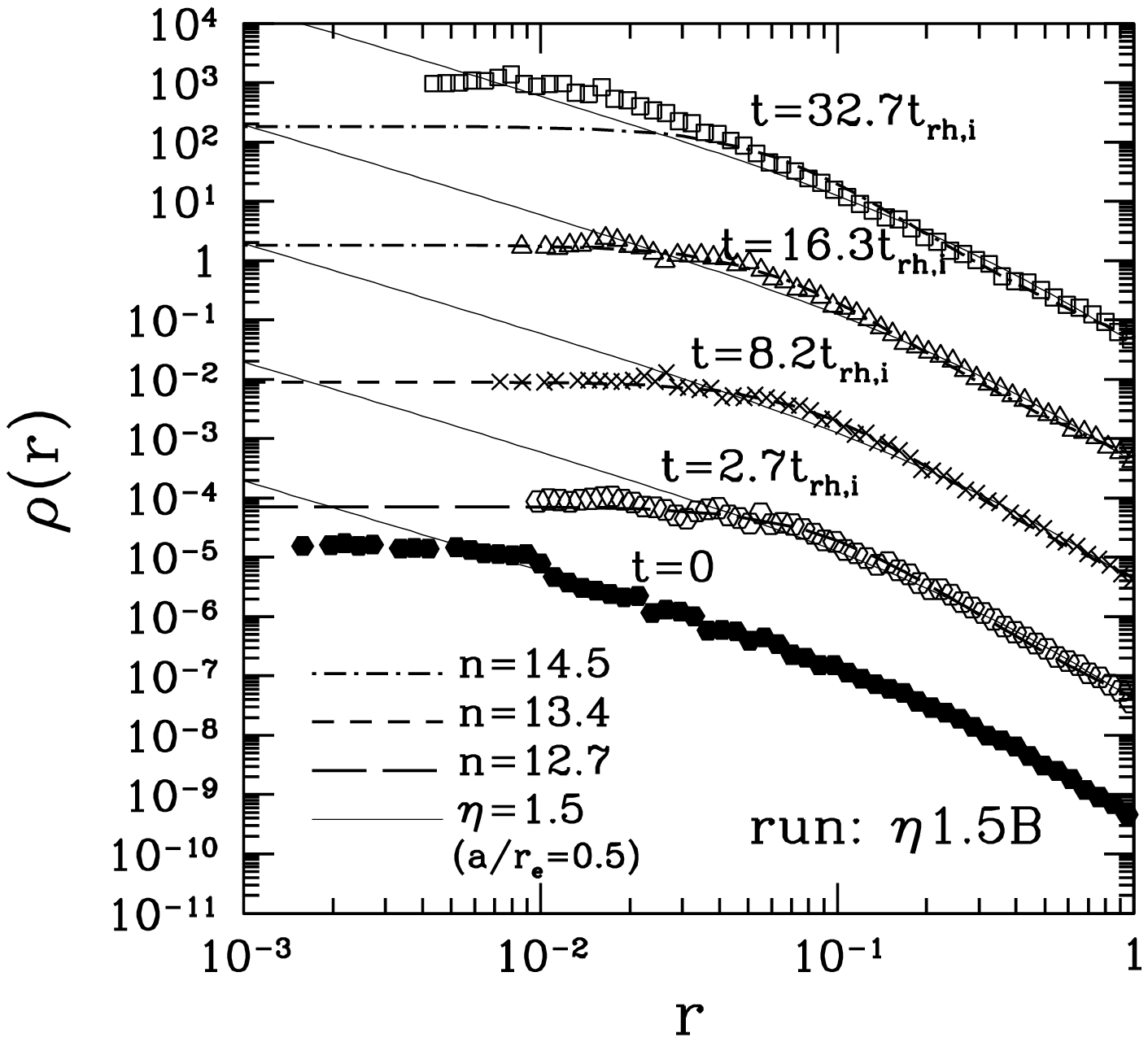}
    \epsfxsize=5.5cm
    \epsfbox{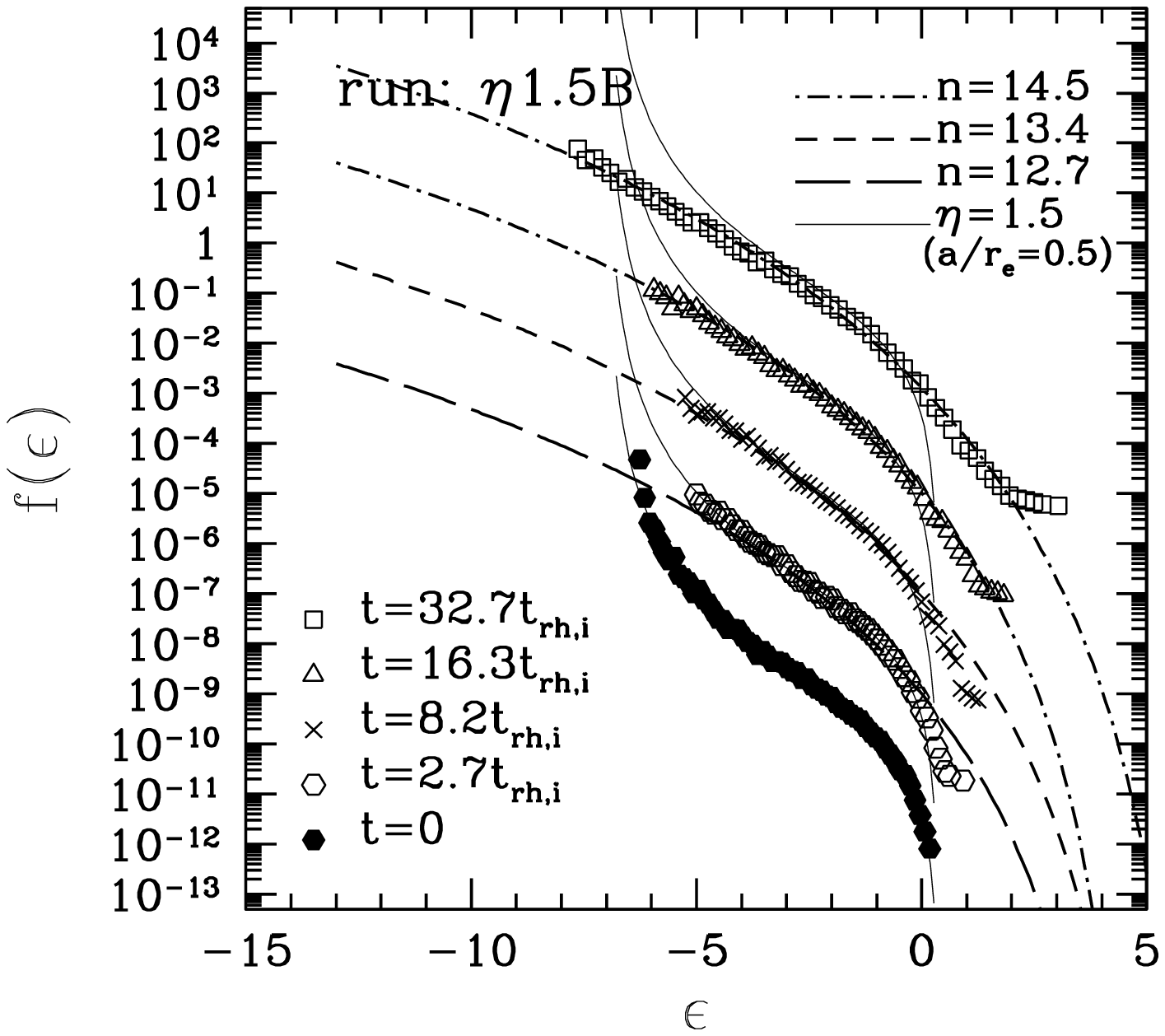}
    \epsfxsize=5.5cm
    \epsfbox{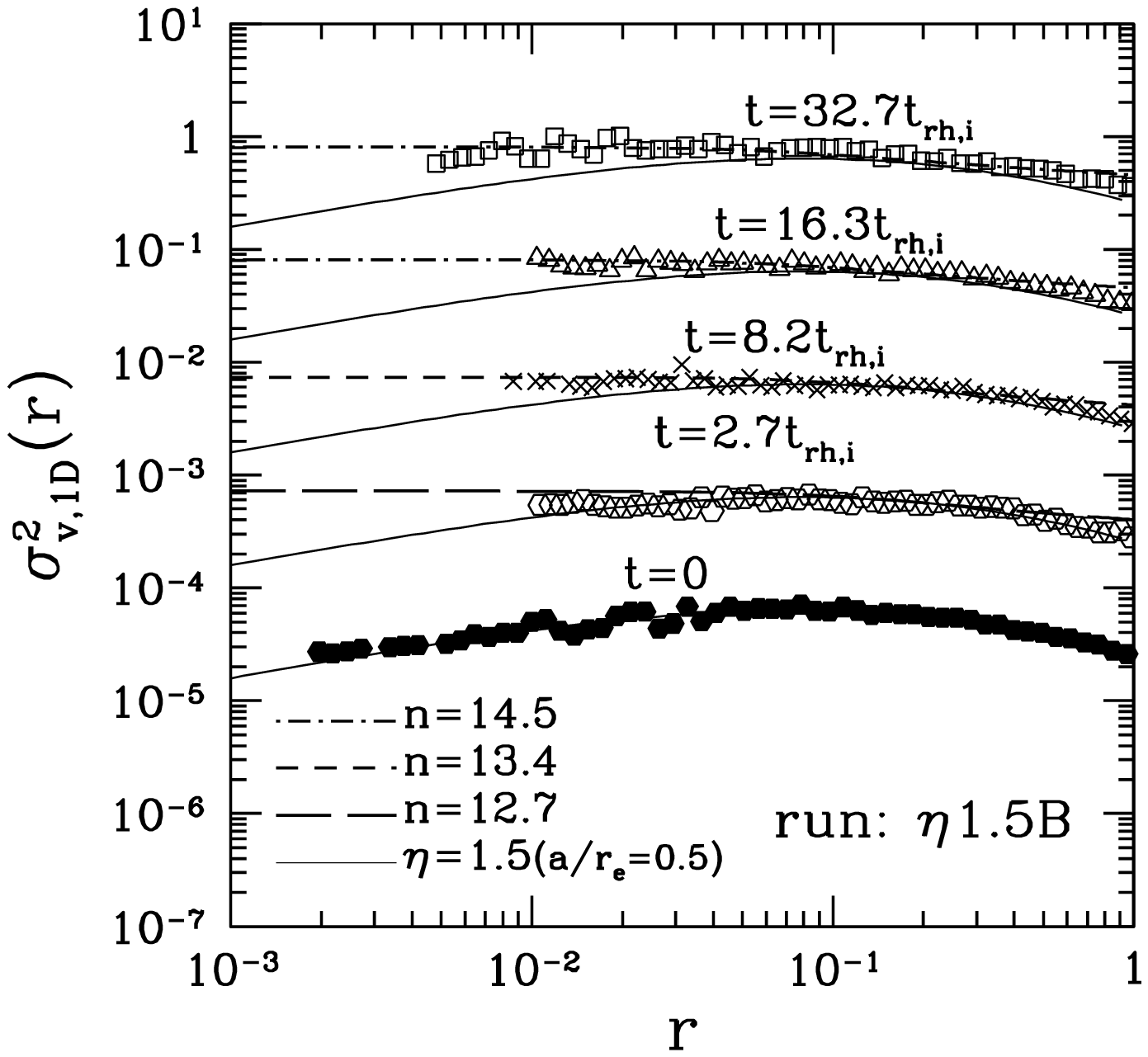}
  \end{center}

\vspace*{-0.2cm}

    \caption{Snapshots of density profile, distribution function and 
	velocity dispersion for run $\eta1.5B$. 
    \label{fig: tremaine_eta1.5B} }
  \begin{center}
    \epsfxsize=5.5cm
    \epsfbox{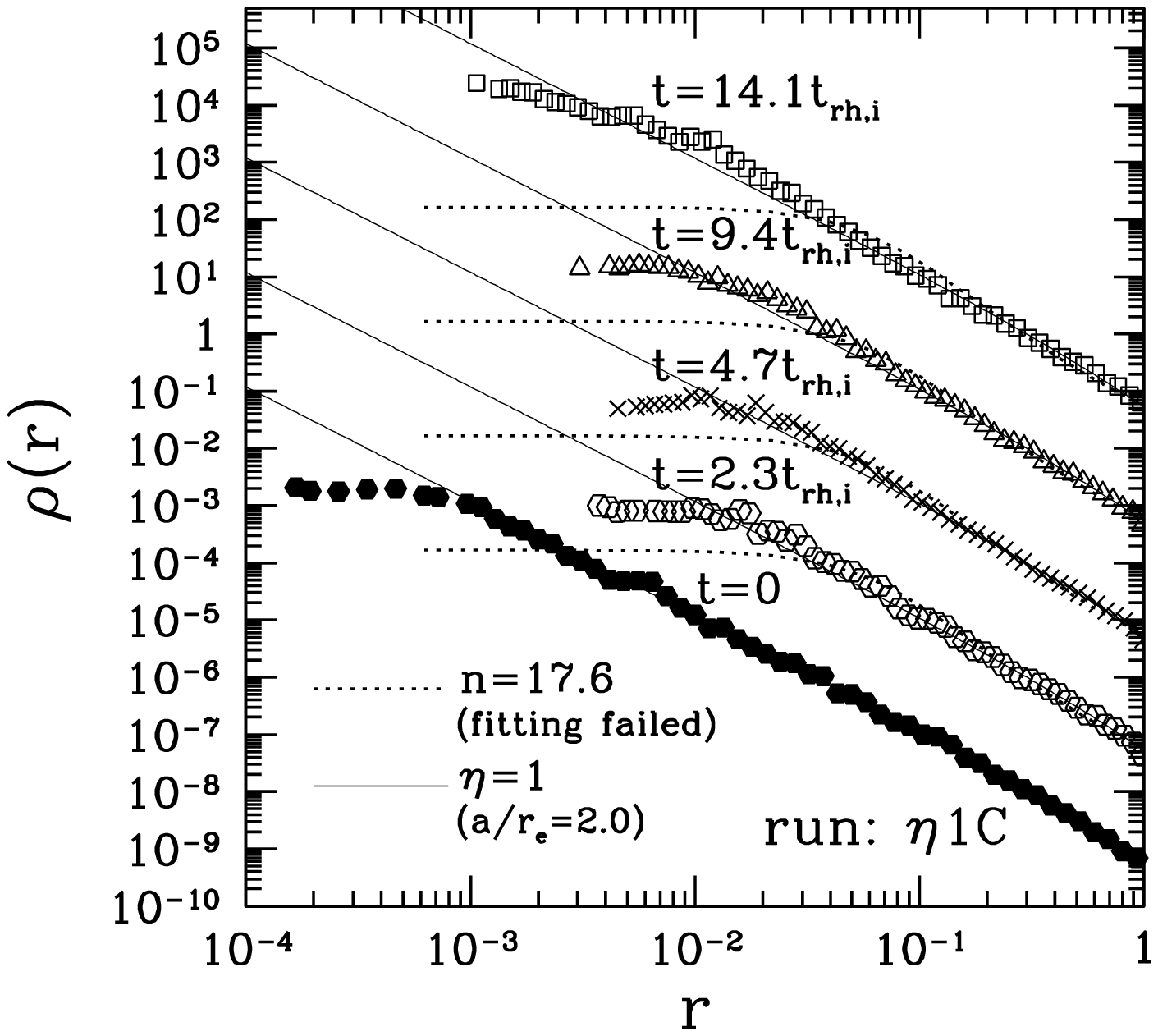}
    \epsfxsize=5.5cm
    \epsfbox{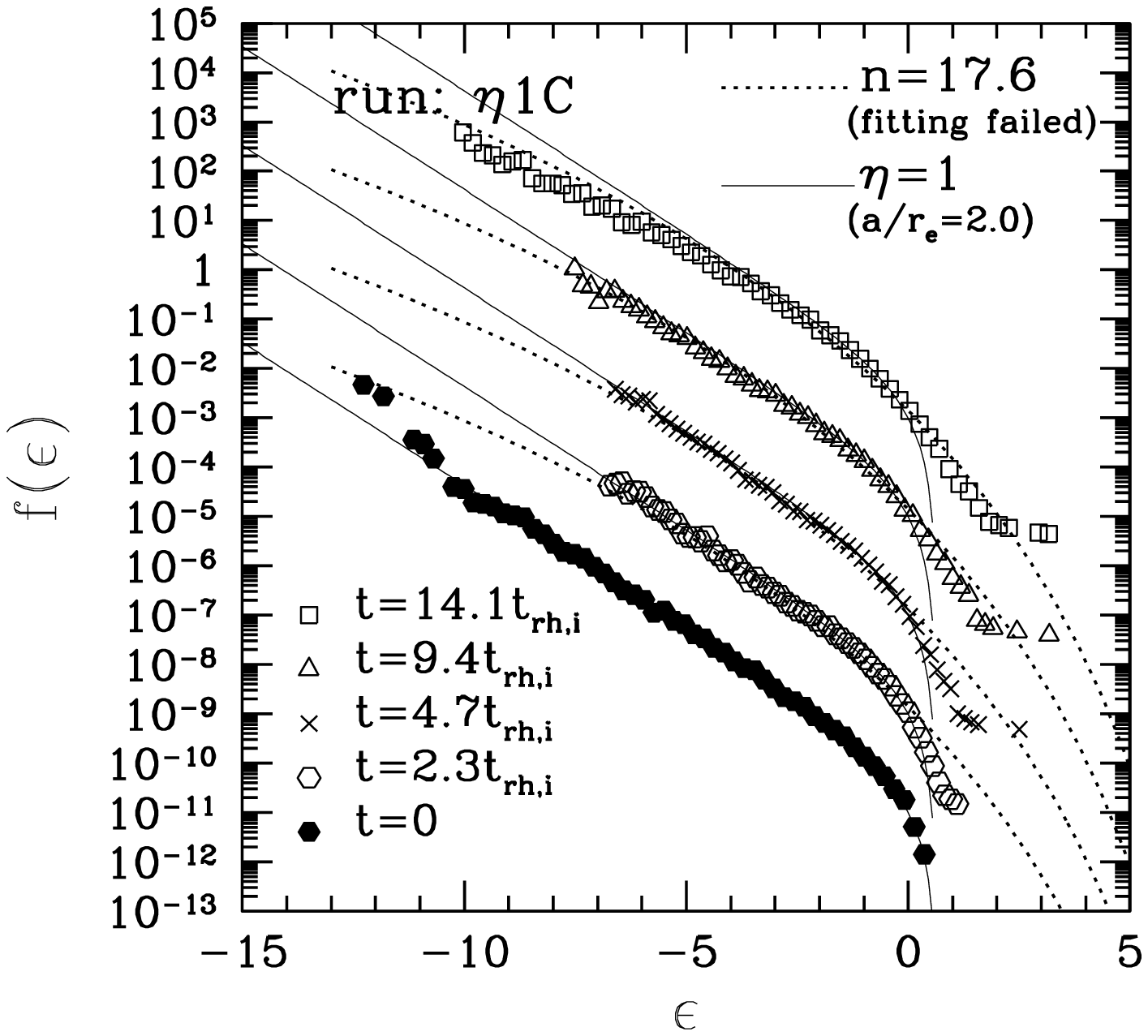}
    \epsfxsize=5.5cm
    \epsfbox{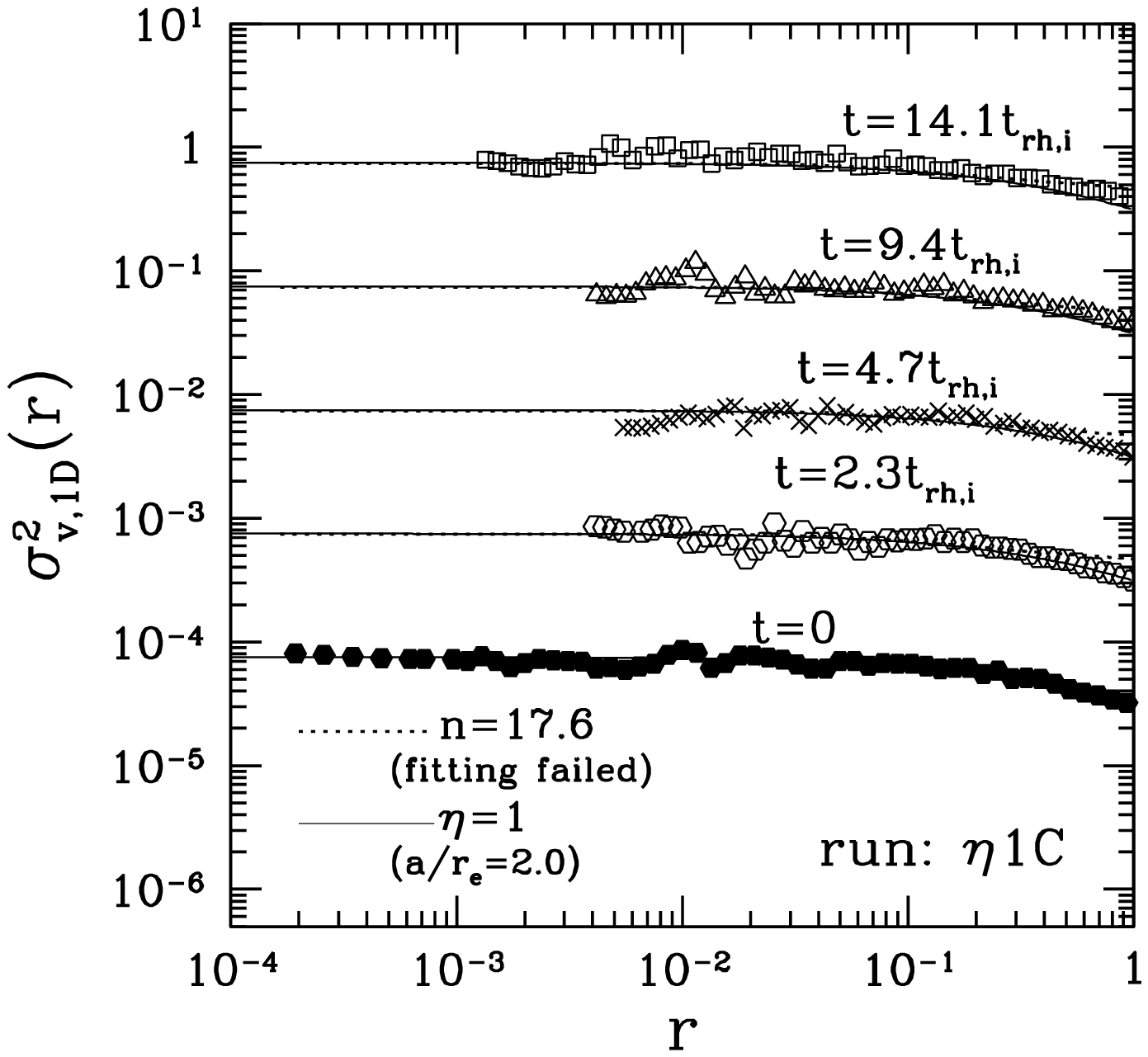}
  \end{center}

\vspace*{-0.2cm}

    \caption{Snapshots of density profile, distribution function and 
	velocity dispersion for run $\eta1C$. 
    \label{fig: tremaine_eta1C} }
 \end{figure}

In figure \ref{fig: s_BG_tremaine}, 
while the time evolution of density 
contrast $D$ shows non-monotonic behaviors, 
the specific entropy $S_{\rm BG}/N$ monotonically increases in time,  
consistent with the law of thermodynamics 
indicated by the Boltzmann H-theorem. The 
evolutionary sequence in the $(D,S_{\rm BG}/N)$-plane depends on the 
initial condition. In the case of the run $\eta2C$, 
the left panel shows that the trajectory is 
located around $S_{\rm BG}\simeq 3.7-3.9$ and the evolution slows down 
after contacting with the trajectory of the 
isothermal distribution. As for the trajectory of the run $\eta1.5B$, it 
first goes across the boundary between the stable and the unstable 
stellar polytropes indicated by the dotted line. Then, it temporarily 
settles down into a stable stellar 
polytropic state with index $n\simeq12-14$ for a long time. At   
that time, the transient states were successfully fitted by the stable 
stellar polytropic distribution (see Fig.\ref{fig: tremaine_eta1.5B}).  
It is interesting to note that the growth of the entropy shown in 
the right panel of Figure \ref{fig: s_BG_tremaine} is 
slightly restrained during the quasi-equilibrium regime, which 
might manifest the minimum entropy production principle in 
non-equilibrium thermodynamics. 
On the other hand, for the trajectory of the run $\eta1C$, 
while it first approaches the stability boundary, the decreases of the 
density contrast eventually terminate at relatively higher value 
$D\simeq10^4$ and accordingly the transients could not be fitted by the 
stable stellar polytropes.

Note also that the transient state of the run $\eta1C$ cannot 
be even fitted by the {\it unstable} stellar polytropes, since 
the density profile of the unstable polytropes shows a 
log-periodic behavior at the outer part
\citep[e.g.,][]{Chandrasekhar1939} 
while no such behavior appears in the $N$-body simulation. 
In this sense, only the quasi-equilibrium state characterized by 
the {\it stable} stellar polytropes might have some 
special meanings. Similar to 
the isothermal distribution, the quasi-equilibrium state has the 
quasi-attractive feature that the system starting from 
some classes of initial conditions tends to approach the polytropic state,  
which may provide an important suggestion for the reality of the 
non-extensive statistics. 
%
%
%
%
%
%
%
%
%
%
%
%
%
%
%
%
%
%
%
%
%
%
%
%
%
%
%
%
%
%
\begin{figure}
  \begin{center}
    \epsfxsize=8.5cm
   \epsfbox{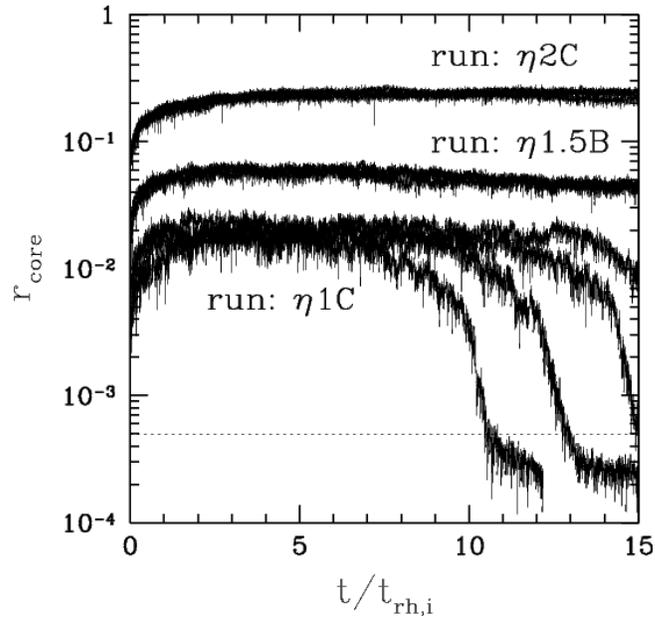}
  \end{center}

\vspace*{-0.2cm}

    \caption{Run-by-Run variation of evolution of core radius 
 among three different realization for the runs $\eta1.5B$ and $\eta2C$, 
 and among four different realization for the run $\eta1C$.  
 The dotted line indicates the length of the potential softening, 
 $\epsilon=4/N\simeq4.9\times10^{-4}$.  
    \label{fig: evolve_r_core} }
\end{figure}
%
%
%
%
%
%
%
%
%
%
%
%
%
%
%
%
%
\begin{table}
\caption{\label{tab:model_tremaine}Model parameters and 
the evolutionary states in cases starting from the stellar 
models by \citet{TRBDFGKL1994}.}

\vspace*{0.0cm}

  \begin{center}
{\large
\begin{tabular}{lcccccc}
\hline
 run $\#$  &  parameters & half-mass radius($r_{\rm h}/r_e$) & $\#$ of particles  & realizations & transient state & final state \\
\hline
 $\eta1$A  & $\eta=1$, $a/r_e=0.5$ & 0.25 & 8k & 2 &      none         & collapse   \\
 $\eta1$B  & $\eta=1$, $a/r_e=0.8$ & 0.3077 & 8k & 6 &      none         & collapse   \\
 $\eta1$C  & $\eta=1$, $a/r_e=2.0$ & 0.4 & 8k & 4 &      none         & collapse   \\
 $\eta1.5$A&$\eta=1.5$, $a/r_e=0.2$ & 0.221 & 8k & 2&      none         & collapse \\
 $\eta1.5$B& $\eta=1.5$, $a/r_e=0.5$ & 0.362 & 8k & 3 & stellar polytrope & collapse \\
 $\eta1.5$C& $\eta=1.5$, $a/r_e=1.0$ & 0.4598 & 8k & 3 & stellar polytrope & collapse \\
 $\eta2$A  & $\eta=2$, $a/r_e=0.2$ & 0.2869 & 8k & 2 &      none         & collapse   \\
 $\eta2$B$^1$& $\eta=2$, $a/r_e=0.5$ & 0.4459 & 8k & 2 & stellar polytrope & collapse   \\
 $\eta2$C$^2$& $\eta=2$, $a/r_e=1.0$ & 0.5469 & 8k & 3 &      none         & isothermal \\
 $\eta3$A  & $\eta=3$, $a/r_e=0.2$ & 0.3907 &  8k & 2 & stellar polytrope & collapse   \\
 $\eta3$B  & $\eta=3$, $a/r_e=0.5$ & 0.5619 & 8k & 2 &      none         & isothermal \\
\hline
\end{tabular}
}

$^1$ initial distribution corresponding to {\it run C1} in \citet{TS2003c} \\
$^2$ initial distribution corresponding to {\it run C2} in \citet{TS2003c} \\
  \end{center}
\end{table}
%
%
%
%
%
%
%
%
%
%
%
%
%
%
%
%
%
%
%
%
%
%
%
%
%
%
\section{Discussion \& Conclusion}
\label{sec: conclusion}
\begin{figure}
  \begin{center}
    \epsfxsize=7cm
    \epsfbox{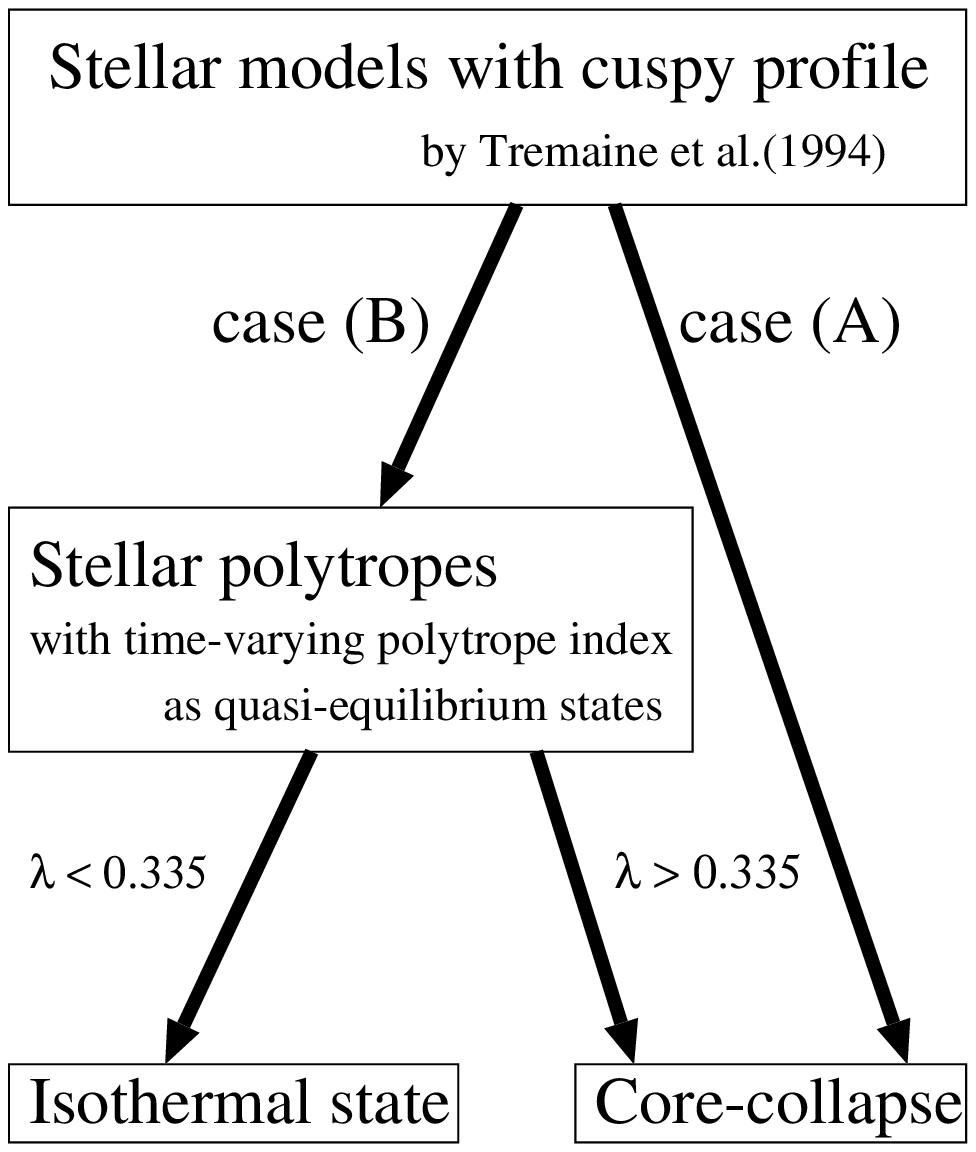}
  \end{center}

\vspace*{-0.2cm}

    \caption{Schematic illustration of the simulation results, which 
 summarizes the evolutionary status shown in figures 
\ref{fig: lambda_poly} and \ref{fig: lambda_tremaine} 
    \label{fig: schematic} }
\end{figure}

%
%
%
%
%
In this paper, we have discussed a possible application of 
non-extensive thermostatistics to the stellar systems and  
attempted to explore its reality. To do this, 
we have numerically investigated the 
quasi-equilibrium properties of the $N$-body systems 
before the core-collapse stage. Particularly focusing on 
the long-term stellar dynamical evolution from the thermostatistical 
point-of-view, we try to characterize the out-of-equilibrium state 
starting with 
various initial conditions in the setup of the so-called Antonov problem.  
We found that the quasi-equilibrium states, 
in which the system evolves gradually on timescales of two-body relaxation,  
appears and the system may follow the quasi-equilibrium sequence for a 
long time if the dimensionless energy 
$\lambda=-Er_e/GM^2$ of the system is not so far from the critical energy 
of the isothermal sphere, $\lambda_{\rm crit}=0.335$. 
The schematic illustration of our basic results is shown in 
figure \ref{fig: schematic}.  
The transient states during the quasi-equilibrium evolution are approximately 
described by the one-parameter family of stellar polytropes as extremum 
states of non-extensive entropy $S_q$ with the 
time-varying polytrope index. The fitted value of the index $n$ 
gradually increases with time 
and the system keeps following a 
sequence of stellar polytropes until reaching the critical index, 
$n_{\rm crit}$. 
In general, the condition for the quasi-equilibrium state 
would depend on the details of the velocity structure in the 
initial conditions, however, within  
a class of initial conditions examined in this paper (i.e., stellar 
polytropes and stellar models with cusped density profiles), the 
out-of-equilibrium states with inner density profiles 
$\rho\propto r^{-\alpha}$ $(\alpha<2)$ (or $\eta\leq1$) exhibit 
the quasi-equilibrium behavior that is attracted to a sequence of 
stellar polytropes.

One may naively think that 
the results obtained here severely depend on the presence of an adiabatic 
wall, since the real stellar systems in absence of the adiabatic wall 
are known to be poorly fitted to the stellar polytropes. 
Recalling the discussion in section \ref{subsubsec:discussion}, however, 
the outer part of the system is expected to be mainly 
affected by the boundary condition, since    
the relaxation timescale at the outer part 
is rather longer than that at the core. 
In other words, as long as the relaxation time at the central part is 
shorter than that at the outer part, the modification of the 
boundary condition only alters the outer part of the system, 
not all of the system.  In fact,  we have seen in 
section \ref{subsubsec:discussion} that the phenomenological King model 
which accounts for the globular clusters affected by the 
Galactic tidal field resembles 
the stellar polytropes at the inner part. In this sense, 
the stellar polytropic system as quasi-equilibrium state would be 
a fundamental stellar model and may sometimes make sense even if 
removing the adiabatic wall \citep[][]{TS2003c}.

The present results may give an interesting suggestion for 
the justification and/or the realization of the non-extensive 
thermostatistics based on the Tsallis entropy. Strictly speaking, however, 
the $N$-body simulations just indicate a reality of the 
stellar polytropes as {\it $q$-analogue} of the Boltzmann-Gibbs distribution. 
Further, the quasi-equilibrium state is 
time-dependent, which cannot be rigorously treated by the 
thermostatistics \citep[][]{CS2004}. Even at this moment, 
it might be also possible to give the interpretation that stellar 
polytrope is merely a dynamical equilibrium state, 
but it can be stable even with respect to a non-linear perturbation 
\citep[][]{Chavanis2003}. In this respect, the present $N$-body results 
are not conclusive and the interpretation of their results 
should be carefully discussed. 
Nevertheless, one may hope that the exploration 
of a connection with non-extensive entropy opens a new window 
to understand the non-equilibrium thermodynamics of 
the long-range systems. In this respect, the analytical treatment 
based on the kinetic theory is an important next step to 
interpret the $N$-body results thermodynamically. 
A crucial task is to estimate the timescales for quasi-equilibrium evolution, 
as well as to determine the generic criteria for quasi-equilibria. 
To investigate this, the Fokker-Planck model for stellar dynamics 
would be helpful \citet{TS2004}. 
A quantitative comparison between a numerical solution of the Fokker-Planck 
model with the $N$-body results based on a more 
sophisticated $N$-body code such as the one developed by Aarseth 
will be presented elsewhere. 
%
%
%
%
%
%
%
%
%
%
%
%
%
%
%
%
%
%
%
%
%
%
%
\section*{Acknowledgments}
%
%
%
We are grateful to T. Fukushige for providing us the GRAPE-6 code and 
for his constant supports and helpful comments. We also thank S. Inagaki 
and K. Takahashi for their comments and suggestions to our future 
works on the Fokker-Planck model for stellar dynamics. 
Numerical computations were carried out at ADAC
(the Astronomical Data Analysis Center) of the National Astronomical 
Observatory of Japan. 
This work was supported by the grand-in-aid for Scientific Research 
of Japan Society of Promotion of Science (No.$14740157$, $15540368$). 
\appendix
%
%
%
%
%
\section*{Appendix A: Convergence test of N-body simulation}
\label{appendix: convergence test}
%
%
%
In a series of our $N$-body simulations, the Plummer softening with parameter 
$\epsilon=1/N$ or $4/N$ is used to avoid the formation of tight binaries. 
For our investigation of the quasi-equilibrium state before 
the core-collapse stage, the fourth order Hermite integration code 
with individual time-step provides a robust numerical method without 
a regularization scheme. However,   
it would be crucial to pursue the gravothermally 
unstable regime. While our primary concern is the non-equilibrium  
evolution before entering the core-collapse phase, 
it is important to note the effect of potential softening 
on the estimation of collapse time.

To quantify this, convergence property of the collapse time 
is investigated by varying the softening parameter $\epsilon$. 
For this purpose,  $N$-body simulations starting with the 
Plummer model, i.e., the stellar polytrope with index $n=5$, 
are used to study the effect of potential softening. The 
corresponding simulations are the runs $n5B$ and $n5D$ 
listed in table \ref{tab:model_poly}.

\begin{figure}
  \begin{center}
    \epsfxsize=7.2cm
    \epsfbox{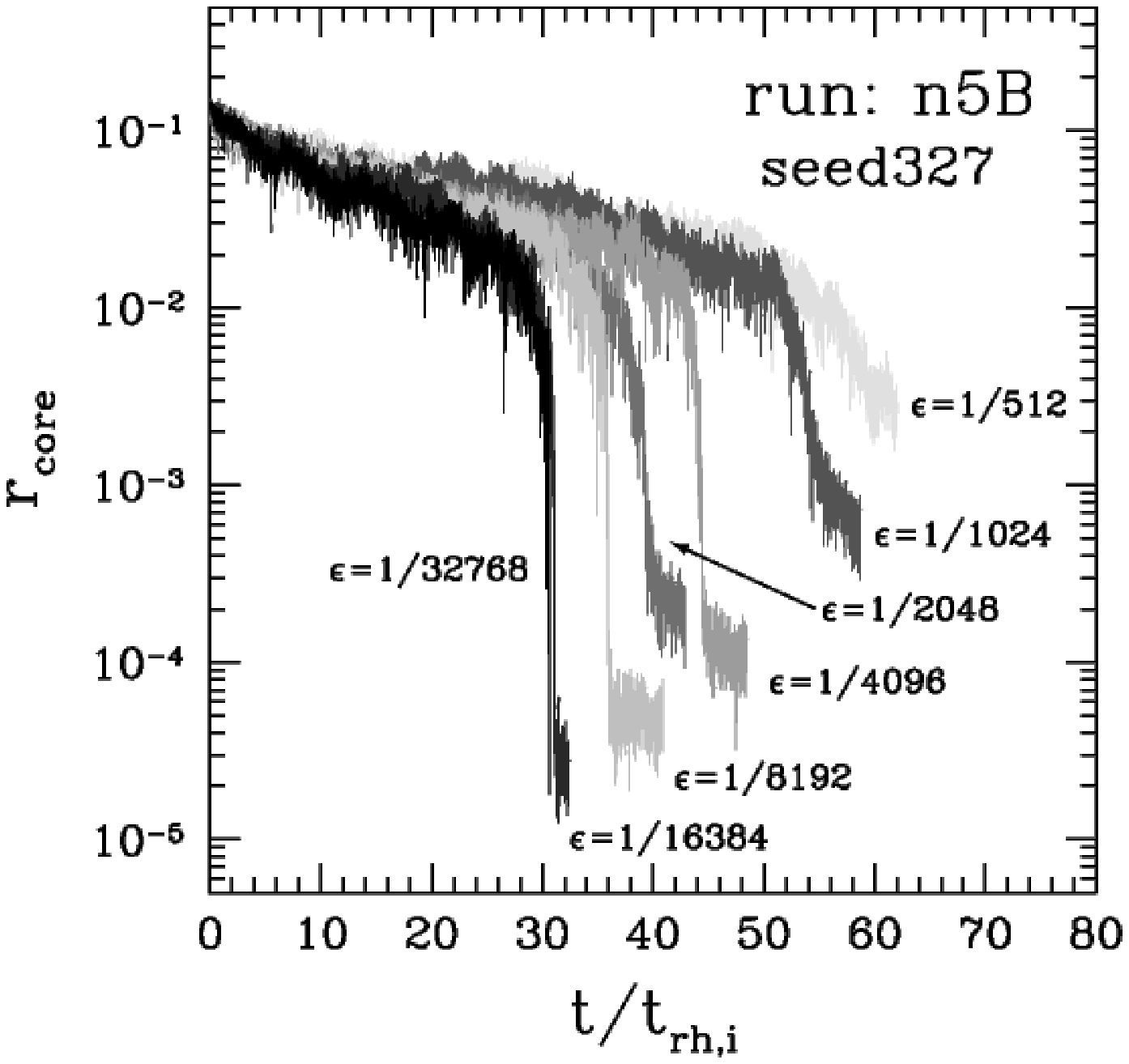}
\hspace*{0.5cm}
    \epsfxsize=7.2cm
    \epsfbox{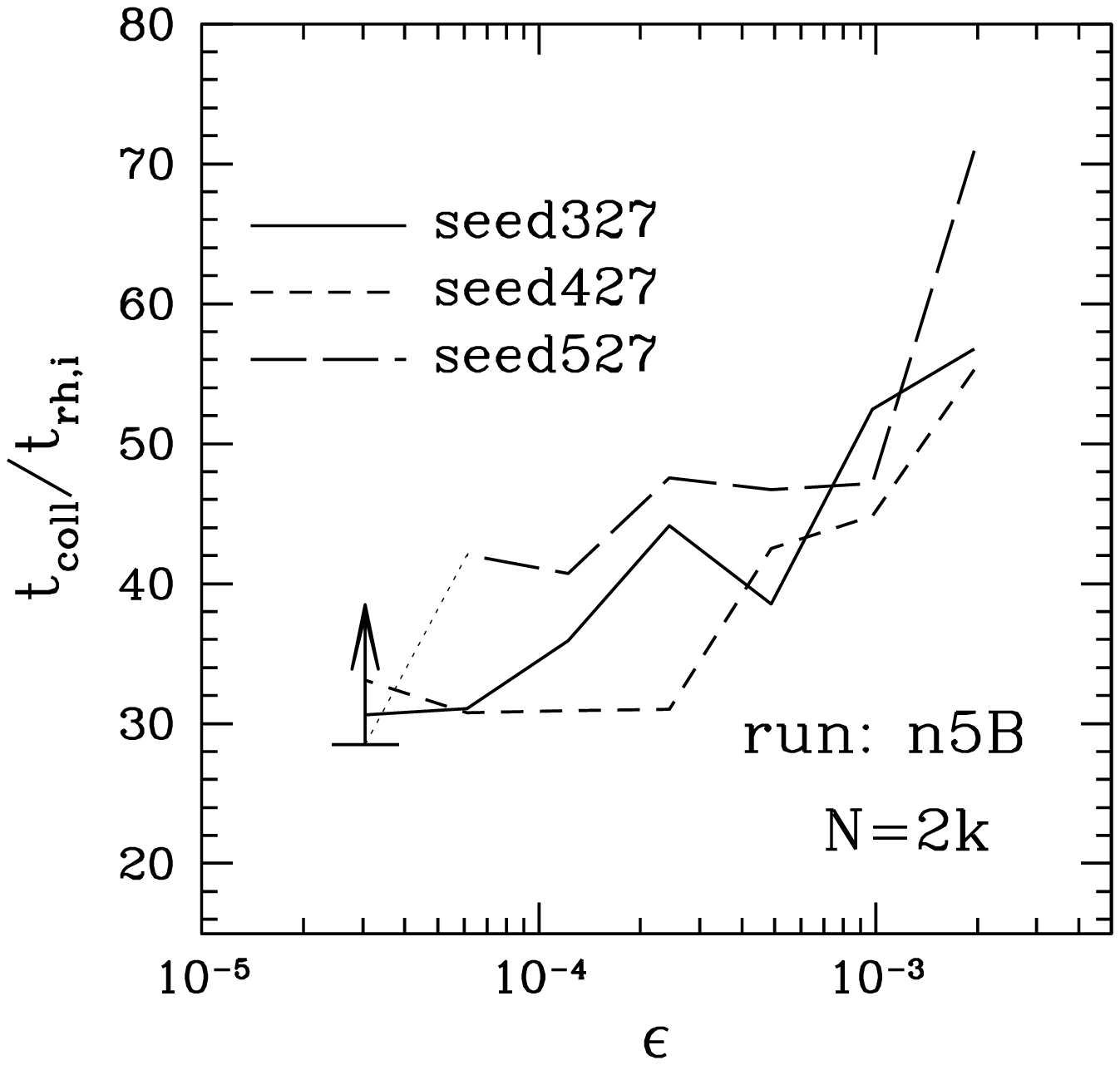}
  \end{center}
    \caption{Convergence properties of collapse time for various choices of 
softening parameter $\epsilon$ in the case of run $n5B$. Here, the 
collapse time is defined as the first passage time when the core radius 
becomes shorter than $3\epsilon$.  
	{\it Left}: evolution of core radius for a realization with 
seed number $327$. 
	{\it Right}: Variation of collapse time as 
 function of softening parameter among three realizations with different 
 random seed numbers.  
    \label{fig: convergence_run_n5B} }

\vspace*{0.2cm}

  \begin{center}
    \epsfxsize=7.2cm
    \epsfbox{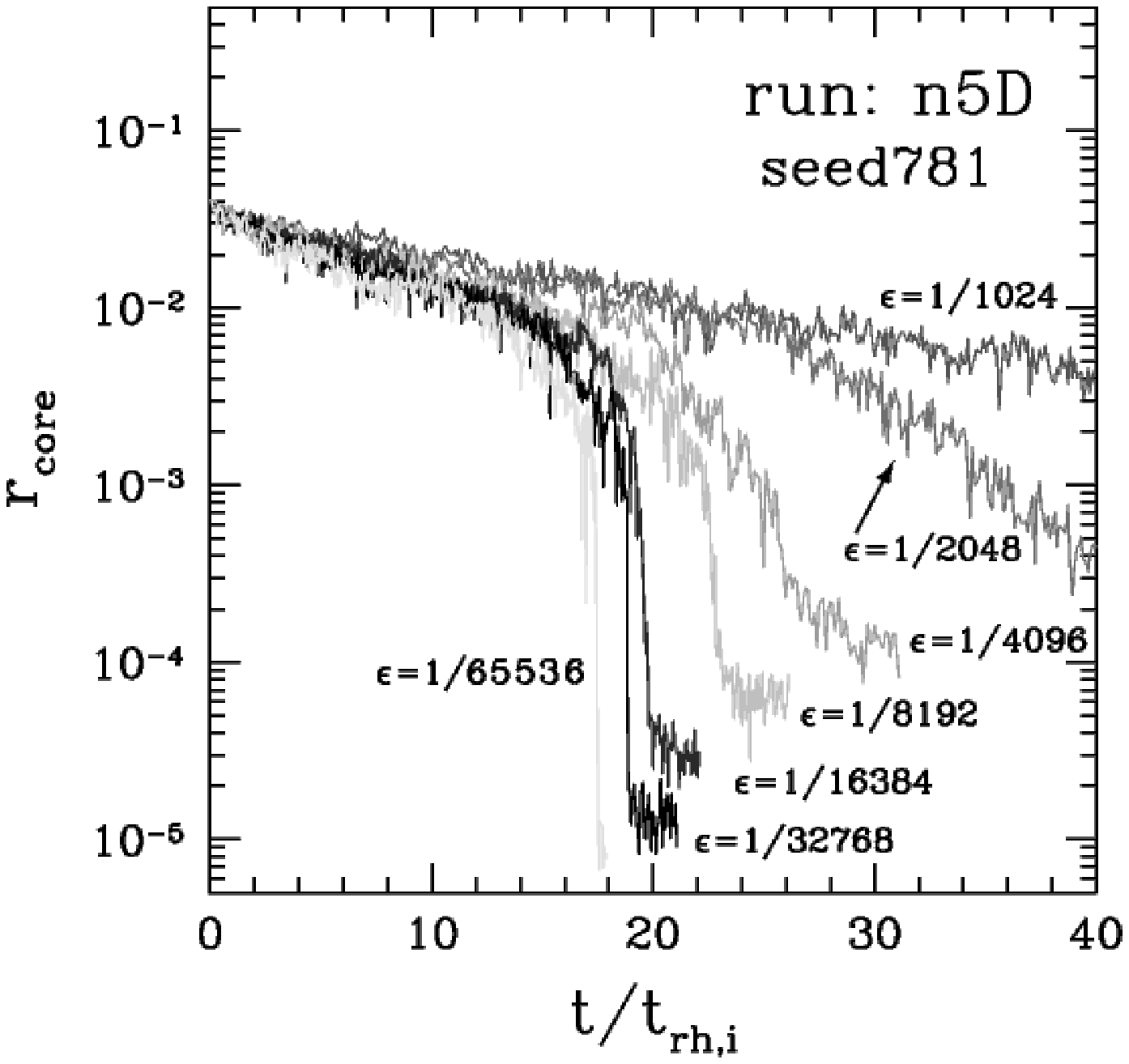}
\hspace*{0.5cm}
    \epsfxsize=7.2cm
    \epsfbox{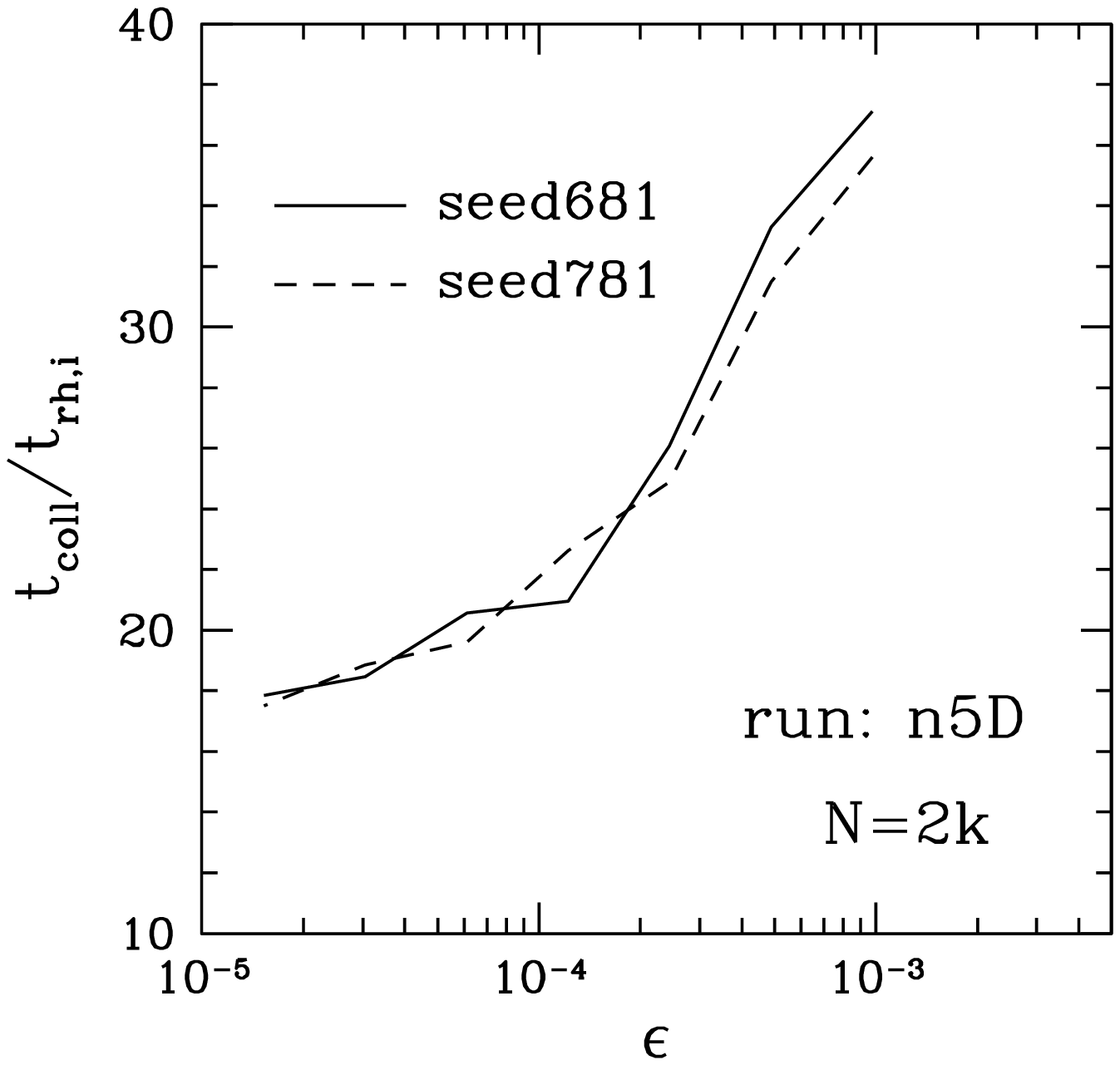}
  \end{center}
    \caption{Same as Fig.\ref{fig: convergence_run_n5B}, but in the cases 
	of run $n5D$. 
    \label{fig: convergence_run_n5D} }
\end{figure}
%
%
%
%
%
%
%
Figure \ref{fig: convergence_run_n5B} 
shows the results of the convergence test based 
on the run $n5B$. Left panel plots the evolution of core radii for 
various choices of softening parameter, while the 
right panel estimate the collapse time as function of softening parameter 
$\epsilon$ for three different realizations of the initial 
condition. Here, the collapse time is defined as the first passage time 
when the core radius becomes shorter than $3\epsilon$. 
Clearly, large value of the softening parameter overestimates 
the estimation of the core-collapse time and the 
transition between the quasi-equilibrium stage to the core-collapse phase
becomes uncertain. For an appropriately small value   
$\epsilon \simlt 1/(4N) \simeq 1.2\times10^{-4}$, 
the core radius sharply  
falls off around the core-collapse time and 
the collapse time seems to converge to 
$t_{\rm coll}\sim30$-$40t_{\rm rh,i}$, 
although there exists a large scatter among three different realizations  
of the initial condition.

In general, the convergence properties of the collapse time 
depend on the initial conditions. Figure \ref{fig: convergence_run_n5D} 
shows the results from 
the run $n5D$, i.e., Plummer model with density contrast $D=10^6$. 
In this case, the scatter becomes even smaller and 
the collapse time converges to 
$t_{\rm coll}\sim18 t_{\rm rh,i}$ at $\epsilon \simlt 6\times 10^{-5}$,  
slower than the case of run $n5B$. These experiments indicate that 
a large softening length affects not only the core-collapse time but 
also the timescales of quasi-equilibrium stage. Further, 
the requirement for the softening parameter 
becomes severe as increasing the 
initial density contrast. Hence, for quantitative estimation of collapse 
time, the softening length should be set to zero with 
implementing a regularization scheme. The use of a small softening 
length is also favorable to investigate the quasi-equilibrium evolution. 
Figures 
\ref{fig: convergence_run_n5B} and \ref{fig: convergence_run_n5D} 
suggest that for initial conditions with moderate range of the initial 
density contrast, $10^2\leq D\leq 10^4$, 
the softening length with $\epsilon\leq1/N$ provides a better 
estimation of the quasi-equilibrium timescales,  
although it still overestimates the collapse time. Hence, in this paper, 
we mainly use the softening length $\epsilon=1/N$. 
%
%
%
%
%
%
%
%
%
%
%
%
%
\section*{Appendix B: A family of stellar model with cusps}
\label{appendix: tremaine}
%
%
%
%
%
%
%
In this appendix, we present analytic formulae for the stellar model 
considered by \citet{TRBDFGKL1994} taking account of the 
adiabatic wall. 

Just for convenience, let us first introduce the following variables: 
\begin{equation}	
A=\eta\,\,\frac{M}{4\pi a^3},~~~~~
B=\frac{GM}{a}\left(\frac{\re+a}{\re}\right)^{\eta}.
\end{equation}
In terms of these, the density profile for stellar models by 
\citet{TRBDFGKL1994} is expressed as 
\begin{equation}
\rho(r)=\frac{A}{(r/a)^{3-\eta}(1+r/a)^{1+\eta}}. 
\label{appen:def_rho}
\end{equation}
The corresponding mass profile and the gravitational potential 
respectively becomes 
\begin{equation}
m(r)=\int_0^r dr' \,\,4\pi\, r^{,2}\,\rho(r') = 
\frac{4\pi a^3 A}{\eta}\left(\frac{r}{r+a}\right)^{\eta},
\end{equation}
and
\begin{eqnarray}
\Phi(r)&=&-G\left\{\frac{m(r)}{r}+\int_0^{\re}dr'\,\,4\pi r'\rho(r')\right\}
\nonumber \\
&=& \left\{
\begin{array}{lcr}
{\displaystyle 
-B\left\{\frac{1}{\eta-1}\left(\frac{\re}{\re+a}\right)^{\eta-1}
-\frac{1}{\eta-1}\left(\frac{r}{r+a}\right)^{\eta-1}+
\frac{(\re/a)^{\eta-1}}{(\re/a+1)^{\eta}}\right\}} & ; & \eta\ne1 
\\
\\
{\displaystyle 
-B\left\{\log\left(\frac{\re}{\re+a}\right)-\log\left(\frac{r}{r+a}\right)+
\frac{1}{\re/a+1}\right\} }& ; & \eta=1
\end{array}
\right.
\label{appen:tremaine_potential}
\end{eqnarray}

For stationary state of the Vlasov equation, the one-dimensional 
velocity dispersion 
profile can be calculated from the density profiles and 
the gravitational potential through the Jeans equation. 
In the case of the isotropic velocity distribution, this gives 
\begin{equation}
\sigma_{\rm v,1D}^2(r)=\frac{P(r)}{\rho(r)},
\end{equation}
with the function $P(r)$ being pressure, determined from 
the hydrostatic equilibrium(Jeans equation), $dP/dr=-\rho~d\Phi/dr$. 
Note that the explicit expression for function $P(r)$ cannot
be expressed in an unified manner. For some values of $\eta$,
we have 
\begin{eqnarray}
P_{\eta=1}(r) &=& AB\left[6\log\left(\frac{y+1}{y}\right)-
\frac{1}{y^2}\left\{6(1+y)-9+\frac{2}{y+1}+\frac{1}{2(y+1)^2}
\right\}\right],
\\
P_{\eta=1.5}(r) &=& AB\left[ 4\log\left(\frac{y}{y+1}\right)+
\frac{1}{y}\left\{4-\frac{2}{y+1}-\frac{2}{3(y+1)^2}
-\frac{1}{3(y+1)^3}\right\} \right],
\\
P_{\eta=2}(r) &=& AB\left[ 4\log\left(\frac{y+1}{y}\right)-
\left\{\frac{1}{y+1}+\frac{1}{2(y+1)^2}
+\frac{1}{3(y+1)^3}+\frac{1}{4(y+1)^4}\right\} \right],
\\
P_{\eta=3}(r) &=& AB\left[ \frac{1}{5(y+1)^5}-\frac{1}{6(y+1)^6}\right], 
\end{eqnarray}
where we defined $y=r/a$. 
Provided the pressure, 
one can also calculate the total energy of the system confined 
in a wall as   
\begin{eqnarray}
E=K+U&=&\int d^6 d^3\xx d^3\vv \left\{\frac{1}{2}v^2+\frac{1}{2}\Phi(r)\right\}f(r,v)
\nonumber\\
&=&\frac{3}{2}\int_0^{\re} dr \,\,4\pi r^2P(r)+
\frac{1}{2}\int_0^{\re} 
dr\,\,4\pi r^2\rho(r)\Phi(r).
\end{eqnarray}
In terms of the dimensionless variable $\lambda=-r_eE/(GM^2)$, 
we obtain
\begin{eqnarray}
\lambda_{\eta=1} &=& \frac{1}{4}
\left(1+5y_e+18y_e^2+12y_e^3\right)-3y_e^2(y_e+1)^2
\log\left(\frac{y_e+1}{y_e}\right),
\\
\lambda_{\eta=1.5} &=& \frac{1}{8}
\left(-3-43y_e-60y_e^2-24y_e^3\right)-3(y_e+1)^3
\log\left(\frac{y_e}{y_e+1}\right),
\\
\lambda_{\eta=2} &=& \frac{1}{12}
\left(29+53y_e+42y_e^2+12y_e^3\right)-(y_e+1)^4
\log\left(\frac{y_e+1}{y_e}\right),
\\
\lambda_{\eta=3} &=& \frac{1}{20y_e^2}
\left(1+7y_e+y_e^2\right)(y_e-1),
\end{eqnarray}
where the variable $y_e$ denotes $\re/a$.

According to the standard text book for stellar dynamics, 
the one-particle distribution function for the isotropic spherical 
stellar model is expressed as a function of the specific energy 
$\epsilon=v^2/2+\Phi(r)$ and can be 
reconstructed from the density profiles through 
the Eddington formula. Introducing the variables 
$\varepsilon=\Phi_0-\epsilon$ and $\psi=\Phi_0-\Phi(r)$, we have
\begin{equation}
f(\epsilon)=\frac{1}{\sqrt{8}\pi^2}\int_0^{\varepsilon}
\frac{d^2\rho}{d\psi^2} \frac{d\psi}{\sqrt{\varepsilon-\psi}}
\label{appen:fedist_tremaine}
\end{equation}
with the regularity condition: 
\begin{equation}
\left(\frac{d\rho}{d\psi}\right)_{\psi=0}=0.
\end{equation}
Note that the numerical constant $\Phi_0$ is determined from 
the above condition. 
With the use of the analytical expressions (\ref{appen:def_rho}) 
and (\ref{appen:tremaine_potential}), after some manipulation, 
the Eddington formula (\ref{appen:fedist_tremaine}) can be 
rewritten with 
\begin{equation}
f(\epsilon)=\frac{1}{\sqrt{8}\pi^2}\frac{A}{B^{3/2}}\,
\int_0^{q(\epsilon)}
\frac{d\psi}{\sqrt{q(\epsilon)-\psi}} 
\frac{12\{s(\psi)\}^2-4(\eta-4)s(\psi)+2(3-\eta)}
{\{s(\psi)\}^{\eta+1}\{1+s(\psi)\}^{3-\eta}}, 
\end{equation}
where the functions $q(\epsilon)$ and $s(\psi)$ are respectively 
given by 
\begin{eqnarray}
q(\epsilon)&=& -\frac{\epsilon}{B}-\frac{1}{\eta-1}\left\{
\left(\frac{\re}{\re+a}\right)^{\eta-1}-1+(\eta-1)
\frac{(\re/a)^{\eta-1}}{(\re/a+1)^{\eta}}\right\},	
\\
s(\psi)&=& \frac{\{1-(\eta-1)\psi\}^{1/(\eta-1)}}
{1-\{1-(\eta-1)\psi\}^{1/(\eta-1)}}
\end{eqnarray}
in cases with $\eta\ne1$ and  
\begin{eqnarray}
q(\epsilon)&=& -\frac{\epsilon}{B}-
\log\left(\frac{\re}{\re+a}\right)-\frac{1}{\re/a+1},
\\
s(\psi)&=&\frac{1}{e^{\psi}-1}
\end{eqnarray}
for $\eta=1$. 
In principle, the distribution function $f(\epsilon)$ is 
obtained from the numerical integration of 
(\ref{appen:fedist_tremaine}). In some specific values of $\eta$, 
however, one can luckily obtain the analytic expressions of  
$f(\epsilon)$: 
\begin{eqnarray}
f_{\eta=1}(\epsilon)&=& \frac{A}{B^{3/2}}\,
\sqrt{\frac{2}{\pi^3}}
\left[\sqrt{\frac{2}{\pi}}F\left(\sqrt{2q}\right)-
\frac{2}{\sqrt{\pi}}F(\sqrt{q})-e^q\,\mbox{erf}(\sqrt{q})+
\frac{e^q}{\sqrt{2}}\,\mbox{erf}\left(\sqrt{2q}\right)\right],
\label{appen:df_eta10}\\
f_{\eta=1.5}(\epsilon)&=& \frac{A}{B^{3/2}}\,
 \frac{1}{\sqrt{8}\pi^2(2-q)^{9/2}}\left[
\frac{3}{2}(3+32q-8q^2)\sin^{-1}\left(\sqrt{\frac{q}{2}}\right)
\right.
\nonumber \\
&&\left.
-\frac{\sqrt{q(2-q)}}{28}\,\left\{63+693q-5670q^2+7410q^3-4488q^4+
1448q^5-240q^6+16q^7\right\}\right],
\label{appen:df_eta15}\\
f_{\eta=2}(\epsilon)&=& \frac{A}{B^{3/2}}\,
\frac{1}{4\sqrt{2}\pi^2}\,\frac{1}{(1-q)^{5/2}}
\left[3\sin^{-1}(\sqrt{q})-\sqrt{q(1-q)}\,(16q^3-24q^2+2q+3)\right],
\label{appen:df_eta20}\\
f_{\eta=3}(\epsilon)&=& \frac{A}{B^{3/2}}\,
\frac{1}{\pi^2(1-2q)}
\left[2\sqrt{2q}\,\,(3-4q)+3(1-2q)\,
\log\left(\frac{1-\sqrt{2q}}{1+\sqrt{2q}}\right)\right].
\label{appen:df_eta30}
\end{eqnarray}
Here, the function $F(x)$ and $\mbox{erf}(x)$ are Dawson's integral 
and the error function: 
\begin{equation}
\mbox{erf}(x)=\sqrt{2}\,\pi\int_0^x dt\,\, e^{-t^2},~~~~~
F(x)=e^{-x^2}\int_0^x dt\,\,e^{t^2}.
\end{equation}
Compared the final expressions 
(\ref{appen:df_eta10})--(\ref{appen:df_eta30}) with those obtained 
by \citet{TRBDFGKL1994}, 
we deduce that the only alternation in the expressions 
of \citet{TRBDFGKL1994} in presence of the 
adiabatic boundary is to replace all the variables $\epsilon$ 
at the right-hand side of equations with the function $q(\epsilon)$ 
defined above. Therefore, 
the above results consistently recover the formulas 
derived by \citet{TRBDFGKL1994} in the limit $r_e\to\infty$.  
%
%
%
%
%
%
%
\section*{Appendix C: Boltzmann-Gibbs entropy for stellar polytropes}
\label{appendix: S_BG_polytropes}
%
%
%
%
%
%
%
%
%
%
%
%
Here, we derive the explicit expression for Boltzmann-Gibbs entropy 
in the case of the stellar polytropic distribution, which is shown in  
theoretical curves of figure \ref{fig: s_BG_tremaine}.

First, we substitute the stellar polytropic distribution 
(\ref{eq:fedist_poly}) into the definition of Boltzmann-Gibbs 
entropy: 
\begin{eqnarray}
S_{\rm BG}^{\rm(poly)}/N &=& -\int d^3\xx d^3\vv 
	\,\,\left(\frac{f^{\rm(poly)}}{N}\right) 
	\ln \left(\frac{f^{\rm(poly)}}{N}\right)
\nonumber\\
&=& -(s_A + s_B + s_C). 
\end{eqnarray}
The quantities $s_A$, $s_B$ and $s_C$ involving  
the phase-space integral are separately evaluated as 
\begin{eqnarray}
s_A &=& \frac{1}{4\sqrt{2}\pi\,\, B(3/2,n-1/2)}\,\,
\ln\left\{\frac{1}{4\sqrt{2}\pi\,B(3/2,n-1/2)}\right\}\,
\int d^3\xx d^3\vv\,\, \frac{\rho/M}{\{(n+1)P/\rho\}^{3/2}}\,
\left\{1-\frac{v^2/2}{(n+1)P/\rho}\right\}^{n-3/2}
\nonumber \\
&=& \ln\left\{\frac{1}{4\sqrt{2}\pi\,B(3/2,n-1/2)}\right\},
\end{eqnarray}
\begin{eqnarray}
s_B &=& \frac{1}{4\sqrt{2}\pi\,\, B(3/2,n-1/2)}\,\,
\int d^3\xx d^3\vv\, \frac{\rho/M}{\{(n+1)P/\rho\}^{3/2}}\,
\left\{1-\frac{v^2/2}{(n+1)P/\rho}\right\}^{n-3/2}\,\,
\ln \left\{1-\frac{v^2/2}{(n+1)P/\rho}\right\}^{n-3/2} 
\nonumber\\
&=& \left(n-\frac{3}{2}\right) \{\psi(n-1/2)-\psi(n+1)\}\,\,
\int d^3\xx \frac{\rho}{M}
\nonumber\\
&=& \left(n-\frac{3}{2}\right) \{\psi(n-1/2)-\psi(n+1)\},
\end{eqnarray}
\begin{eqnarray}
s_C &=& \frac{1}{4\sqrt{2}\pi\,\, B(3/2,n-1/2)}\,\,
\int d^3\xx d^3\vv, \frac{\rho/M}{\{(n+1)P/\rho\}^{3/2}}\,
\left\{1-\frac{v^2/2}{(n+1)P/\rho}\right\}^{n-3/2}\,\,
\ln \frac{\rho/M}{\{(n+1)P/\rho\}^{3/2}}
\nonumber\\
&=& \int d^3\xx \frac{\rho}{M}\,\ln\frac{\rho/M}{\{(n+1)P/\rho\}^{3/2}}.
\end{eqnarray}
Here, the functions $\psi(z)$ and $B(a,b)$ are the digamma and the 
beta functions, respectively. 
Collecting the above terms,  the Boltzmann-Gibbs entropy becomes 
\begin{eqnarray}
S_{\rm BG}^{\rm(poly)}/N = 
\ln\{4\sqrt{2}\,\pi\,\,B\left(\frac{3}{2},\,n-\frac{1}{2}\right)\} 
- \left(n-\frac{3}{2}\right)\left\{\psi\left(n-\frac{1}{2}\right)-
\psi\left(n+1\right)\right\}-\int d^3\xx\,\frac{\rho}{M}\,\ln
\left\{\frac{\rho/M}{\{(n+1)P/\rho\}^{3/2}}\right\}.
\nonumber
\end{eqnarray}

For the remaining integral in the above expression, 
a further reduction is possible by 
repeating the integration by part. 
Using the equations of hydrostatic equilibrium,  
a straightforward calculation yields 
\begin{eqnarray}
\int d^3\xx \frac{\rho}{M}\,\ln\frac{\rho/M}{\{(n+1)P/\rho\}^{3/2}}
&=& -\ln M\left\{\frac{(n+1)P_e}{\rho_e^{1+1/n}}\right\}^{3/2}+
\nonumber\\
&&+
\frac{n-3/2}{n}\,\left[
\ln \rho_e - \frac{n}{n+1} \frac{GM\rho_e}{P_er_e}-
\frac{8\pi r_e^3 \rho_e}{M}+6 +\frac{n}{(n+1)^2} \int_0^{r_e} 
dr\left(\frac{\rho}{P}\right)^2\,\frac{G^2m^3}{Mr^3}\right]. 
\end{eqnarray}
Note that the subscript $_e$ denotes the quantities evaluated at 
the wall, $r=r_e$.  
In terms of the homology invariants $(u,v)$ defined in equations 
(\ref{eq: def_u}) and (\ref{eq: def_v}),  the entropy 
$S_{\rm BG}$  is finally expressed as: 
\begin{eqnarray}
&S_{\rm BG}^{\rm(poly)}/N &= - 6 \left(\frac{n-3/2}{n}\right) 
-\left(n-\frac{3}{2}\right)\left\{\psi\left(n-\frac{1}{2}\right)
-\psi\left(n+1\right)\right\}
\nonumber\\
&&+\left(\frac{n-3/2}{n+1}\right)\,v_e + 2\left(\frac{n-3/2}{n}\right)\,u_e 
-\ln\left(\frac{u_ev_e^{3/2}}{4\pi}\right) +\frac{3}{2} \ln (2\pi GMr_e)
\nonumber\\
&&+\ln \left\{2\pi (n+1)^{3/2}\,
B\left(\frac{3}{2},n-\frac{1}{2}\right)\right\}
- \frac{n-3/2}{(n+1)^2}\,\int_0^{r_e}dr \left(\frac{\rho}{P}\right)^2
\frac{G^2m^3}{Mr^3}. 
\label{eq:S_BG_poly}
\end{eqnarray}
In the limit $n\to\infty$, we correctly recover the Boltzmann-Gibbs entropy 
for the isothermal state:  
\begin{equation}
S_{\rm BG}^{\rm(iso)}/N = -\frac{9}{2} +v_e + 2u_e -
\ln\left(\frac{u_ev_e^{3/2}}{4\pi}\right) +\frac{3}{2} \ln (2\pi GMr_e).  
\label{eq:S_BG_iso}
\end{equation}
%
%
%
%
%
%
%
%
%


\label{lastpage}
\end{document}